\documentclass[]{elsarticle}
\usepackage[dvipsnames]{xcolor}

\makeatletter
\def\ps@pprintTitle{%
 \let\@oddhead\@empty
 \let\@evenhead\@empty
 \def\@oddfoot{}%
 \let\@evenfoot\@oddfoot}
\makeatother

\usepackage{moreverb}

\usepackage[colorlinks,bookmarksopen,bookmarksnumbered,citecolor=red,urlcolor=red]{hyperref}

\newcommand\BibTeX{{\rmfamily B\kern-.05em \textsc{i\kern-.025em b}\kern-.08em
\kern-.1667em\lower.7ex\hbox{E}\kern-.125emX}}

\usepackage{geometry}
\usepackage{graphicx}
\graphicspath{{figures/}}
\usepackage{subcaption}
\usepackage{algorithm}
\usepackage{algpseudocode}
\usepackage[numbers]{natbib}
\usepackage{import} 
\usepackage{siunitx}
\usepackage[textsize=tiny]{todonotes}
\usepackage{amssymb}
\usepackage{amsthm}
\usepackage{mathtools} 
\usepackage{amsmath}
\usepackage{booktabs}
\usepackage{longtable}
\usepackage{cleveref}
\usepackage{lineno}
\usepackage{txfonts} 
\usepackage{listings} 
\usepackage{lmodern} 
\usepackage{tensorNotation}

\usepackage{lineno}

\usepackage{soul}
\colorlet{Reviewer1}{black}
\colorlet{Reviewer2}{black}

\let\vec\mathbf

\colorlet{Reviewer12}{black}

\colorlet{Reviewer22}{black}

\colorlet{Reviewer32}{black}

\begin{document}

\title{An unstructured geometrical un-split VOF method for viscoelastic two-phase flows}
\author[add1]{Matthias Niethammer} 
\ead{niethammer@tu-darmstadt.de}

\author[add1]{Muhammad Hassan Asghar} 
\ead{asghar@mma.tu-darmstadt.de}

\author[add1]{Dieter Bothe} 
\ead{bothe@mma.tu-darmstadt.de}

\author[add1]{Tomislav Maric\corref{corr}} 
\cortext[corr]{Corresponding author}
\ead{maric@mma.tu-darmstadt.de}

\address[add1]{Mathematical Modeling and Analysis, Technische Universit\"{a}t Darmstadt, Germany}

\begin{abstract}
\textbf{This is the preprint version of the published manuscript \href{https://doi.org/10.1016/j.cpc.2024.109475}{https://doi.org/10.1016/j.cpc\\.2024.109475}: please cite the published manuscript when refering to the contents of this document.}

Since viscoelastic two-phase flows arise in various industrial and natural processes, developing accurate and efficient software for their detailed numerical simulation is a highly relevant and challenging research task. We present a geometrical unstructured Volume-of-Fluid (VOF) method for handling two-phase flows with viscoelastic liquid phase, where the latter is modeled via generic rate-type constitutive equations and a one-field description is derived by conditional volume averaging of the local instantaneous bulk equations and interface jump conditions. The method builds on the plicRDF-isoAdvector geometrical VOF solver that is extended and combined with the modular framework DeboRheo for viscoelastic computational fluid dynamics (CFD). A piecewise-linear geometrical interface reconstruction technique on general unstructured meshes is employed for discretizing the viscoelastic stresses across the fluid interface. DeboRheo facilitates a flexible combination of different rheological models with appropriate stabilization methods to address the high Weissenberg number problem.

\noindent{\bf PROGRAM SUMMARY}
\begin{small}
\noindent
{\em Program Title:} DeboRheo                                         \\
{\em CPC Library link to program files:} (to be added by Technical Editor) \\
{\em Developer's repository link:} \url{https://gitlab.com/deborheo/deborheorelease/} \\
{\em Code Ocean capsule:} (to be added by Technical Editor)\\
{\em Licensing provisions:} GPLv3 \\
{\em Programming language:} C++\\
{\em Nature of problem:} DNS of viscoelastic two-phase flows encounters major challenges due to abrupt changes of physical properties and rheological behaviors of the two phases at the fluid interface, and viscoelastic flows characterized with high Weissenberg numbers introduce additional numerical challenges.\\
{\em Solution method:} A geometrical unstructured Volume-of-Fluid (VOF) method for handling two-phase flows with a viscoelastic liquid phase, where the latter is modeled by generic rate-type constitutive equations. Appropriate stabilization techniques are included to address the High Weissenberg Number Problem (HWNP). 
\end{small}

\end{abstract}


\maketitle

\section{Introduction}\label{sec:intro}

Two-phase flows with at least one viscoelastic phase are encountered in a wide range of industrial and natural processes, including pharmaceutical manufacturing, plastics technology, and food processing. One example is laminar and dispersive mixing operations, such as industrial emulsification processes, where well-defined shear and elongational flows subject large drops to deformation and are fragmented, resulting in the formation of smaller drops \cite{Fischer2007,Windhab2005}. A thorough understanding and accurate simulation of such viscoelastic two-phase flows are crucial for adjusting product quality, ensuring safety properties, and optimizing industrial mixing processes.

A fundamental issue in emulsion technology is the study of a single drop's deformation under shear \cite{Taylor1934,Guido2004,VanPuyvelde2008,Guido2011}, as seen in shear mixing processes where large drops deform and fragment. This phenomenon is critical in predicting drop size distributions and rheological properties of emulsions \cite{Tucker2002}.
These properties hold significant implications for various industrial applications, such as pharmaceutical manufacturing, where they profoundly impact the performance and stability of the final products.\ Similarly, in plastics technology, the properties of paints and varnishes are crucially determined by the deformation and fragmentation behavior of droplets in the emulsions.

In Direct Numerical Simulations (DNS) of viscoelastic two-phase flows, major challenges arise from the abrupt change of physical properties and rheological behaviors of the two fluid phases at the fluid interface, as well as from the proper incorporation of viscoelastic constitutive equations into the computational framework. Additionally, the presence of high Weissenberg numbers, associated with strong elastic components in the viscoelastic phase, introduces additional numerical difficulties~\cite{Joseph1985, Keunings1986} that demand specific numerical treatment~\cite{Niethammer2018}. Furthermore, the deformation of a drop in simple shear flow is characterized by strongly deformed fluid interfaces and potential drop breakup, which can impact the stability and computational efficiency of DNS. Under these conditions, the Volume-of-Fluid (VOF) \cite{Hirt1981} method emerges as an appropriate numerical modeling approach that has been previously employed with success in addressing viscoelastic drop deformation in both 2D \cite{Chinyoka2005} and 3D shear flows \cite{Khismatullin2006,Afkhami2009,Verhulst2009,Figueiredo2016,Lopez-Herrera2019}.\ Moreover, a generic closure for the viscoelastic constitutive equations, supporting various rheological models and stabilization techniques for high Weissenberg numbers, was systematically derived in \cite{Niethammer2019} through conditional volume averaging \cite{Anderson1967, Slattery1967, Whitaker1967}. This framework has demonstrated the ability to provide high-fidelity predictions of viscoelastic two-phase flow phenomena \cite{Bothe2022} and is the foundation for the developments addressed in the present study. 
While the extension of the structured geometrical VOF method to viscoelastic flows has recently successfully been achieved \citep{Lopez-Herrera2019},  this work extends the unstructured Finite Volume framework for complex fluids `DeboRheo' \cite{Niethammer2019, Niethammer2019c} with the unstructured geometrical VOF method `plicRDF-isoAdvector' \cite{Scheufler2019, Roenby2016} from the `TwoPhaseFlow' library \cite{Scheufler2023, Scheufler2022}. This integration results in a new unstructured geometrical VOF approach extended for viscoelastic two-phase flows, which is introduced and validated below.

The VOF method is an interface-capturing method widely employed for direct numerical simulations of two-phase flows, because of its inherent global and local volume conservation and ability to handle strong deformations, coalescence, and breakup of the fluid interface.\ Additionally, the VOF method can be applied to unstructured domain discretization and provides a straightforward parallel computation model.
In the VOF method, the sharp interface between phases is captured by a phase indicator function. The governing equations are given in a one-fluid formulation, utilizing a single set of equations for both phases throughout the entire domain and incorporating appropriate source terms to account for the fluid interface \cite{Tryggvason2005}.

Two types of VOF methods can be categorized based on the specific advection scheme utilized to propagate the sharp interface: 

\textit{Algebraic VOF methods} \cite{Zalesak1979, Muzaferija1999, Ubbink1999} advect the volumetric phase fraction algebraically through numerical schemes by solving a scalar transport equation. Over the last decades, algebraic VOF methods have seen widespread application on general unstructured grids due to their flexibility and high level of serial and parallel computational efficiency. Despite these advantages, the advection schemes of algebraic VOF methods struggle to keep the interface sharp and free of oscillations. The abrupt change of the volume fraction field at the interface causes large interpolation errors and, subsequently, artificial diffusion.

\textit{Geometric VOF methods} \citep{Maric2020} utilize geometrical operations to approximate the solution of the volume fraction advection equation rather than algebraic calculations. Contemporary geometrical VOF methods reconstruct the interface as a piecewise-planar surface within multi-material cells. Subsequently, the reconstructed interface is advected by computing the phase-specific volumes that are fluxed across the boundary of the finite volume.\ Un-split geometrical VOF methods \cite{Maric2020} facilitate the use of unstructured meshes to manage complex geometrical domains and improve the overall solution accuracy, which motivated the extension of the `DeboRheo' computational framework with the `plicRDF-isoAdvector' geometrical VOF method \citep{Scheufler2019}.

The objective of this paper is threefold: (i) to model an extension of the geometrical VOF for viscoelastic fluids; (ii) to demonstrate the numerical stability and robustness of the method at a high Weissenberg number for technically relevant geometries; (iii) to study the deformation of a single drop in shear flow using transient simulations in both 2D and 3D, both for scenarios involving a viscoelastic drop in a Newtonian matrix as well as a Newtonian drop in a viscoelastic matrix.

\section{Mathematical model}\label{sec:model}

The extended VOF equations for viscoelastic fluids can be derived rigorously from the local instantaneous conservation equations in both bulk phases $\Omega^\pm(t)$ and the interfacial jump conditions on the interface $\Sigma(t)$ by Conditional Volume Averaging (CVA). Readers interested in a comprehensive derivation of the extended VOF equations are referred to Niethammer (2019)~\cite{Niethammer2019}. The outcome of this derivation is a one-field formulation, where a single set of equations is valid for both phases across the entire domain. The one-field formulation is summarized below under the assumptions of laminar, isothermal flows with constant mass densities in the bulk phases.

\begin{figure}[tb]
	\centering
	\captionsetup{position=top}
	\def\svgwidth{0.5\textwidth}
	{
		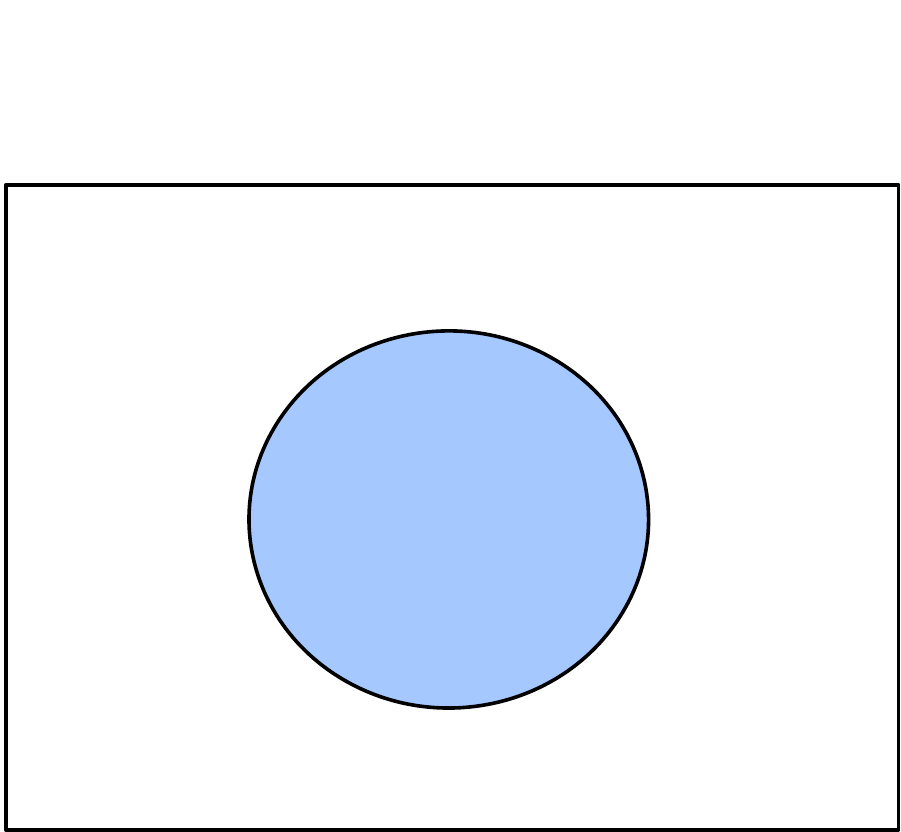
	}
	\vspace{0.5em}
\caption{Schematic diagram of a two-phase system in a domain $\Omega$.}
\label{fig:schematic-diagram-of-multiphase-domain}
\end{figure}

A physical domain, represented by $\Omega \subset \mathbb{R}^3$, is separated by a time-dependent interface, represented by a sharp surface $\Sigma(t)$, into two distinct subdomains (bulk phases) $\Omega^\pm(t)$, as depicted in~\cref{fig:schematic-diagram-of-multiphase-domain}.\ These sub-domains are filled with different types of fluids having constant densities. The interface unit normal pointing towards the phase $\Omega^-(t)$ is denoted as $\vec{n}_\Sigma$. The \textit{phase-indicator function}, 
\begin{equation}
    \chi(\textbf{x}, t) =\begin{cases}
    1 & \textbf{x} \in \Omega^+(t) \\
    0 &  \ \textnormal{otherwise},
\end{cases}
\label{eq:phase-indicator}
\end{equation}
is utilized to distinguish the sub-domains $\Omega^{\pm}$. The sub-domain $\Omega^+$ can therefore be written as
\begin{equation}
    \Omega^+(t) := \{\textbf{x} \in\Omega: \chi(\textbf{x}, t) =1\}.
    \label{eq:indicated-omegaminus}
\end{equation}

The phase-indicator function describes the distribution of the bulk phases in $\Omega$, with its one-sided limits defined on $\Sigma(t)$.

In CVA, the phase-indicator function is applied to the local instantaneous bulk equations for mass and momentum prior to performing volume averaging over the selected volume. This technique requires defining volume averaged quantities, which are introduced below. The volumetric phase fraction, or \textit{volume fraction}, is defined as the volume average of the phase indicator function
\begin{equation}
\alpha(\textbf{x},t) :=\frac{1}{|V|} \int_{V} \chi(\textbf{x}+\textbf{y}, t) \,d\textbf{y}\, .
\label{eq:def-of-volume-fraction}
\end{equation}
In this definition, the centroid of the averaging volume is indicated by the position vector $\textbf{x}$, while the integration over the volume $V$ is performed in terms of the components of the relative position vector $\textbf{y}$.
The volume fraction describes the ratio of the volume $\Omega^+(t)$ occupied by the phase $+$ in the shifted set $(\textbf{x} + V)$ at time $t$ to the total volume of $V$ and is constrained by $0 \leq \alpha \leq 1$. Alternatively, the volume fraction can be defined as the average
\begin{equation}
\alpha(\textbf{x},t) :=\frac{1}{|V|} \int_{(\textbf{x} + V) \, \cap \, \Omega^+(t)} 1 \,d\textbf{y}\, = \frac{\left\vert (\textbf{x} + V) \, \cap \, \Omega^+(t) \right\vert}{|V|}
\label{eq:def-of-volume-fraction-b}
\end{equation}
over the \textit{phase-specific volume} $(\textbf{x} + V) \, \cap \, \Omega^{+}(t)$. From~\cref{eq:def-of-volume-fraction-b} it follows for a scalar or tensor variable $\PhiT(\textbf{x}, t)$ that
\begin{equation}
\begin{aligned}
\frac{1}{|V|} \int_{V} \chi(\textbf{x}+\textbf{y}, t)\, \PhiT(\textbf{x}+\textbf{y}, t) \,d\textbf{y}\,
&=
\frac{|(\textbf{x} + V) \, \cap \, \Omega^+(t)|}{|V|} \frac{1}{|(\textbf{x} + V) \, \cap \, \Omega^+(t)|} \int_{(\textbf{x} + V) \, \cap \, \Omega^+(t)} \PhiT(\textbf{y}, t) \,d\textbf{y}\,\\
&=
\alpha \, \overline{\PhiT(\textbf{x}, t)}^{+}\, ,
\end{aligned}
\label{eq:product-cva}
\end{equation}
where $\overline{\PhiT(\textbf{x}, t)}^{+}$ denotes the intrinsic \textit{phase average} over the volume $(\textbf{x} + V) \, \cap \, \Omega^{+}(t)$. This phase average is defined for both phases according to
\begin{equation}
\overline{\PhiT(\textbf{x}, t)}^{\pm}
=  \frac{1}{|(\textbf{x} + V) \, \cap \, \Omega^\pm(t)|} \int_{(\textbf{x} + V) \, \cap \, \Omega^\pm(t)} \PhiT(\textbf{y}, t) \,d\textbf{y}\, .
\label{eq:phaseaverage}
\end{equation}

With these definitions, the volume-averaged mass and momentum conservation equations can be written as
\begin{align}
\label{mixContieq00}
\nabla \dprod {\velocity} &= 0, \\
\label{mixUeq00}
\partial_t {\rho} {\velocity} 
+ ({\velocity} \dprod \nabla) {\rho} {\velocity}
&=
- {\nabla p}
+ {\nabla \dprod {\stressE}}
+ {\rho} \vec{b}
+ \vec{f}_\sigma,
\end{align}
where $p$ is the pressure, $\vec{b}$ represents external body forces and $\vec{f}_\sigma$ denotes the surface tension forces. Note that the one-field representation \cref{mixContieq00,mixUeq00} is formulated in terms of mixture quantities, namely, the mixture density 
\begin{equation}
\rho = \alpha \rho^+ + (1 - \alpha) \rho^- \, ,
\label{eq:def-rho-cva}
\end{equation}
the mixture velocity
\begin{equation}
\velocity = \frac{\alpha \rho^+ \overline{\velocity}^+\, + \, (1 - \alpha) \rho^-  \overline{\velocity}^-}{\rho} \, ,
\label{eq:def-velocity-cva}
\end{equation}
and the mixture extra stress
\begin{equation}
{\stressE} = {\alpha \overline{\stressE}^+\, + \, (1 - \alpha) \overline{\stressE}^-} \, .
\label{eq:def-stress-cva}
\end{equation}
Note also that the term $\vec{f}_\sigma$ in \cref{mixUeq00} originally is a surface averaged surface tension force as this results from the CVA process together with the jump conditions at the interface. However, in the numerical treatment, we use the CSF model~\cite{Brackbill1992} for the surface tension force, which corresponds to $\vec{f}_\sigma = \sigma \kappa \nabla \alpha$, where $\sigma$ denotes the surface tension and $\kappa$ represents twice the mean curvature.

Additional closures are required for the stress term on the right-hand side of the momentum equation (\ref{mixUeq00}). For the extra stress ${\stressE}$, we employ the solvent-polymer stress splitting, i.e.
\begin{align}
\label{spss01}
{\stressE} = {\stressS} + {\stressP},
\end{align}
where ${\stressS}$ is the solvent stress and ${\stressP}$ corresponds to the polymer stress. The viscous solvent stress obeys Newton's constitutive relation, given by
\begin{align}
\label{soventstress01}
{\stressS} &= 2 {\svisc} \DT = {\svisc} \big(\grad{{\velocity}} + \trans{\grad{{\velocity}}}\big),
\end{align}
with 
\begin{equation}
{\svisc} = \alpha {\eta}^+_s + (1 - \alpha) {\eta}^-_s \, 
\label{eq:def-svisc-cva}
\end{equation}
denoting the mixture solvent viscosity. The \cref{soventstress01,eq:def-svisc-cva} constitute a complete closure for the solvent stress contribution in \cref{spss01}.\ Nonetheless, considering that the stress appears in a stress divergence term in the momentum equation (\ref{mixUeq00}), additional manipulations can enhance the numerical evaluation of this specific term.\ Incorporating \cref{soventstress01,mixContieq00} into \cref{mixUeq00}, the solvent stress divergence can be reformulated according to
\begin{align}
\label{divsoventstress01}
{\nabla \dprod \stressS} &= \nabla \dprod \left({\svisc} \grad{{\velocity}}\right) + \nabla \dprod \left({\svisc} \trans{(\grad{{\velocity}})}\right) \\
\label{divsoventstress02}
&= \nabla \dprod \left({\svisc} \grad{{\velocity}}\right) + \grad{{\velocity}} \dprod \nabla {\svisc},
\end{align}
where the elliptic (Laplacian-type) term in \cref{divsoventstress02} is discretized implicitly within the present work's finite volume framework. Furthermore, it is important to note that the discretization of \cref{divsoventstress02} requires an evaluation of the mixture viscosity ${\svisc}$ at cell faces using the phase indicator approximated by the geometrical VOF method, necessary to ensure numerical stability for phases with strongly different densities~\citep{Liu2023}.\ This scheme enables the computation of the terms in \cref{divsoventstress02} from the face-reconstructed volume fractions instead of the standard interpolation of cell-centered values to faces. Incorporating the Piecewise Linear Interface Calculation (PLIC) reconstruction scheme in evaluating the mixture solvent viscosity at the cell faces ensures a compact computational stencil, reducing potential sources of interpolation errors. This numerical procedure is described in \cref{sec:method}.

The polymer stress $\stressP$ in \cref{mixUeq00} can be represented as a linear function of the polymer conformation tensor $\CT$~\cite{Bird1977, Larson1988}. This relationship can be expressed as
\begin{align}
\label{eq:stressdef02}
\stressP = G \left(\CT - \IT\right),
\end{align}
where the mixture modulus $G$ is defined as $G =  \alpha G^{\phaseOne} + \left(1 - \alpha \right) G^{\phaseTwo}$ with the phase-specific constants 
$G^{\pm} = \frac{\eta_{p}^{\pm}}{\lambda^{\pm}(1 - \zeta^{\pm})}$ and $\IT$ is the unit tensor. Here, $\pvisc$ is the polymer viscosity and $\lambda$ is the relaxation time. 
The `slip' parameter $\zeta$ describing non-affine motion of the polymer chains is given in Table \ref{tab:constitutiveFull}.
The mixture conformation tensor $\CT$ in \cref{eq:stressdef02} is defined as
\begin{equation}
\CT = \frac{\alpha G^+ \overline{\CT}^+\, + \, (1 - \alpha) G^-  \overline{\CT}^-}{G} \, .
\label{eq:def-conformation-cva}
\end{equation}
Note that the conformation tensor is symmetric and positive definite by definition. We exploit the diagonalizability of the conformation tensor, along with its three real and positive eigenvalues, as a key element in the numerical stabilization approach for high Weissenberg numbers. For this reason, the mathematical model is formulated with the conformation tensor rather than the polymer stress.

To describe the behavior of the conformation tensor, the introduction of rheological \textit{constitutive equations} is required.\ In this study, the focus is placed on rate-type partial differential equations~\cite{Bird1977,Larson1988,Oldroyd1950}. Up to this point, the equations presented are applicable to scenarios with two viscoelastic phases. However, for the polymer stress, a closure is provided specifically for the case where only one phase exhibits viscoelasticity. Under this condition, \cref{eq:stressdef02} reduces to
\begin{align}
\label{eq:stressdef03}
\stressP = \alpha \, G^+ \left( \overline{\CT}^+\, - \IT\right).
\end{align}
Then, a generic constitutive equation for $\overline{\CT}^+$ is given by~\cite{Niethammer2019}
\begin{align}
\label{eq:constitutivepolymer01}
\partial_t \overline{\CT}^+ + \velocity \dprod \nabla \overline{\CT}^+ - \alpha \, \overline{\LT} \dprod \overline{\CT}^+ - \alpha \, \overline{\CT}^+ \dprod \trans{\overline{\LT}}
=
\frac{\alpha}{\lambda^+} \cvec{P} ( \overline{\CT}^+ ),
\end{align}
where $\overline{\LT} := \trans{\overline{\nabla \velocity}} - \zeta \overline{\DT}$ and the material parameter $\zeta$ characterizing the non-affine response of polymer chains to an imposed deformation. For $\zeta = 0$ the motion becomes affine and the dericative in (\ref{eq:constitutivepolymer01}) reduces to the upper-convected derivative.

The function $\cvec{P}(\overline{\CT}^+)$ in \cref{eq:constitutivepolymer01} is a real analytic and isotropic tensor function of the conformation tensor.\ As a consequence, we can express $\cvec{P}(\overline{\CT}^+)$ using the well-known tensor representation theorems~\cite{RivlinEricksen1955, Rivlin1955} as
\begin{equation}
\label{Pfunc}
\cvec{P}(\overline{\CT}^+ ) = g_{0} \IT + g_{1} \overline{\CT}^+ + g_{2} {\overline{\CT}^+}^2,
\end{equation}
where the coefficients $g_0$, $g_1$, and $g_2$ in \cref{Pfunc} are isotropic scalar functions of the three principal invariants of the conformation tensor. Thus, the specific expressions for $g_0$, $g_1$, and $g_2$ in \cref{Pfunc} determine the viscoelastic behavior of the material. Table~\ref{tab:constitutiveEx} lists suitable constitutive expressions for the coefficients used in this study, while `DeboRheo' \cite{DeboRheo} offers an extensive list of constitutive expressions listed in~\Cref{tab:constitutiveFull}, that we through this publication make available to the community. Constitutive relations from~\Cref{tab:constitutiveFull} have been extensively verified and validated in our previous works for single-phase flows and two-phase flows with the algebraic VOF method. As a part of our future work, we will validate the combination of DeboRheo and TwoPhaseFlow's geometrical VOF method including more constitutive expressions from~\Cref{tab:constitutiveFull} using in-house experiments.

\begin{table}[h!]
\caption{Scalar coefficients for the generic constitutive \cref{eq:constitutivepolymer01,Pfunc} used in the validation study.}
\label{tab:constitutiveEx}
\begin{center}
\begin{tabular}{@{}lcccccc@{}}
\toprule
constitutive model & \ $g_{0}$ \ & \ $g_{1}$ \ & \ $g_{2}$ \  \\ 
\midrule
Maxwell/Oldroyd-B & $1$ & $-1$ & $0$ \\
Giesekus & $1 - \alpha$ & $2 \alpha - 1$ & $ - \alpha$ &  \\
\bottomrule
\end{tabular}
\end{center}
\end{table}

\begin{table}[h!]
\caption{Scalar coefficients for the generic constitutive \cref{eq:constitutivepolymer01,Pfunc} for selected models available in DeboRheo.}
\label{tab:constitutiveFull}
\begin{center}
\begin{tabular}{ccccc}
\toprule
\textbf{Constitutive Model} & $\zeta$ & $g_{0}$ & $g_{1}$ & $g_{2}$ \\ 
\midrule
Maxwell/Oldroyd-B & 0 & 1 & $-1$ & 0 \\ 
Leonov & 0 & $\frac{1}{2}$ & $\frac{1}{6}(I_{1} - I_{2})$ & $-\frac{1}{2}$ \\ 
Giesekus & 0 & $1 - \alpha$ & $2\alpha - 1$ & $-\alpha$ \\ 
Johnson-Segalman & $\in [0, 2]$ & 1 & $-1$ & 0 \\ 
LPTT & $\in [0, 2]$ & $1 + \frac{\varepsilon}{1 - \zeta}(\text{tr } \mathbf{C} - 3)$ & $-g_{0}$ & 0 \\ 
EPTT & $\in [0, 2]$ & $\exp \left[\frac{\varepsilon}{1 - \zeta}(\text{tr } \mathbf{C} - 3)\right]$ & $-g_{0}$ & 0 \\
\bottomrule
\end{tabular}
\end{center}
\end{table}

\noindent \Cref{eq:stressdef03,eq:constitutivepolymer01,Pfunc} provide a closure for the polymer stress term in the VOF equations, specifically when phase $\Omega^+$ is known to be viscoelastic, while phase $\Omega^-$ is considered purely viscous. The challenge of finding an appropriate closure when both phases are viscoelastic remains, to the best of the authors' knowledge, an unresolved issue in the field of CVA and is not within the scope of the present study.

\section{Stabilization}
\label{subsec:stab}
To address the high Weissenberg number problem (HWNP)~\cite{Joseph1985, Keunings1986}, numerical stabilization techniques are necessary to enhance the numerical stability of the method at moderate and high levels of fluid elasticity.
Fattal and Kupferman~\cite{Fattal2004} demonstrated that using a logarithmic transformation for the conformation tensor equation can circumvent the HWNP. Balci et al.~\cite{Balci2011} proposed a square root representation for the conformation tensor, eliminating the need for diagonalization. Afonso et al.~\cite{Afonso2012} generalized these approaches by introducing generic kernel transformation functions. The present work employs a unified mathematical and numerical stabilization framework proposed by Niethammer et al.~\cite{Niethammer2018, Niethammer2019c} and extended to two-phase flows within an algebraic VOF framework~\cite{Niethammer2019}. This section briefly summarizes the general aspects of 
the stabilization framework in the context of the geometric VOF method.

The generic constitutive equation for the conformation tensor (\cref{eq:constitutivepolymer01}) can be transformed through a change of variable to obtain an evolution equation for a real-analytic tensor function ${\cvec{F}(\overline{\CT}^+)}$. Assuming that ${\cvec{F}(\overline{\CT}^+)}$ is an \textit{isotropic} tensor-valued function of the conformation tensor, the volume averaged change-of-variable representation is given by
\begin{equation}
\label{evolutionConf3}
\partial_t {\cvec{F}(\overline{\CT}^+)}
+ (\velocity \dprod \nabla) {\cvec{F}(\overline{\CT}^+)}
=
\alpha \left(
{
{2 \BT \dprod \overline{\UpsilonT}^+ \dprod \overline{\CT}^+}
+
\OmegaT \dprod {\cvec{F}(\overline{\CT}^+)} - {\cvec{F}(\overline{\CT}^+)} \dprod \OmegaT
+
\frac{1}{\lambda} \overline{\UpsilonT}^+ \dprod \cvec{P} (\overline{\CT}^+)
}
\right).
\end{equation}
A comprehensive derivation of this representation (\cref{evolutionConf3}) can be found in~\cite{Niethammer2019}. In \cref{evolutionConf3}, the deformation terms present in the convective derivative are decomposed into the first three terms on the right-hand side, employing the tensors $\BT$ and $\OmegaT$.\ This local decomposition was first proposed in~\cite{Fattal2004} and has been applied to the volume-averaged VOF equations in~\cite{Niethammer2019}.\ Such a decomposition is generally necessary for transforming the constitutive equation for the conformation tensor equation (\ref{eq:constitutivepolymer01}) into an evolution equation for the real-analytic tensor function ${\cvec{F}(\overline{\CT}^+)}$ by a change of variable.\ The tensors $\BT$ and $\OmegaT$ in \cref{evolutionConf3} are defined as mixture quantities that are dependent on the gradient of the mixture velocity. Their functional relationship with the velocity gradient is presented in \ref{asec:definitionBOmega}.

The generic constitutive equation (\ref{evolutionConf3}) reduces to the classical logarithm or root conformation representations (LCR or RCR for short), depending on the choice of ${\cvec{F}(\overline{\CT}^+)}$ and $\overline{\UpsilonT}^+$.\ Table~\ref{tab:JJ} shows the specific terms corresponding to different conformation representations.
\begin{table}[htbp]
\caption{Function ${\cvec{F}(\overline{\CT}^+)}$ and $\UpsilonT$ terms in \cref{evolutionConf3} for the conformation tensor representation, the $k$-th root of $\CT$ representation, and the logarithm of $\CT$ to base $a$ representation.}
\renewcommand{\arraystretch}{1.35}
\begin{center}
\linespread{1.15}
\begin{tabular}{rlll}
\toprule
\ & $\cvec{F}(\CT)$ & \ $\UpsilonT$ \ \\
\midrule
conformation tensor \ & $\CT$ & \ $\IT$ \ \\
$k$-th root of $\CT$ \ & $\RT = \QT \dprod \DT^{\frac{1}{k}} \dprod \trans{\QT}$ & \ $\frac{1}{k} \RT^{1-k}$ \ \\
logarithm of $\CT$ to base $a$ \ & $\ST = \QT \dprod \loga{(\DT)} \dprod \trans{\QT}$ & \ $\frac{1}{\operatorname{ln}(a)} a^{-\ST}$ \ \\
\bottomrule
\end{tabular}
\end{center}
\renewcommand{\arraystretch}{1.0}
\label{tab:JJ}
\end{table}
In the present work, we focus on the RCR, which reduces \cref{evolutionConf3} to a more concise form according to:
\begin{align}
\label{evolutionConfRTb}
\DDt \RT
&=
\frac{2}{k} \BT \dprod \RT
+
\OmegaT \dprod {\RT} - {\RT} \dprod \OmegaT
+
\frac{1}{k \lambda} \left(g_0 \RT^{1-k} + g_1 \RT + g_2 \RT^{1+k}\right).
\end{align}
For the scope of the present study, the square root formulation with $k=2$ turned out to be stable across a range of Deborah numbers from $0$ to $16$.
\section{Numerical method}\label{sec:method}

We utilize the unstructured geometrical VOF method~\citep{Maric2020} because its discretization of the fluid interface consistently matches with the discretization of \cref{eq:stressdef03}.

\subsection{The geometrical Volume-of-Fluid method}
\label{subsec:geoVOF}

The flux-based geometrical VOF method updates volume fractions defined in \cref{eq:def-of-volume-fraction} by integrating the flux of the phase $\Omega^+(t)$ through the boundary  $\partial\Omega_c$ of the fixed control volume $\Omega_c$ over the time step $[t^n, t^{n+1}]$, namely
\begin{equation}
    \alpha_{c}(t^{n+1}) = 
        \alpha_{c}(t^n) - \frac{1}{|\Omega_c|} \int_{t^n}^{t^{n+1}}
            \int_{\partial\Omega_c}
                \chi(\textbf{x},t)  \mathbf{v}\cdot\mathbf{n}
            ~dS
        ~dt.
\label{eq:volfraction-integral}
\end{equation}
The reader is directed to~\citep{Maric2020} for a derivation of \cref{eq:volfraction-integral}. Let us also note that we have chosen to use $\Omega_c$ for the shifted averaging volume to have a closer connection to the Finite Volume notation. Moreover, as we only use a single centered value $\alpha_c$ for each cell as opposed to a full field of the averaged quantity, no spatial variable is left in $\alpha_c$ in \cref{eq:volfraction-integral} and below. The boundary $\partial\Omega_c$ of the cell $\Omega_c$ in \cref{eq:volfraction-integral} is a union of surfaces (faces) bounded by line segments (edges), namely
\begin{equation}
    \partial\Omega_c := \bigcup_{f\in F_c} S_f.
    \label{eq:cellboundary}
\end{equation}
Using \cref{eq:cellboundary}, \cref{eq:volfraction-integral} is reformulated according to 
\begin{equation}
    \alpha^{n+1}_{c} = \alpha^n_{c} - \frac{1}{|\Omega_c|} \sum_{f \in F_{c}} \int_{t^n}^{t^{n+1}} \int_{S_f} \chi(\textbf{x},t)  \mathbf{v}\cdot\mathbf{n}~dS~dt .
\label{eq:volfrac-final}
\end{equation}
\Cref{eq:volfrac-final} is an exact identity provided that $\partial \Omega  = \cup_{c \in C} (\partial \Omega \bigcap \partial \Omega_c)$, and because \cref{eq:volfrac-final} does not involve any approximations.\ It is important to note that \cref{eq:volfrac-final} utilizes the sharp phase indicator $\chi(\mathbf{x},t)$ to obtain $\alpha^{n+1}_c:=\alpha_c(t^{n+1})$.\ Using $\chi(\mathbf{x},t)$ is advantageous for discretizing the equations for two-phase viscoelastic flows because it is also used for the conditional volume averaging in the derivation of the single-field formulation of two-phase Navier-Stokes equations for viscoelastic flows~\citep{Niethammer2019}. 

The second-order accurate Unstructured Finite Volume Method (UFVM)~\citep{Jasak1996error} utilizes second-order accurate centroid averages 
\begin{equation}
    \phi_D = \frac{1}{|D|}\int_D \phi(\mathbf{x}) dx = \phi(\mathbf{x}_c) + \frac{1}{|D|}\int_D \nabla\nabla \phi : (\mathbf{x} - \mathbf{x}_c)\otimes (\mathbf{x} - \mathbf{x}_c) \, d\textbf{x}
\end{equation}
where $D$ can be the control volume $\Omega_c$ or one of its faces $S_f$, in which case the integral is understood as a surface integral, i.e., $dx$ is the surface measure $dS$, and $x_c$ is their centroid. 
Therefore, velocity and the face-area-normal vector in \cref{eq:volfrac-final} are averaged, resulting in 
\begin{equation}
\begin{aligned}
    \alpha^{n+1}_{c} & = \alpha^n_{c} - \frac{1}{|\Omega_c|} \sum_{f \in F_{c}} \int_{t^n}^{t^{n+1}}  \mathbf{v}_f \cdot \frac{\mathbf{S}_f}{|\mathbf{S}_f|} \int_{S_f} \chi(\textbf{x},t)~dS~dt + e_{\alpha_h}(h^2) \\
    & = \alpha_{c}(t^n) - \frac{1}{|\Omega_c|} \sum_{f \in F_{c}} \int_{t^n}^{t^{n+1}}  F_f(t) |A_f(t)| ~dt + e_{\alpha_h}(h^2)
\label{eq:volfrac-fvm}
\end{aligned}
\end{equation}
with $|A_f|:=\int_{S_f} \chi(\textbf{x},t)~dS$ defining the phase-specific face area, and $F_f:=\int_{S_f} \mathbf{v}_f \cdot \mathbf{n} dS$ is the volumetric flux. The integral on the r.h.s.~of \cref{eq:volfrac-fvm} represents the volume of the phase $\Omega^+$ (\emph{phase-specific volume}) fluxed through the face $S_f$ over the time interval $[t^n, t^{n+1}]$. 

Each flux-based unstructured geometrical VOF method computes the fluxed phase-specific volume differently~\citep{Maric2020}; we use specifically the plicRDF-isoAdvector method~\citep{Scheufler2019} because of its computational efficiency.

\subsection{The plicRDF-isoAdvector Volume-of-Fluid method}
\label{subsec:plic}

The plicRDF-isoAdvector method~\citep{Scheufler2019} approximates the phase indicator function with $\tilde{\chi}(\mathbf{x},t)\approx \chi(\mathbf{x},t)$ as 
\begin{equation}
    \tilde{\chi}_c(\mathbf{x},t) = \begin{cases}
        1, \quad P_c(\mathbf{x},t) \le 0, \\
        0, \quad P_c(\mathbf{x},t) > 0,
    \end{cases}
    \label{eq:chitilde}
\end{equation}
with
\begin{equation}
    P_c(\mathbf{x},t) = a_{c,0}(t) + a_{c,1}(t) x + a_{c,2}(t) y + a_{c,3}(t) z,
    \label{eq:plane}
\end{equation}
defined in such a way that, given $\mathbf{n}_c = [a_{c,1}, a_{c,2},a_{c,3}]$, the value $a_{c,0}$ at some time $t$ is found that satisfies
\begin{equation}
    \alpha_c(t) = \frac{1}{|\Omega_c|}\int_{\Omega_c}\tilde{\chi}(\mathbf{x},t) \, dV,
\end{equation}
the so-called interface positioning problem that ensures volume conservation~\citep{Maric2021,Kromer2022} of the phase-indicator function approximation. The plicRDF-isoAdvector~\citep{Scheufler2019} relies on the reconstructed distance function to iteratively improve the initial estimate of $\mathbf{n}_c$.\ Cell-specific $P_c(\mathbf{x},t)$ makes the approximation of the fluid interface piecewise linear.

\begin{figure}[htb]
    \centering

    \begin{minipage}{0.49\textwidth}
        \begin{subfigure}{\columnwidth}
	\includegraphics[width=\columnwidth]{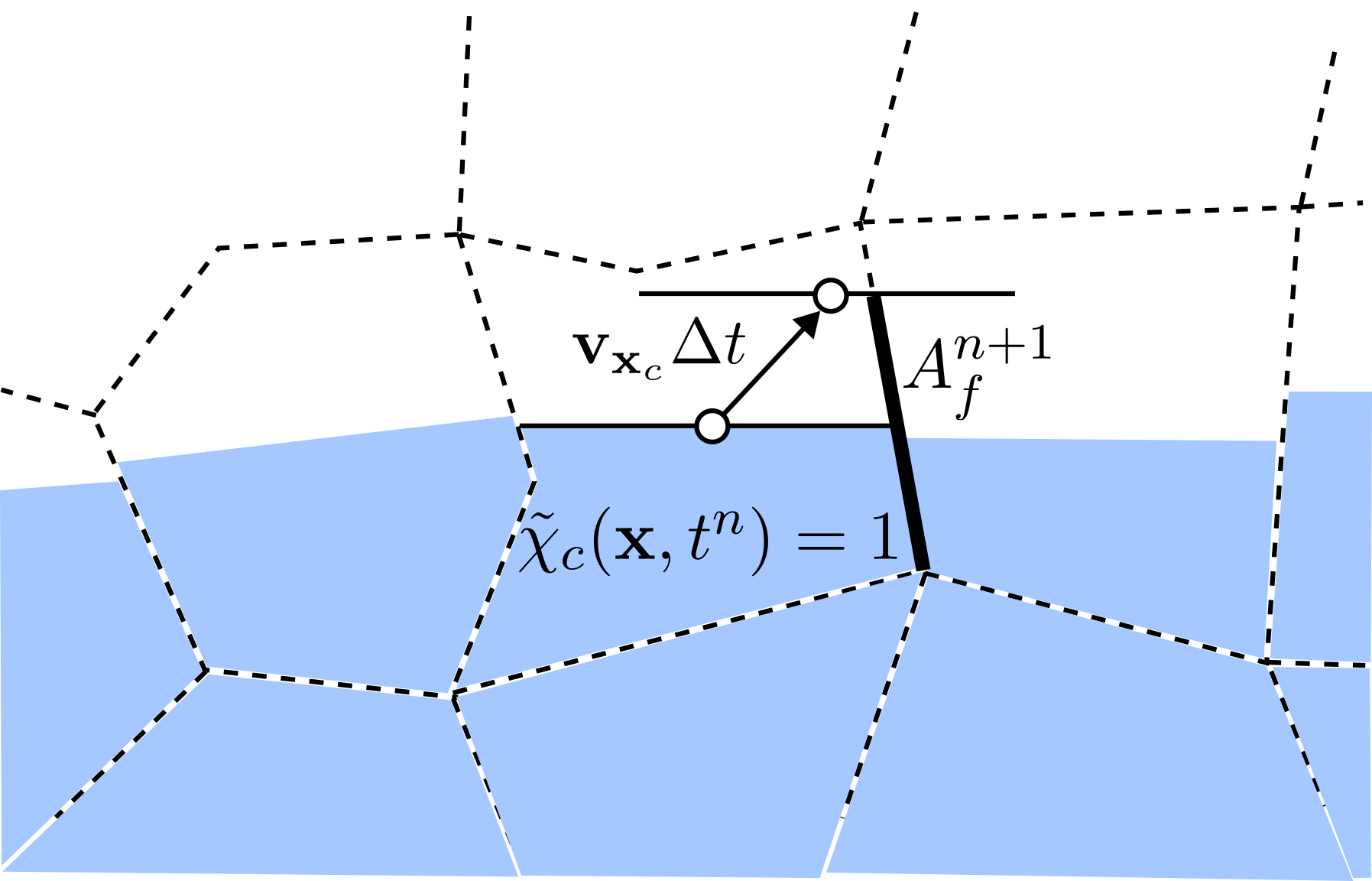}
    \subcaption{Upwind $A_f^{n+1}$ computed in \cref{eq:plic-isoadvector} during $\alpha_c^{n+1}$ update with $\mathbf{n}_c^n$.}
    \label{fig:afisoadvect}
 \end{subfigure}
    \end{minipage}\hfill
    \begin{minipage}{0.49\textwidth}
         \begin{subfigure}{0.9\columnwidth}
	\includegraphics[width=\columnwidth]{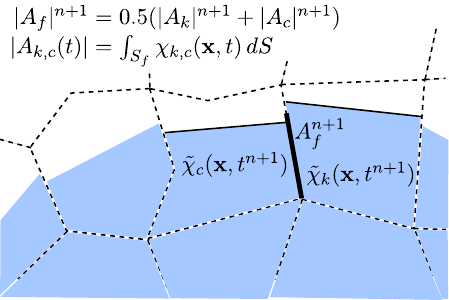}
     \subcaption{$A_f^{n+1}$ reconstructed using $\mathbf{n}_c^{n+1}$ and $\alpha_c^{n+1}$.}
     \label{fig:afrecon}
 \end{subfigure}
    \end{minipage}

    \caption{Different ways to compute $A_f^{n+1}$ for $\alpha_f$ weighting of the polymer stress $\tau_p$ in the pressure Poisson equation (\ref{eq:peqn}).}
\label{fig:af}
\end{figure}

The plicRDF-isoAdvector further approximates \cref{eq:volfrac-fvm} into
\begin{equation}
\begin{aligned}
    \alpha^{n+1}_{c} 
    & \approx \alpha_{c}(t^n) - \frac{1}{|\Omega_c|} \sum_{f \in F_{c}} 0.5 (F_f^n + F_f^{n+1}) \int_{t^n}^{t^{n+1}}  |A_f(t)| ~dt,
\label{eq:plic-isoadvector}
\end{aligned}
\end{equation}
computing $\int_{t^n}^{t^{n+1}}  |A_f(t)| ~dt$ geometrically, by approximating $|A_f|^{n+1}$ by translating the plane $P_c(\mathbf{x},t^n)$ along $n_c^n$ with the displacement given by $\mathbf{v}_{\mathbf{x}_c} \Delta t$, with the time step $\Delta t := t^{n+1} - t^n$, as schematically shown in \cref{fig:afisoadvect}.\ This dimensional reduction in computing the fluxed phase-specific volume via $\int_{t^n}^{t^{n+1}}  |A_f(t)| ~dt$ instead of reconstructing the 3D fluxed phase-specific volume~\citep{Maric2018,Maric2020} ensures a high degree of computational efficiency for the plicRDF-isoAdvector method.


\subsection{Solution algorithm}
\label{subsec:solution}

The key to simulating a viscoelastic fluid phase with the geometrical VOF method is \cref{eq:stressdef03} and its use in the Poisson equation for the pressure in the segregated solution algorithm of \cref{mixUeq00,mixContieq00,eq:volfrac-fvm} and \cref{evolutionConf3} with a representation of the conformation tensor selected from \cref{tab:JJ}. Without diving into a detailed derivation of segregated solution algorithms \textcolor{Reviewer22}{(cf. ~\citep{Tolle2020,Liu2023} for details), we focus on the role of the viscoelastic stress in the solution algorithm for the one-field formulation of Navier-Stokes equations for two-phase flows. We start with the discretized one-field formulation of the momentum conservation equation \cref{mixUeq00} and omit the time step (outer iteration) information for conciseness, giving 
\begin{equation}
  a_c \mathbf{v}_c + \sum_{n \in N_c} a_n \mathbf{v}_n = - (\nabla p)_c - (\nabla \rho)_c (\mathbf{g} \cdot \mathbf{x})_c + (\nabla \cdot \mathbf{\tau}_p)_c + (\mathbf{f}_\sigma)_c.
  \label{eq:momdiscr}
\end{equation}
The divergence of the polymer stress $\nabla\cdot{\tau}_p$ from \cref{eq:stressdef03} that appears on the right-hand side of the momentum conservation equation \cref{mixUeq00}, integrated by the unstructured Finite Volume method over the control volume $\Omega_c$, becomes
\begin{equation}
\begin{aligned}
   \int_{\Omega_c} \nabla \cdot \tau_p \, dV 
   & = \sum_{f \in F_c} \int_{S_f} \chi \tau_p^- \cdot \mathbf{n} \, dS 
   = \sum_{f \in F_c} \frac{|\mathbf{S}_f|}{|\mathbf{S}_f|} (\tau_p^- \cdot \mathbf{n}_f) \int_{S_f} \chi  \, dS + O(h^2) \\
  &  = 
  \sum_{f \in F_c} (\tau_p^- \cdot \mathbf{S}_f)  \frac{1}{|\mathbf{S}_f|} \int_{S_f} \chi  \, dS + O(h^2)
  = \sum_{f \in F_c} (\tau_p^- \cdot \mathbf{S}_f)  \alpha_f + O(h^2).
\end{aligned}
\label{eq:divtau}
\end{equation}
The final term in \cref{eq:divtau} represents the phase-specific volumetric force density resulting from the fact that we only consider one viscoelastic phase. Another discrete divergence of this term further appears in the derivation of the discrete pressure Poisson equation, derived by dividing \cref{eq:momdiscr} by $a_c$ and taking the discrete unstructured FV divergence of the cell-centered values from \cref{eq:momdiscr}, i.e. $\nabla_c \cdot \Psi \approx \frac{1}{|\Omega_c|} \sum_{f \in F_c} \Psi \cdot \mathbf{S}_f$. The discrete pressure Poisson equation for the coupling between the viscoelastic flow and two-phase flows then becomes
\begin{equation}
\begin{aligned}
    \sum_{f \in F_c} \left( \frac{1}{a_c} \right)_f (\nabla p)_f \cdot \mathbf{S}_f 
     & =  \sum_{f \in F_c}  \frac{1}{(a_c)_f} \left[\sum_{f' \in F_k}  \alpha_f' (\tau_p)_f' \cdot \mathbf{S}_f'\right]_f \cdot\mathbf{S}_f + 
     \sum_{f \in F_c}  \frac{1}{(a_c)_f} 
     \mathbf{H}(F_f, \velocity_f)_f
     \cdot\mathbf{S}_f \\
     & + 
      \sum_{f \in F_c} \velocity_f \cdot \mathbf{S}_f
      + \sum_{f \in F_c} \mathbf{S}_p(\velocity_c)_f \cdot \mathbf{S}_f.
    \label{eq:peqn}
\end{aligned}
\end{equation}}
\textcolor{Reviewer22}{The first term on the r.h.s. of \cref{eq:peqn} we compute using geometrically evaluated area fractions $\alpha_f'$. The index $f'$ ensures that the internal divergence term is computed first, for all cells $\Omega_k$, and then interpolated to faces $f$.}
The consistency between $\alpha$ and $\alpha_c$ when the finite volume $\Omega_c$ is used for conditional volume averaging makes $\alpha_f$ in \cref{eq:peqn} defined as 
\begin{equation}
    \alpha_f := \frac{1}{|S_f|} \int_{S_f} \chi(\mathbf{x},t) \, dS = \frac{|A_f(t)|}{|S_f|},
\end{equation}
numerically consistent with the geometrical VOF method. Although the plicRDF-isoAdvector method relies only on $|A_f|$ to update $\alpha_c^{n+1}$, other geometrical VOF methods that compute $|A_f|$ as a part of the approximation of the fluxed phase-specific volume in \cref{eq:volfrac-fvm} are equivalently numerically consistent with \cref{eq:stressdef03}. 

The segregated solution algorithm is outlined in \cref{Algorithm} and it ensures \cref{mixUeq00,mixContieq00,eq:volfrac-fvm,evolutionConf3} are satisfied at $t^{n+1}$, therefore, we require $\alpha_f^{n+1} = \frac{|A_f^{n+1}|}{|S_f|}$ in \cref{eq:peqn}. While uniquely defined for each $S_f$, the obvious choice of the upwind $|A_f^{n+1}|$ from the plicRDF-isoAdvector \cref{eq:plic-isoadvector} is not the right choice (cf. \cref{fig:afisoadvect}).\ Namely, when the plicRDF-isoAdvector method computes $|A_f^{n+1}|$ in \cref{eq:plic-isoadvector}, it does so by translating the interface plane given by \cref{eq:plane} in the direction of the interface-normal at the \emph{old time step}, namely $\mathbf{n}_c^n$, thus computing $A_f^{n+1}$ as a first-order accurate estimate in time, which is acceptable because in \cref{eq:plic-isoadvector} the integration over $\Delta t$ recovers second-order accuracy in time. However, upwind $A_f^{n+1}$ results from the old interface normal $\mathbf{n}_c^n$; only after $\alpha_c^{n+1}$ has been updated using \cref{eq:plic-isoadvector}, and the interface has been again reconstructed at $t^{n+1}$ by optimizing $\mathbf{n}_c^{n+1}$ and positioning the interface by finding $a_{c,0}$, do the area fractions $\alpha_f^{n+1}$ correspond to the phase indicator approximation $\tilde{\chi}_c(\mathbf{x},t)$ at $t^{n+1}$.\ Since the solution algorithm must satisfy equation \cref{eq:peqn} at $t^{n+1}$,  $|A_f^{n+1}|$ should be defined by $\chi_{c,k}(\mathbf{x},t^{n+1})$, where $c,k$ denote indices of cells that share $S_f$. We therefore advect $\alpha_c^{n+1}$ using the plicRDF-isoAdvector scheme to solve \cref{eq:plic-isoadvector}, then reconstruct the interface to obtain $\tilde{\chi}_c(\mathbf{x},t^{n+1})$, as shown schematically in \cref{fig:afrecon}.\ The issue in computing the reconstructed $|A_f^{n+1}|$ is the piecewise planarity of $\tilde{\chi}$, resulting in two candidate values from cells $\Omega_{k,c}$, namely
\begin{equation}
    |A_{k,c}(t)| = \int_{S_f} \tilde{\chi}_{k,c}(\mathbf{x}, t) \, dS,
\end{equation}
and we solve the issue of uniqueness of $A_f^{n+1}$ by averaging 
\begin{equation}
    |A_f^{n+1}| = 0.5(|A_k|^{n+1} + |A_c|^{n+1}),
\end{equation}
ensuring sufficient mesh resolution for the plicRDF-isoAdvector scheme to reduce the discontinuity of $\tilde{\chi}$ on $\partial \Omega_c$.

\begin{algorithm}[!htb]
\caption{Solution Algorithm} \label{Algorithm}
\begin{algorithmic}[1]
\For{$o = 1$  to $N_{\text{outer\_iterations}}$}
    \State Solve the conformation tensor equation (\ref{evolutionConf3}) for $(\tau_p)^o_c$.
    \State Assemble (solve) the momentum equation (\ref{mixUeq00}) for $\mathbf{v}_c^o$.
    \State Solve the volume fraction equation (\ref{eq:plic-isoadvector}).
    \State Reconstruct the interface to obtain $\tilde{\chi}_c^{o}$.
    \State Compute reconstructed area fractions $\alpha_f^{o}$.
    \State Assemble the viscoelastic stress term (\ref{eq:stressdef03}) for the pressure equation (\ref{eq:peqn}).
    \For{$i = 1$  to  $N_{\text{inner\_iterations}}$}
        \State Solve the pressure equation (\ref{eq:peqn}).
        \State Update the cell-centered velocity $\mathbf{v}_c^i$ and flux $F_f^i$.
    \EndFor
\EndFor
\end{algorithmic}
\end{algorithm}


\textcolor{Reviewer22}{The main advantage of this approach is the sharp representation of the fluid interface provided by \cref{eq:chitilde}, which remains sharp in combination with the volume fraction $\alpha_c$ as both are advected by the geometrical VOF method. Algebraic methods tend to diffuse volume fractions, which leads to an artificial numerical diffusion of the interface region. This artificial diffusion is suppressed in algebraic VOF methods by a counter-diffusion term, making the governing equation model inconsistent with the one-field formulation of Navier-Stokes equations for incompressible two-phase flows without phase change, which is built upon a mathematically sharp phase indicator. In other words, a non-zero compression adds a diffusion term to the volume fraction equation, which is not present when this equation is derived from the mass conservation law. By combining DeboRheo with the geometrical VOF method, we provide a numerical methodology and an open-source software for simulating viscoelastic two-phase flows with a higher degree of accuracy.} 

\textcolor{Reviewer22}{It is crucial for this approach that the geometrical VOF method provides a second-order convergent reconstruction \citep{Maric2020}, which minimizes the discrepancy between cell-centered interface normals, bringing area-fractions from adjacent cells close to each other, and significantly reducing the errors when computing the unique area fraction $|A_f^{n+1}| = 0.5(|A_k|^{n+1} + |A_c|^{n+1})$ used to weight the contribution of the polymer stress. For the calculation of $A_f$, it is imperative to perform one additional reconstruction of $\tilde{\chi}^{n+1}$ after advecting for $\alpha_c^{n+1}$ (step 5 in \cref{Algorithm}). We are discretizing the momentum equation using a segregated solution algorithm, which relies on outer iterations to facilitate a coupled solution of the mass (volume) conservation, momentum conservation, and the pressure Poisson equations (more details are provided in \citep{Tolle2020}). The consequence of the segregated solution approach is the need to evaluate the divergence of the polymer stress on the r.h.s. for the outer iteration, that eventually converges to $n+1$, hence the need for a unique $A_f^{n+1}$ at face centers and an additional interface reconstruction. In our simulations, we have not observed any significant computational expense increase by adding the reconstruction, as it only requires a fraction of a percent w.r.t. the solution of one-field Navier-Stokes equations.}

\section{Results}\label{sec:results}

\subsection{2D drop deformation in simple shear flow}
A two-dimensional single drop deforming under simple shear flow is studied as an initial test of the method.\ This flow was studied numerically by several researchers, see for example~\cite{Pillapakkam2001, Chinyoka2005, Chung2008}, for different combinations of the droplet's material properties and the surrounding fluid matrix.\ Here, we are interested in the configurations drop/matrix: Newtonian/Newtonian (NN), Newtonian/viscoelastic (NV), and viscoelastic/Newtonian (VN).

\subsubsection{Problem Description}
\begin{figure}[h!]
\centering
    {
        \includegraphics[width=.5\textwidth]{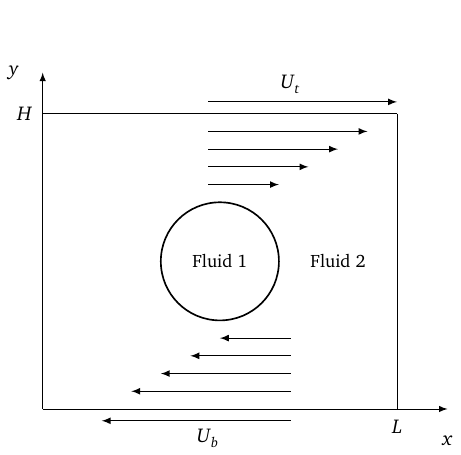}
    }
\caption{Schematic diagram of the two-dimensional drop deformation in shear flow.}
\label{fig:sketchSD}
\end{figure}

A two-dimensional circular drop (fluid 1) is immersed in the center of a rectangular domain, as shown in \cref{fig:sketchSD}.\ 
The channel height equals $H = 8 r$, where $r$ is the initial drop radius. The horizontal channel length is $L = 20 r$. Periodic boundary conditions are imposed at the two outer channel boundaries.\ At the channel walls, Dirichlet boundary conditions are prescribed with a constant value for the velocity.\ The (t)op and (b)ottom walls move with a speed $U_t = -U_b$ to generate a shear flow. For the stress and the generic constitutive variable, a zero gradient boundary condition is defined for the channel walls. The simulation is initialized by a fully developed shear flow velocity field, while the other fields are initially zero.

The total viscosities of the fluid $i = 1,2$ are defined as $\eta_i = \svisc{_i} + \pvisc{_i}$, where $\svisc{_i}$ is the solvent dynamic viscosity and $\pvisc{_i}$ the polymer viscosity. The relaxation time is denoted as $\lambda_i$, the density $\rho_i$, and the surface tension is $\sigma$.\ The constant shear rate reads $\dot{\gamma} = (U_t - U_b)/H$. With these definitions, the system is governed by the following dimensionless numbers: Reynolds number $\operatorname{Re}$, Deborah number $\operatorname{De}$, Capillary number $\operatorname{Ca}$ and viscosity ratio $\Pi$ of drop (fluid 1) to the matrix (fluid 2) defined by
\begin{align}
\label{eq:dimensionlessnumbers}
\operatorname{Re} = \frac{\rho_2 \dot{\gamma} r^2}{\eta_2}, \quad
\operatorname{De} = \lambda \dot{\gamma}, \quad
\operatorname{Ca} = \frac{\eta_2 \dot{\gamma} r}{\sigma}, \quad
\Pi = \frac{\eta_1}{\eta_2}.
\end{align}
For the present study, the parameters read $\operatorname{Re} = 0.0003$, $\Pi = 1$ and for the respective viscoelastic phase $\svisc{_i} = \pvisc{_i}$. The Deborah number and the Capillary number are varied. The drop deformation is studied by means of the deformation parameter~\cite{Taylor1934}
\begin{equation}
\label{eq:deformationparameter}
D = \frac{r_{\textnormal{max}} - r_{\textnormal{min}}}{r_{\textnormal{max}} + r_{\textnormal{min}}},
\end{equation}
where $r_{\textnormal{max}}$ and $r_{\textnormal{min}}$ are the largest and shortest distances from the drop center to its boundary, respectively.\ The orientation angle $\theta$ is defined as the angle between the principal axes of the deformed drop and the flow direction.

\subsubsection{Results and discussion}
\begin{figure}[h!]
    \centering
    \includegraphics[height=0.4\textwidth]{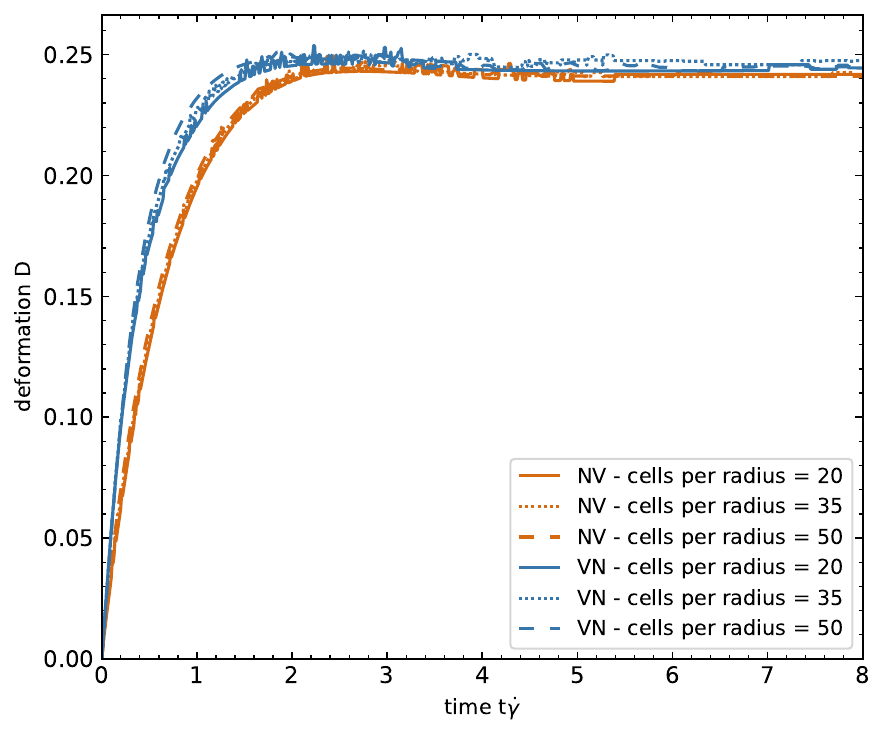}
    \caption{Drop deformation as a function of time with varying mesh refinement for the VN and NV systems at $\operatorname{Ca}=0.24$ and $\operatorname{De}=0.4$.
    }
    \label{fig:meshConvergence2D}
\end{figure}

A mesh convergence study is carried out to verify the algorithm and determine the required mesh cells per initial drop radius to achieve mesh-independent results.
\Cref{fig:meshConvergence2D} displays the deformation for various mesh resolutions over time. The coarsest mesh configuration has $20$ cells per drop radius, totaling $64000$ cells. In contrast, the finest mesh consists of $50$ cells per drop radius, summing up to $400000$ cells. The VN system appears to have a more evident mesh sensitivity than the NV system. However, convergence towards the finest resolution is achieved for both systems.\ A steady state solution is reached by the time $t\dot{\gamma} = 8$. Using the densest mesh as the reference solution, the deformation error on the coarsest configuration, with $20$ cells per drop radius, remains under $1 \%$ in the steady state. This can be seen from Table~\ref{tab:resolutionDTheta}, which presents the deformation parameter across different mesh resolutions. During the transient deformation phase, the maximum error exceeds that of the steady state. For subsequent analyses, a mesh resolution of $50$ cells per drop radius is selected to ensure the highest accuracy during transient and steady state deformations.
\begin{table}[h!]
    \centering
\caption{Comparison of drop deformation $D$ and orientation angle $\theta$ at time $t = 8~\dot{\gamma}^{-1}$. NV and VN systems for $\operatorname{Ca} = 0.24$ and $\operatorname{De} = 0.4$ on different mesh resolutions.}
\begin{tabular}{@{}lllllllll@{}}
\toprule
   & \multicolumn{2}{l}{cells per radius $ = 20$} & & \multicolumn{2}{l}{cells per radius $ = 35$} & & \multicolumn{2}{l}{cells per radius $ = 50$} \\ \cmidrule(lr){2-3} \cmidrule(lr){5-6} \cmidrule(l){8-9} 
   & $D$         & $\theta$        &  & $D$                              & $\theta$ &  & $D$                           & $\theta$ \\ \midrule
NV & 0.2418       & 25.36          &  & 0.2408                           & 25.14    &  & 0.241                        & 25.97    \\
VN & 0.2444       & 33.3          &  & 0.2473                           & 31.21    &  & 0.2448                        & 32.10    \\ \bottomrule
\end{tabular}
\label{tab:resolutionDTheta}
\end{table}

\begin{figure}[h!]
    \centering
    \includegraphics[height=0.4\textwidth]{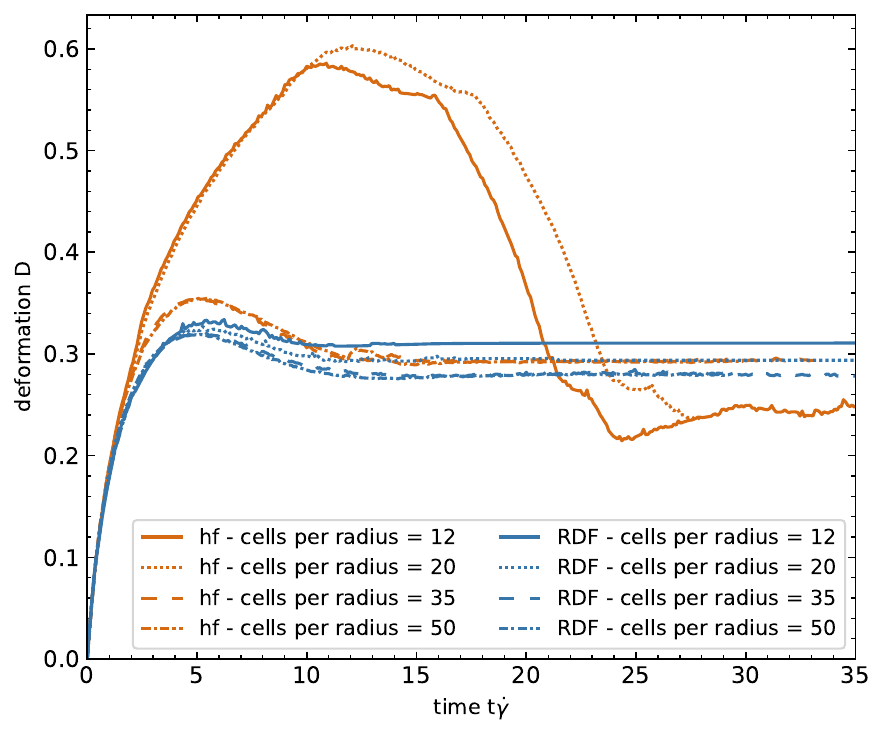}
\caption{Drop deformation as a function of time for the VN system at $\operatorname{Ca}=0.6$ and $\operatorname{De}=0.4$ with the \textit{height function} (hf) and \textit{RDF} curvature models.}
    \label{fig:stm_comparison}
\end{figure}

A further mesh convergence study is performed using two different curvature models: \textit{RDF} and \textit{height function}, both of which are available in the TwoPhaseFlow OpenFOAM project~\cite{Scheufler2022}. More details on these models can be found in~\cite{Scheufler2023}. Figure~\ref{fig:stm_comparison} shows a comparison of the drop deformation as a function of time, using these curvature models across various mesh resolutions. The findings suggest that the \textit{RDF} curvature model offers superior convergence and accuracy compared to the \textit{height function}. As a result, the \textit{RDF} curvature model is selected for use in this research. Recall that the height function approach is highly accurate for Cartesian meshes, its application to a general unstructured mesh, as used here, significantly complicates the implementation. In its current state, the \textit{RDF} curvature model is slightly more accurate, but we intend to enhance the accuracy of the height function implementation in future developments.
%


The results from the finest mesh with 50 cells per drop radius are validated against simulation data from the literature in Table \ref{tab:cpSD}. This comparison covers the Newtonian drop in a Newtonian matrix as well as the viscoelastic NV and VN scenarios for $\operatorname{Ca} = 0.24$ and $\operatorname{De} = 0.4$. The data refer to the time step at $t = 3\,\dot{\gamma}$, at which a steady state is not yet reached completely in all configurations, which can be seen from the comparison with Table \ref{tab:resolutionDTheta}. Notably, there are more substantial discrepancies among all sources at $t = 3\,\dot{\gamma}$ for the NV and VN systems than for the Newtonian case. The results of the present study indicate a reduced drop deformation in the viscoelastic cases compared to the Newtonian case, which is consistent with the existing literature. 
The smallest deformation parameter $D$ is observed in the NV combination, confirming previous research findings. Moreover, the values of the deformation parameter closely correspond with those reported in~\cite{Chung2008} across all the systems: NN, NV, and VN, albeit slightly smaller in our case.
The orientation angle $\theta$ lies within the range reported in references~\cite{Chinyoka2005} and~\cite{Chung2008} for both Newtonian and viscoelastic cases. Consistent with prior research, the orientation angle in the VN system is larger than that in the NN system, while the NV system exhibits the smallest value for $\theta$. 

\begin{table}[h!]
    \centering
\caption{Comparison of drop deformation $D$ and orientation angle $\theta$ at time $t = 3~\dot{\gamma}^{-1}$. NN, NV and VN systems for $\operatorname{Ca} = 0.24$ and $\operatorname{De} = 0.4$.}
\begin{tabular}{@{}lllllllll@{}}
\toprule
   & \multicolumn{2}{l}{This work} & & \multicolumn{2}{l}{~\cite{Chinyoka2005}} & & \multicolumn{2}{l}{~\cite{Chung2008}} \\ \cmidrule(lr){2-3} \cmidrule(lr){5-6} \cmidrule(l){8-9} 
   & $D$         & $\theta$        &  & $D$                              & $\theta$ &  & $D$                           & $\theta$ \\ \midrule
NN & 0.2559       & 31.63          &  & 0.2878                           & 32.26    &  & 0.2674                        & 28.90    \\
NV & 0.2441       & 26.88          &  & 0.2656                           & 28.15    &  & 0.2550                        & 25.38    \\
VN & 0.247       & 32.03          &  & 0.2799                           & 32.53    &  & 0.2582                        & 30.32    \\ \bottomrule
\end{tabular}
\label{tab:cpSD}
\end{table}

\begin{figure}[h!]
    \centering
    \begin{minipage}{0.49\textwidth}
        \begin{subfigure}[b]{\textwidth}
            \caption{VN,$\; \operatorname{Ca} = 0.24,\, 0.6$.}
            \includegraphics[width=\textwidth]{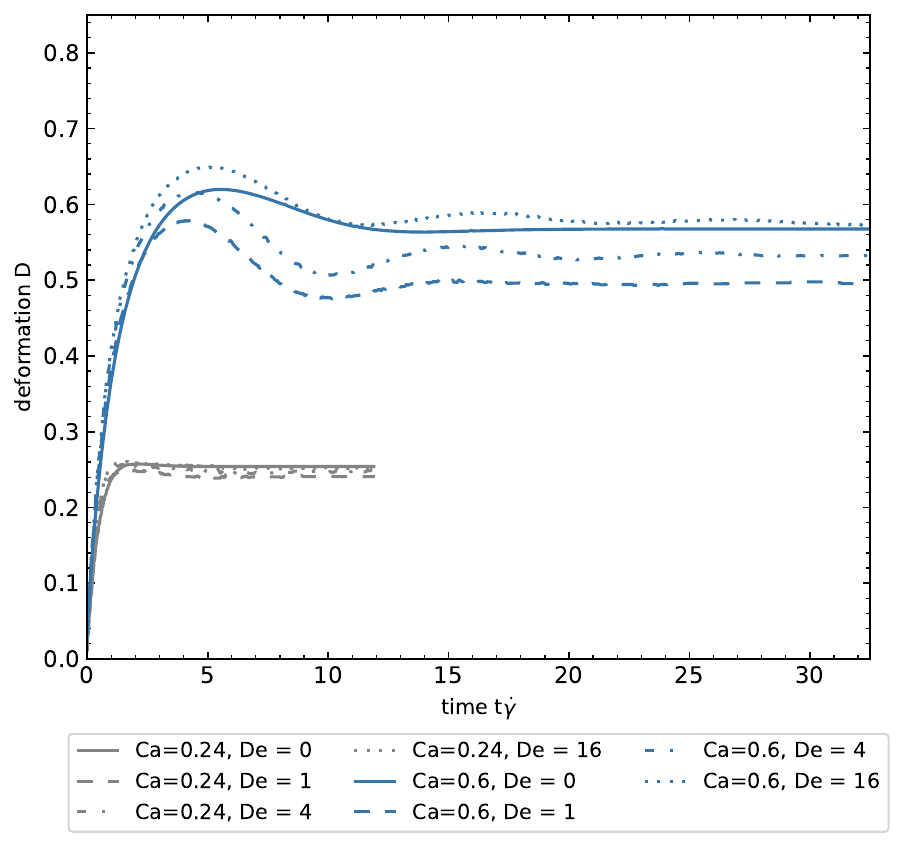}
            \label{subfig:2D_VN}
        \end{subfigure}
    \end{minipage}\hfill
    \begin{minipage}{0.49\textwidth}
        \begin{subfigure}[b]{\textwidth}
            \caption{NV,\; $\operatorname{Ca} = 0.24,\, 0.6$.}
            \includegraphics[width=\textwidth]{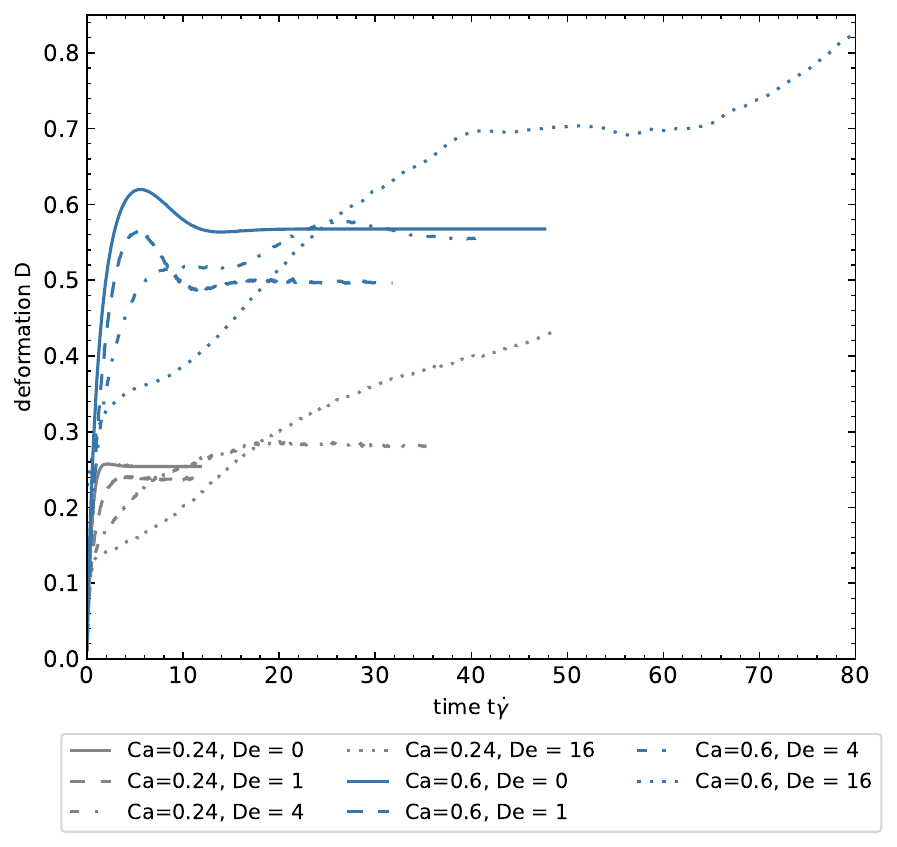}
            \label{subfig:2D_NV}
        \end{subfigure}
    \end{minipage}
    \begin{minipage}{0.49\textwidth}
        \begin{subfigure}[b]{\textwidth}
            \caption{VN,$\; \operatorname{Ca} = 0.24,\, 0.6$.}            \includegraphics[width=\textwidth]{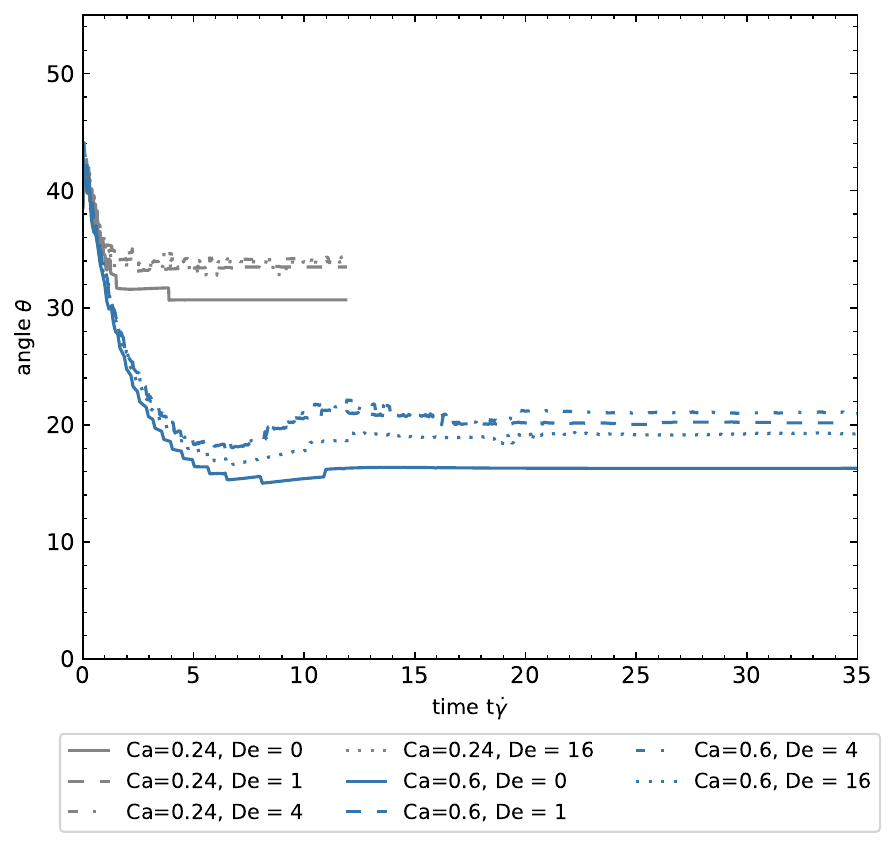}
            \label{subfig:2D_VN_angle}
        \end{subfigure}
    \end{minipage}\hfill
    \begin{minipage}{0.49\textwidth}
        \begin{subfigure}[b]{\textwidth}
            \caption{NV,$\; \operatorname{Ca} = 0.24,\, 0.6$.}
            \includegraphics[width=\textwidth]{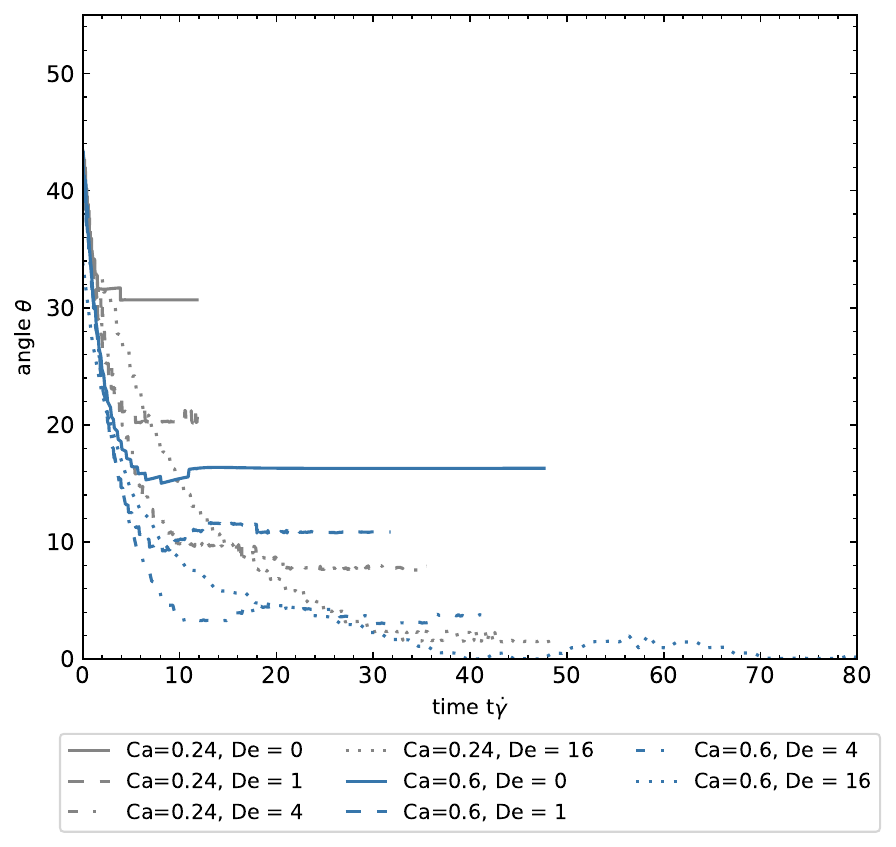}
            \label{subfig:2D_NV_angle}
        \end{subfigure}
    \end{minipage}
    \caption{Drop deformation parameter $D$ and orientation angle $\theta$ as a function of time for the NN, VN, and NV systems at $\operatorname{Ca}=0.24$ and $\operatorname{Ca}=0.6$ and varying Deborah numbers.}
    \label{fig:deformationTransient}
\end{figure}

Figure~\ref{fig:deformationTransient} displays the transient deformation for varying $\operatorname{Ca}$ and $\operatorname{De}$ values. 
At lower Capillary and Deborah numbers, a steady state is achieved more rapidly in terms of physical time. The drop exhibits prolonged dynamic changes with increasing $\operatorname{De}$ and $\operatorname{Ca}$. Moreover, at $\operatorname{Ca} = 0.6$, the deformation significantly increases across all scenarios compared to $\operatorname{Ca} = 0.24$. Comparing \Cref{subfig:2D_VN,subfig:2D_NV} indicates that an increase in $\operatorname{De}$ has a more pronounced effect on the NV system, where the matrix fluid exhibits viscoelastic properties. Specifically, within the range of $\operatorname{De} = 4$ to $16$, the NV system shows a significantly more pronounced deformation behavior, with the steady state solution not being reached even after a simulation time of $t\dot{\gamma} = 80$. This can be attributed to the matrix fluid's inherent ability to accumulate and release elastic stress over time, which is observable even at considerable distances from the deforming fluid interface. 

\Cref{subfig:2D_VN_angle,subfig:2D_NV_angle} display the transient orientation angle $\theta$. As the Capillary numbers increase, there is a notable decrease in the value of the orientation angle. In the VN system, increasing the Deborah number to $\operatorname{De} = 2$ results in increased $\theta$ values. At Deborah numbers above this threshold, the orientation angle slightly reduces, yet the values remain substantially higher than those observed in the Newtonian drop. Conversely, in the NV system, the orientation angle decreases as the Deborah numbers increase. The effect of varying Deborah numbers is more pronounced in the NV system compared to the VN system. Specifically, at $\operatorname{Ca} = 0.6$ and $\operatorname{De} = 16$, the orientation angle approaches almost zero.

\begin{figure}[h!]
    \centering

    \begin{minipage}{0.49\textwidth}
        \begin{subfigure}[b]{\textwidth}
            \caption{VN,$\; \operatorname{Ca} = 0.24,\, 0.6$.}
            \includegraphics[width=\textwidth]{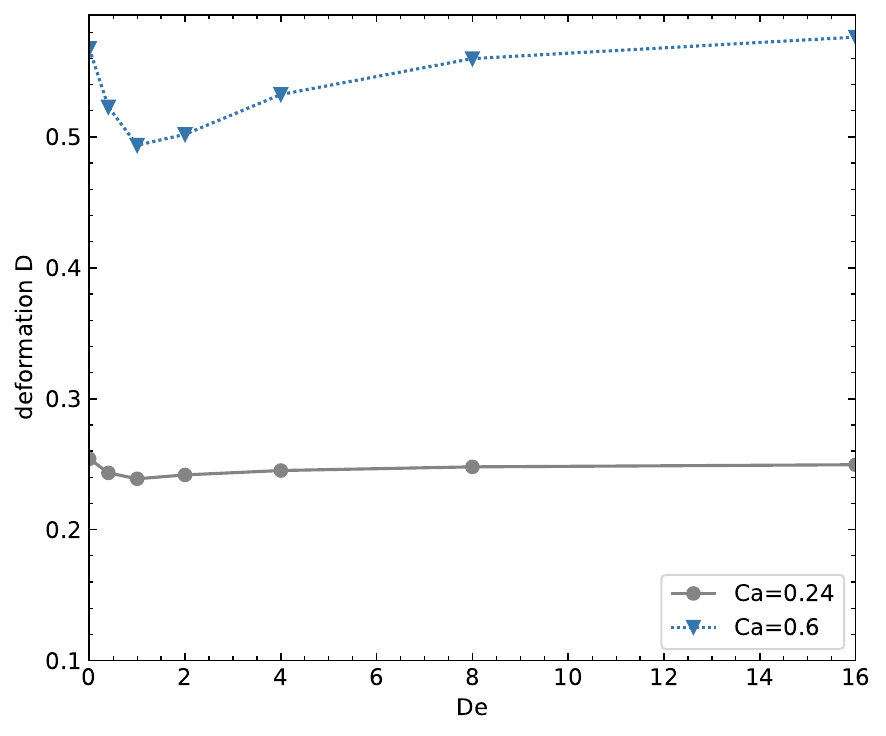}
            \label{subfig:2D_VN_D_vs_De}
        \end{subfigure}
    \end{minipage}\hfill
    \begin{minipage}{0.49\textwidth}
        \begin{subfigure}[b]{\textwidth}
            \caption{NV,$\; \operatorname{Ca} = 0.24,\, 0.6$.}
            \includegraphics[width=\textwidth]{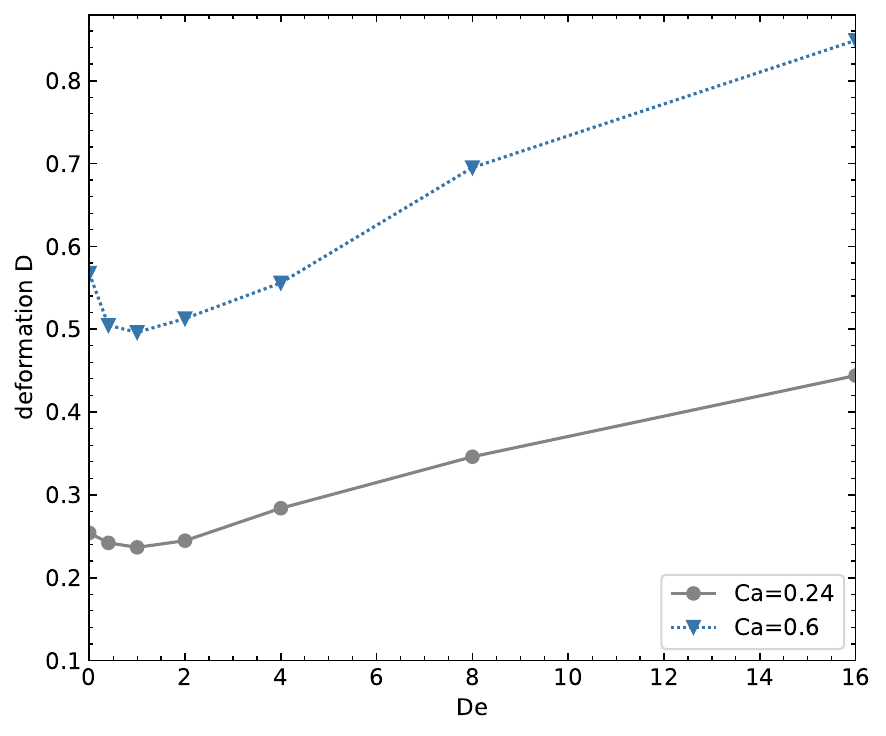}
            \label{subfig:2D_NV_D_vs_De}
        \end{subfigure}
    \end{minipage}
    \caption{Drop deformation as a function of Deborah numbers for the NN, VN, and NV systems at $\operatorname{Ca}=0.24$ and $\operatorname{Ca}=0.6$. The deformation parameter D value corresponds to the latest D values of \cref{fig:deformationTransient}.}
    \label{fig:D_vs_De}
\end{figure}

Further analysis reveals a distinct pattern in the steady state deformation, as shown in~\cref{fig:D_vs_De}, there is a decline within the range of $\operatorname{De} = 0$ to $\operatorname{De} = 1$, followed by an increase for higher Deborah numbers, specifically when $\operatorname{De} > 1$. This behavior aligns with previous observations made by Yue~\cite{Yue2005} and Aggarwal and Sarkar~\cite{Aggarwal2008} for a Deborah number range of up to $\operatorname{De} \leq 2$. The present study confirms these earlier findings and demonstrates a sustained increase in the deformation parameter in the range $1 < \operatorname{De} \leq 16$. The impact of the Deborah number on steady state deformation is considerably more pronounced in the NV system. As shown in~\cref{fig:D_vs_De}, the deformation surpasses the Newtonian reference state when $\operatorname{De} > 4$. Hence, depending on the particular value of the Deborah number, viscoelasticity can either suppress or enhance droplet deformation relative to a Newtonian drop in a Newtonian matrix.
We emphasize that no steady state is reached within $t = 80~\dot{\gamma}^{-1}$ for $\operatorname{De} = 16$; the plot shows the final simulation values in that case. For future research, exploring whether a critical Deborah number exists beyond which a drop can no longer maintain a stationary shape would be valuable. Previously, a critical Capillary number $\operatorname{Ca}_c$ was identified at around $0.875$~\cite{Zhou1993} beyond which periodic drop oscillations occur. However, higher critical Capillary numbers around $1.0$ have been reported~\cite{Chung2007}. Our simulations suggest that both the Deborah number and the Capillary number play significant roles. At lower Capillary numbers, oscillatory behavior may emerge when the Deborah number is sufficiently high.
%

%
%
%
%

\begin{figure}[h!]
    \centering

    \begin{minipage}{0.49\textwidth}
        \begin{subfigure}[b]{\textwidth}
            \caption{VN,$\; \operatorname{Ca} = 0.24$.}
            \includegraphics[width=\textwidth]{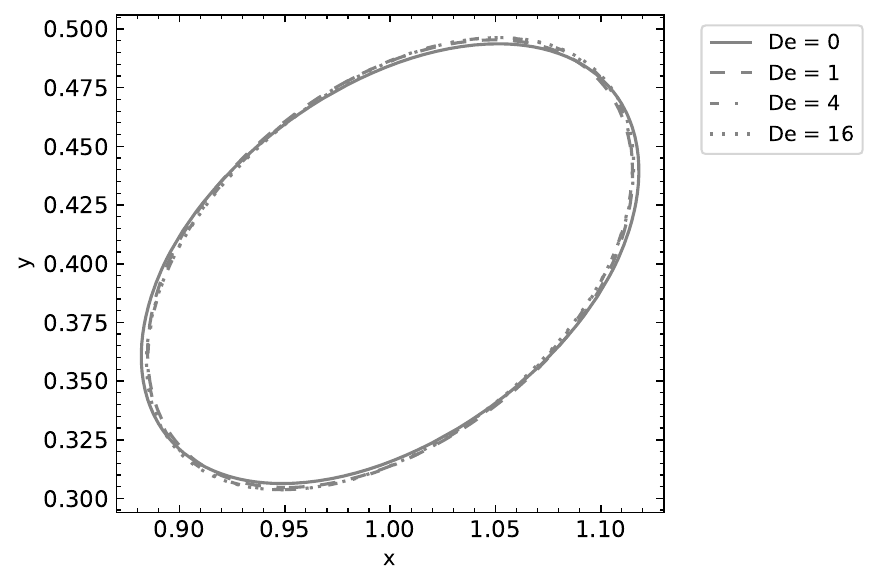}
            \label{subfig:2D_VN_1_tshape}
        \end{subfigure}
    \end{minipage}\hfill
    \begin{minipage}{0.49\textwidth}
        \begin{subfigure}[b]{\textwidth}
            \caption{NV,$\; \operatorname{Ca} = 0.24$.}
            \includegraphics[width=\textwidth]{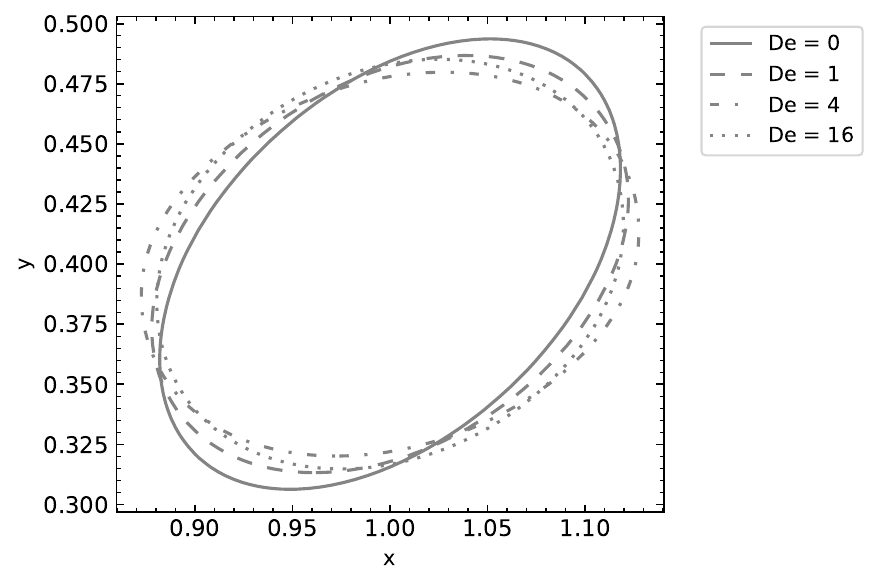}
            \label{subfig:2D_NV_1_tshape}
        \end{subfigure}
    \end{minipage}

    \begin{minipage}{0.49\textwidth}
        \begin{subfigure}[b]{\textwidth}
            \caption{VN,$\; \operatorname{Ca} = 0.6$.}
            \includegraphics[width=\textwidth]{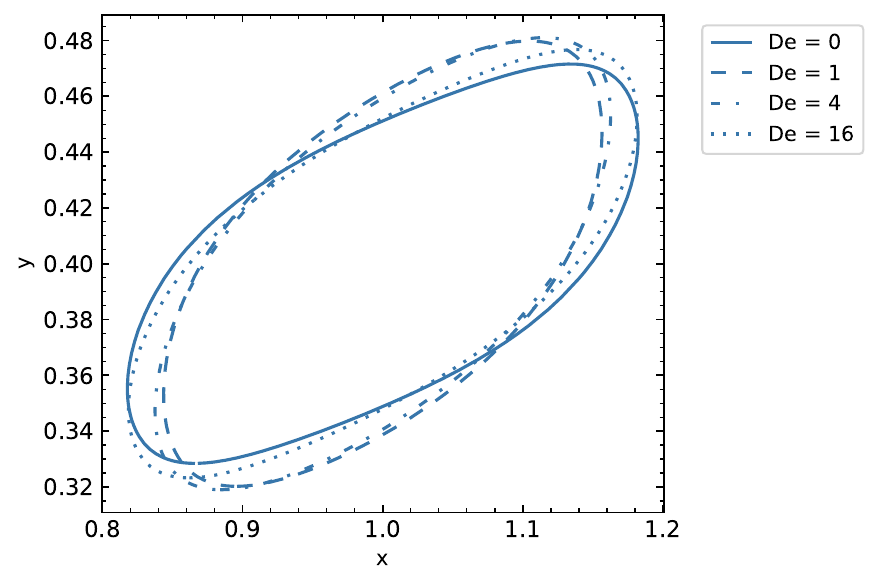}
            \label{subfig:2D_VN_2_tshape}
        \end{subfigure}
    \end{minipage}\hfill
    \begin{minipage}{0.49\textwidth}
        \begin{subfigure}[b]{\textwidth}
            \caption{NV,$\; \operatorname{Ca} = 0.6$.}
            \includegraphics[width=\textwidth]{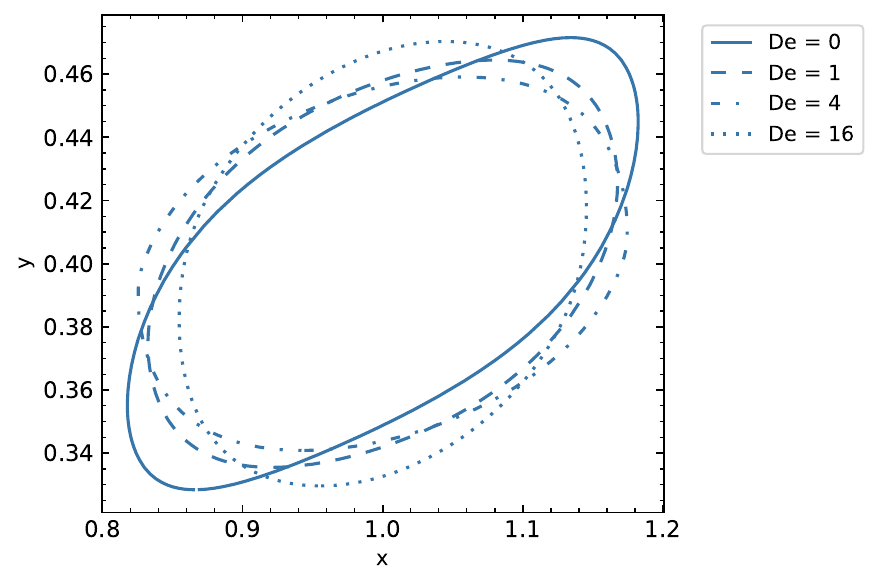}
            \label{subfig:2D_NV_2_tshape}
        \end{subfigure}
    \end{minipage}

    \caption{Drop shapes at $t=10\,\dot{\gamma}^{-1}$.}
    \label{fig:dropShapes}
\end{figure}

\begin{figure}[h!]
    \centering

    \begin{minipage}{0.49\textwidth}
        \begin{subfigure}[b]{\textwidth}
            \caption{VN,$\; \operatorname{Ca} = 0.24$.}
            \includegraphics[width=\textwidth]{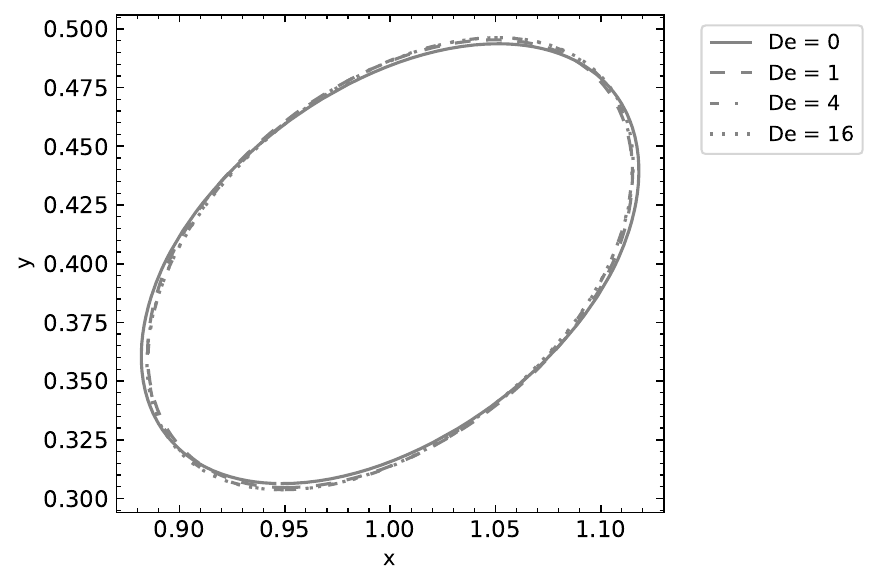}
            \label{subfig:2D_VN_1_eqshape}
        \end{subfigure}
    \end{minipage}\hfill
    \begin{minipage}{0.49\textwidth}
        \begin{subfigure}[b]{\textwidth}
            \caption{NV,$\; \operatorname{Ca} = 0.24$.}
            \includegraphics[width=\textwidth]{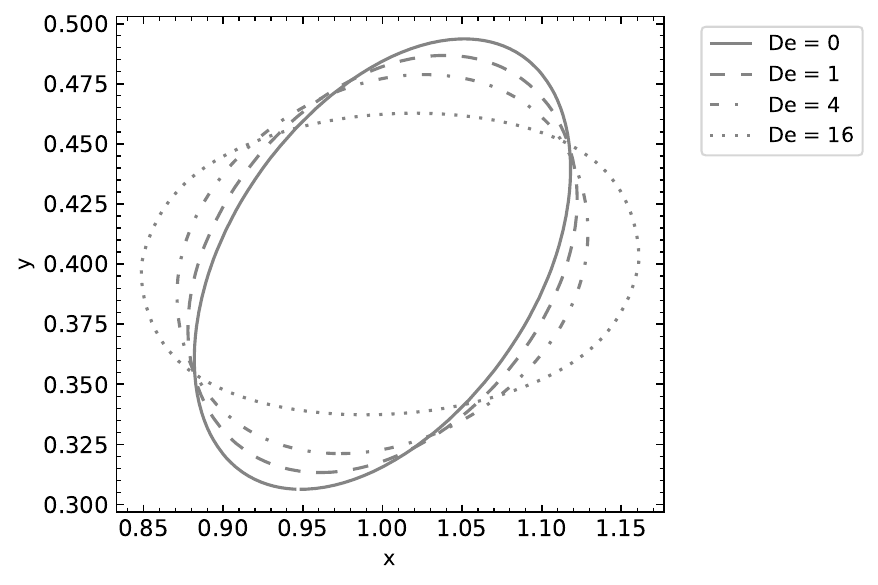}
            \label{subfig:2D_NV_1_eqshape}
        \end{subfigure}
    \end{minipage}

    \begin{minipage}{0.49\textwidth}
        \begin{subfigure}[b]{\textwidth}
            \caption{VN,$\; \operatorname{Ca} = 0.6$.}
            \includegraphics[width=\textwidth]{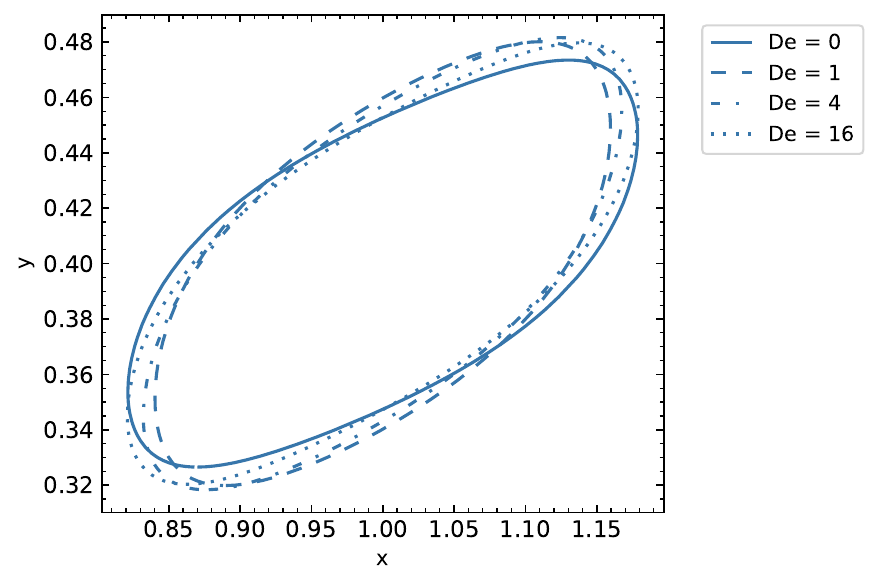}
            \label{subfig:2D_VN_2_eqshape}
        \end{subfigure}
    \end{minipage}\hfill
    \begin{minipage}{0.49\textwidth}
        \begin{subfigure}[b]{\textwidth}
            \caption{NV,$\; \operatorname{Ca} = 0.6$.}
            \includegraphics[width=\textwidth]{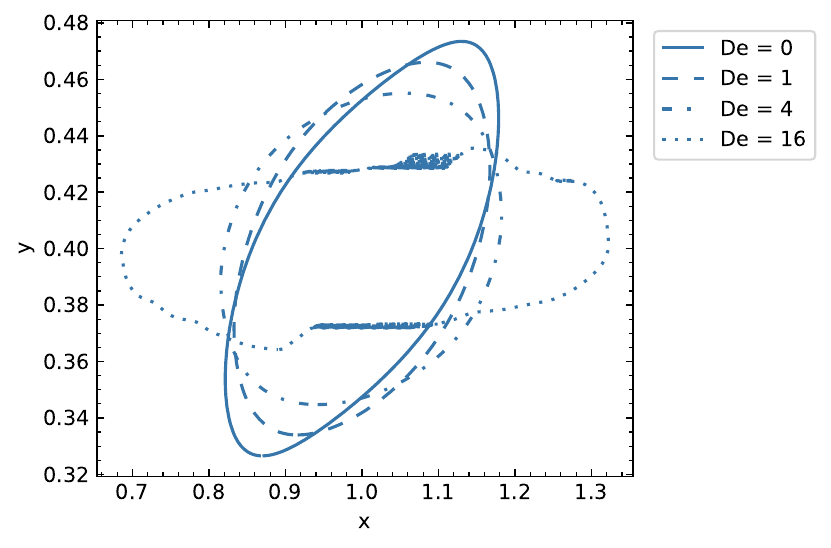}
            \label{subfig:2D_NV_2_eqshape}
        \end{subfigure}
    \end{minipage}

    \caption{Drop shapes at the latest time corresponding to \cref{fig:deformationTransient}.}
    \label{fig:equilibriumDropShape}
\end{figure}

The drop shapes are visualized in \cref{fig:dropShapes} at $t=10\,\dot{\gamma}^{-1}$ and for the latest times up to $t=80\,\dot{\gamma}^{-1}$ in \cref{fig:equilibriumDropShape}. \Cref{fig:dropShapes} serves as a reference for comparing with literature results where simulations were halted prior to reaching a steady state~\cite{Chinyoka2005}. For the lower Capillary number, the drops have a more ellipsoidal shape. At $\operatorname{Ca} = 0.6$, the stretching is significantly higher, which corresponds to the results of the deformation parameter reported above. For the Newtonian matrix fluid, the drop continues to deform and change its orientation, as evident when comparing \cref{subfig:2D_NV_1_tshape,subfig:2D_NV_2_tshape} to~\cref{subfig:2D_NV_1_eqshape,subfig:2D_NV_2_eqshape}. In particular, with an increase in the Deborah number, the drops tend to align more closely with the direction of the flow, resulting in lower orientation angles. At $\operatorname{Ca} = 0.6$ and $\operatorname{De} = 16$, the drop shape differs significantly from an ellipsoidal shape. A wave-like structure is formed on the upper and lower sides of the interface. We assume that this is related to the local extrema of the first normal stress difference observed near the interface, which has an impact on the flow field.
%
%

\begin{figure}
    \centering

    \begin{minipage}{0.49\textwidth}
        \begin{subfigure}[b]{\textwidth}
            \caption{VN.}
            \includegraphics[width=\textwidth]{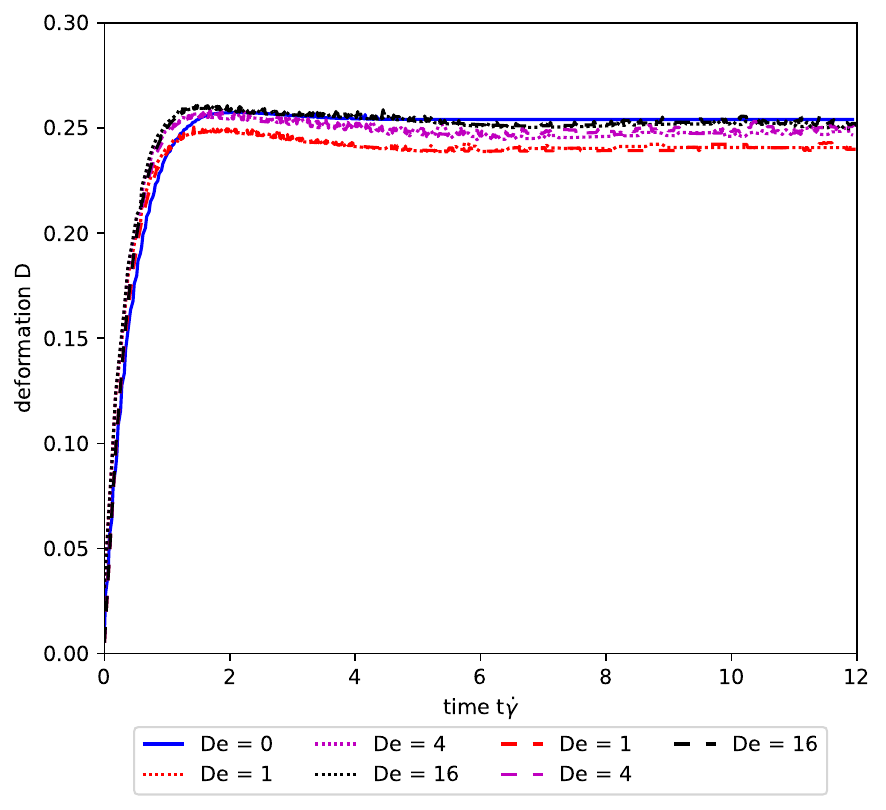}
            \label{subfig:2D_VN_RheoModelComp}
        \end{subfigure}
    \end{minipage}\hfill
    \begin{minipage}{0.49\textwidth}
        \begin{subfigure}[b]{\textwidth}
            \caption{NV.}
            \includegraphics[width=\textwidth]{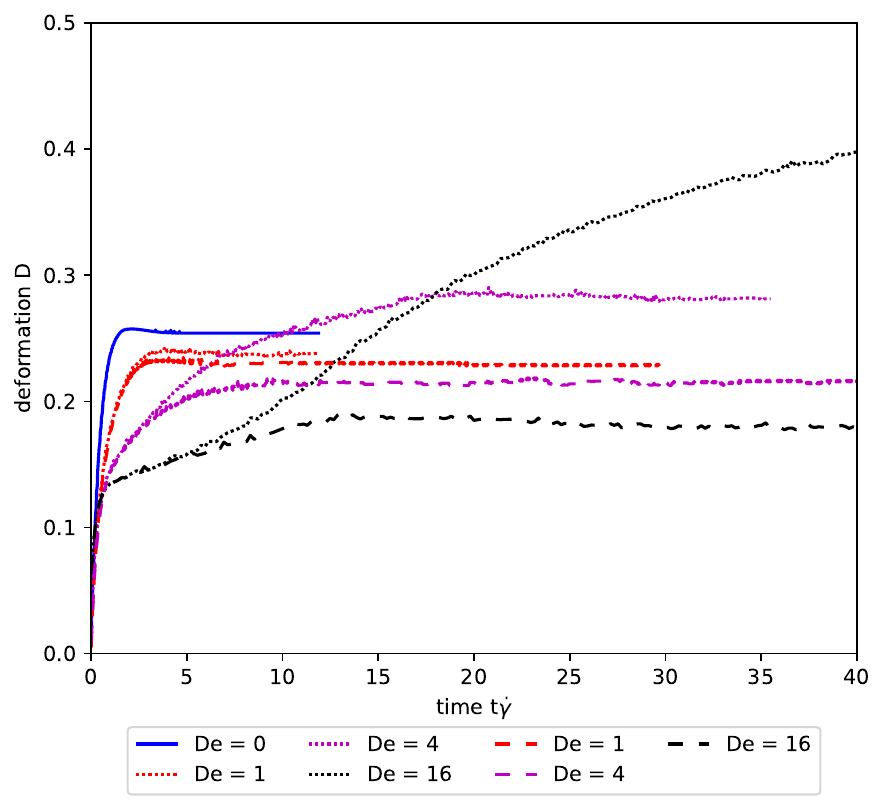}
            \label{subfig:2D_NV_RheoModelComp}
        \end{subfigure}
    \end{minipage}
    \caption{Comparison of rheological models: Oldroyd-B (dotted) and the Giesekus model with $\alpha = 0.03$ (dashed) for droplet deformation as a function of time for NN, VN, and NV systems at $\operatorname{Ca}=0.24$ and varying Deborah numbers.}
    \label{fig:rheoModelComp}
\end{figure}

Next, we examine the effect of different rheological models. In this context, simulations are carried out using the Giesekus model, which introduces an additional nonlinear term compared to the Oldroyd-B model. The mobility parameter $\alpha$ in the Giesekus model, an empirical constant, determines the degree of nonlinear features to be considered. When $\alpha$ is set to zero, the Oldroyd-B model is recovered. A slightly increased $\alpha$ value of $0.03$ already significantly impacts the flow behavior, as shown in \cref{fig:rheoModelComp}.\ While the deformation remains nearly unchanged in the VN system, a significant deviation occurs in the NV system. \Cref{subfig:2D_NV_RheoModelComp} shows that the deformation decreases with a mobility parameter $\alpha = 0.03$. This effect becomes more pronounced as the Deborah number increases. In contrast to the Oldroyd-B model, the Giesekus model shows a progressive reduction in the deformation parameter as the Deborah number increases. Additionally, the size and duration of the droplet dynamics are significantly smaller. The Giesekus matrix fluid at $\operatorname{De} = 16$ approaches a steady state deformation within the interval $t\dot{\gamma} = 40$. For the Oldroyd-B model, this time interval is insufficient, and it remains unclear whether a steady state solution exists.

\begin{figure}
    \centering

    \begin{minipage}{0.49\textwidth}
        \begin{subfigure}[b]{\textwidth}
            \includegraphics[width=\textwidth]{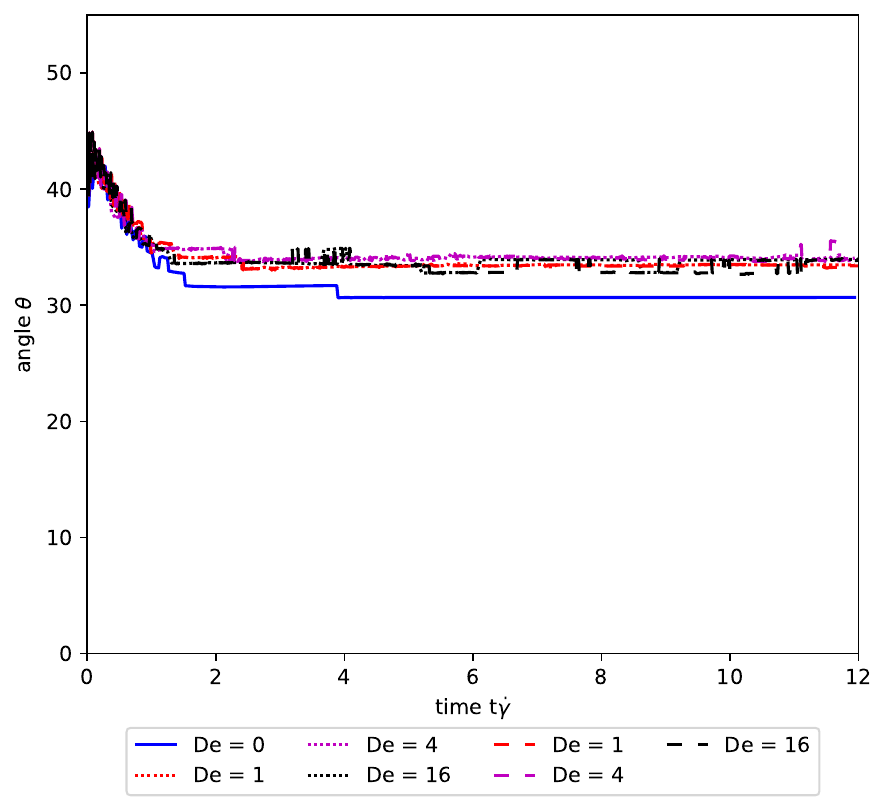}
            \caption{VN.}
            \label{subfig:2D_VN_RheoModelComp_angle}
        \end{subfigure}
    \end{minipage}\hfill
    \begin{minipage}{0.49\textwidth}
        \begin{subfigure}[b]{\textwidth}
            \includegraphics[width=\textwidth]{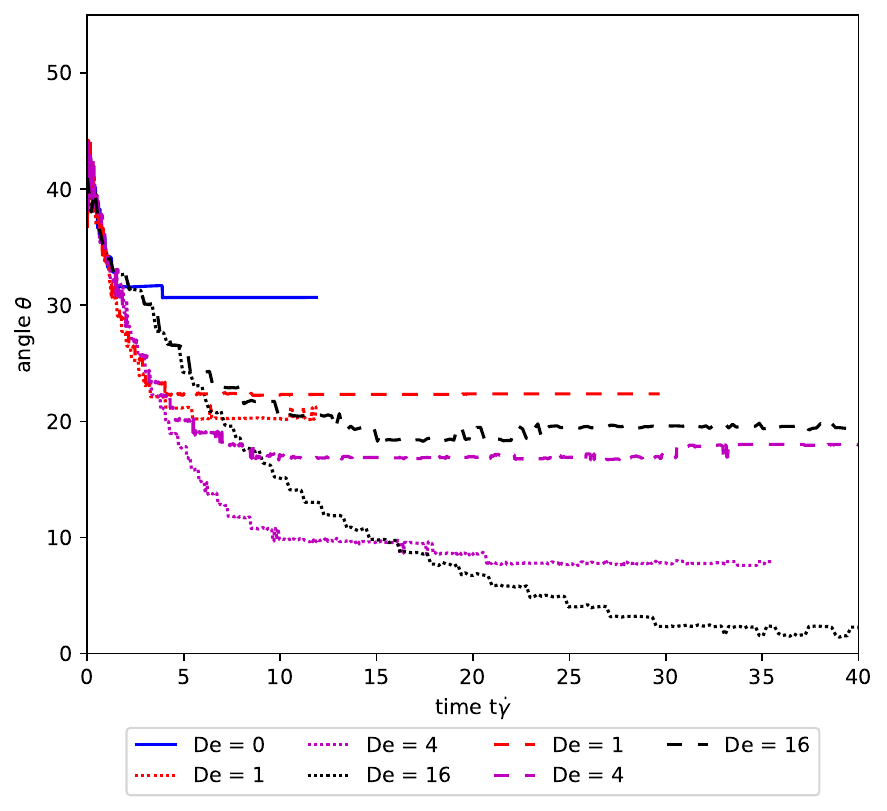}
            \caption{NV.}
            \label{subfig:2D_NV_RheoModelComp_angle}
        \end{subfigure}
    \end{minipage}
    \caption{Comparison of rheological models: Oldroyd-B (dotted) and the Giesekus model (dashed) for droplet deformation angle as a function of time for NN, VN, and NV systems at $\operatorname{Ca}=0.24$ and varying Deborah numbers.}
    \label{fig:rheoModelComp_angle}
\end{figure}

\Cref{fig:rheoModelComp_angle} shows the corresponding drop orientation angles $\theta$ for the Giesekus and the Oldroyd-B models. In the VN system, the orientation angles are similar for both models. However, as shown in \cref{subfig:2D_NV_RheoModelComp_angle}, the orientation angles in the Giesekus matrix fluid are consistently higher compared to those in the Oldroyd-B fluid. This implies that in the Giesekus fluid, the drop's alignment with the flow direction is less pronounced. Notably, the minimum $\theta$ is observed at a Deborah number of $4$. Beyond this, with increasing $\operatorname{De}$, the orientation angle tends to rise in the Giesekus fluid.

\begin{figure}[h!]
    \centering

    \begin{minipage}{0.49\textwidth}
        \begin{subfigure}[b]{\textwidth}
            \caption{VN.}
            \includegraphics[width=\textwidth]{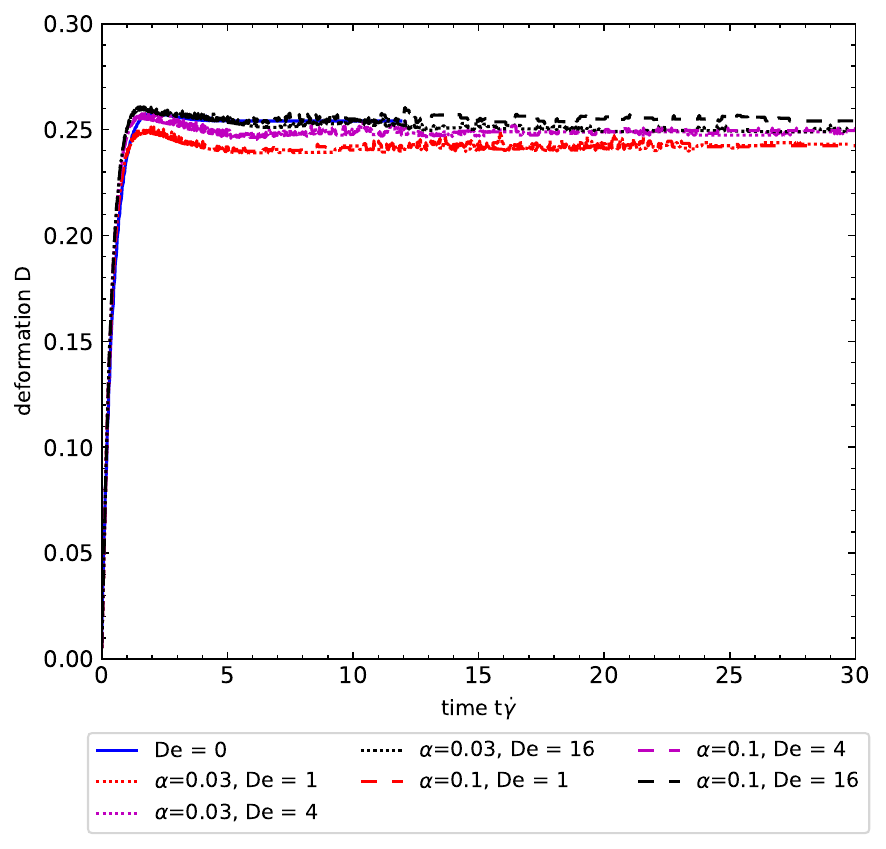}
            \label{subfig:2D_VN_Giesekus_alpha}
        \end{subfigure}
    \end{minipage}\hfill
    \begin{minipage}{0.49\textwidth}
        \begin{subfigure}[b]{\textwidth}
            \caption{NV.}
            \includegraphics[width=\textwidth]{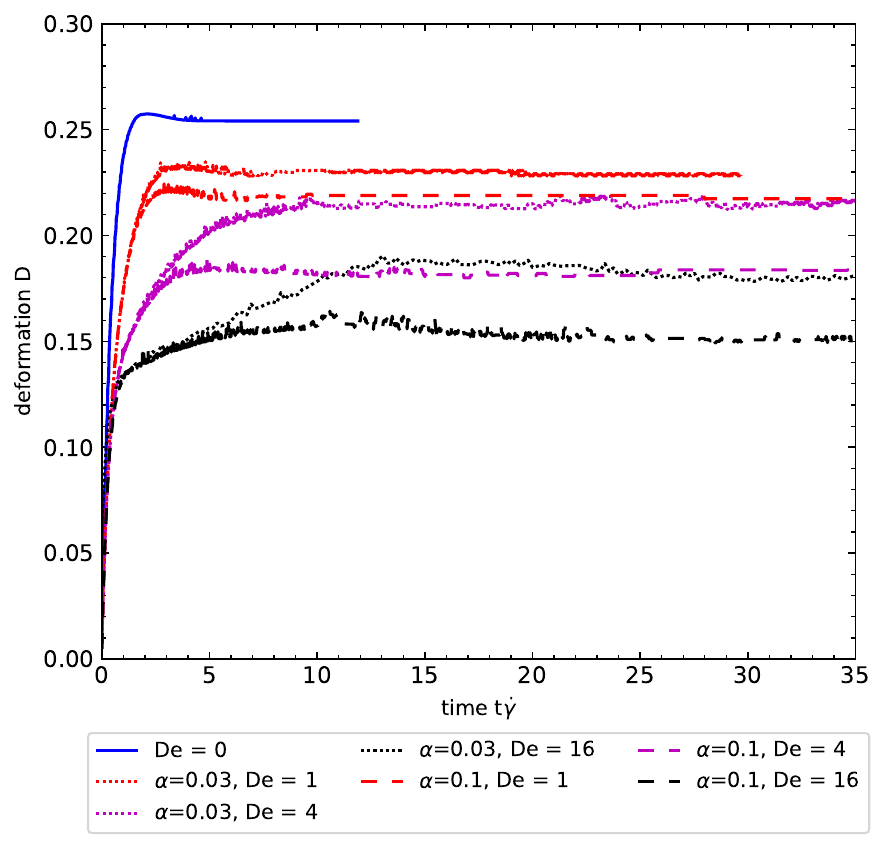}
            \label{subfig:2D_NV_Giesekus_alpha}
        \end{subfigure}
    \end{minipage}
    \caption{Drop deformation as a function of time for NN, VN, and NV at $\operatorname{Ca}=0.24$. Comparison of the Giesekus fluid for the model parameters $\alpha = 0.03$ (dotted) and $\alpha = 0.1$ (dashed) at varying Deborah numbers.}
    \label{fig:giesekus_model}
\end{figure}

\Cref{fig:giesekus_model} illustrates the effect of further increasing the mobility parameter. Raising $\alpha$ from $0.03$ to $0.1$ results in marginally increased deformations in the VN system and substantially reduced deformations in the NV system. In addition, with the increase of the Giesekus parameter, the drop dynamics are further reduced in the NV system.

\begin{figure}[h!]
    \centering

    \begin{minipage}{0.49\textwidth}
        \begin{subfigure}[b]{\textwidth}
            \caption{VN, Oldroyd-B model.}
            \includegraphics[width=0.96\textwidth]{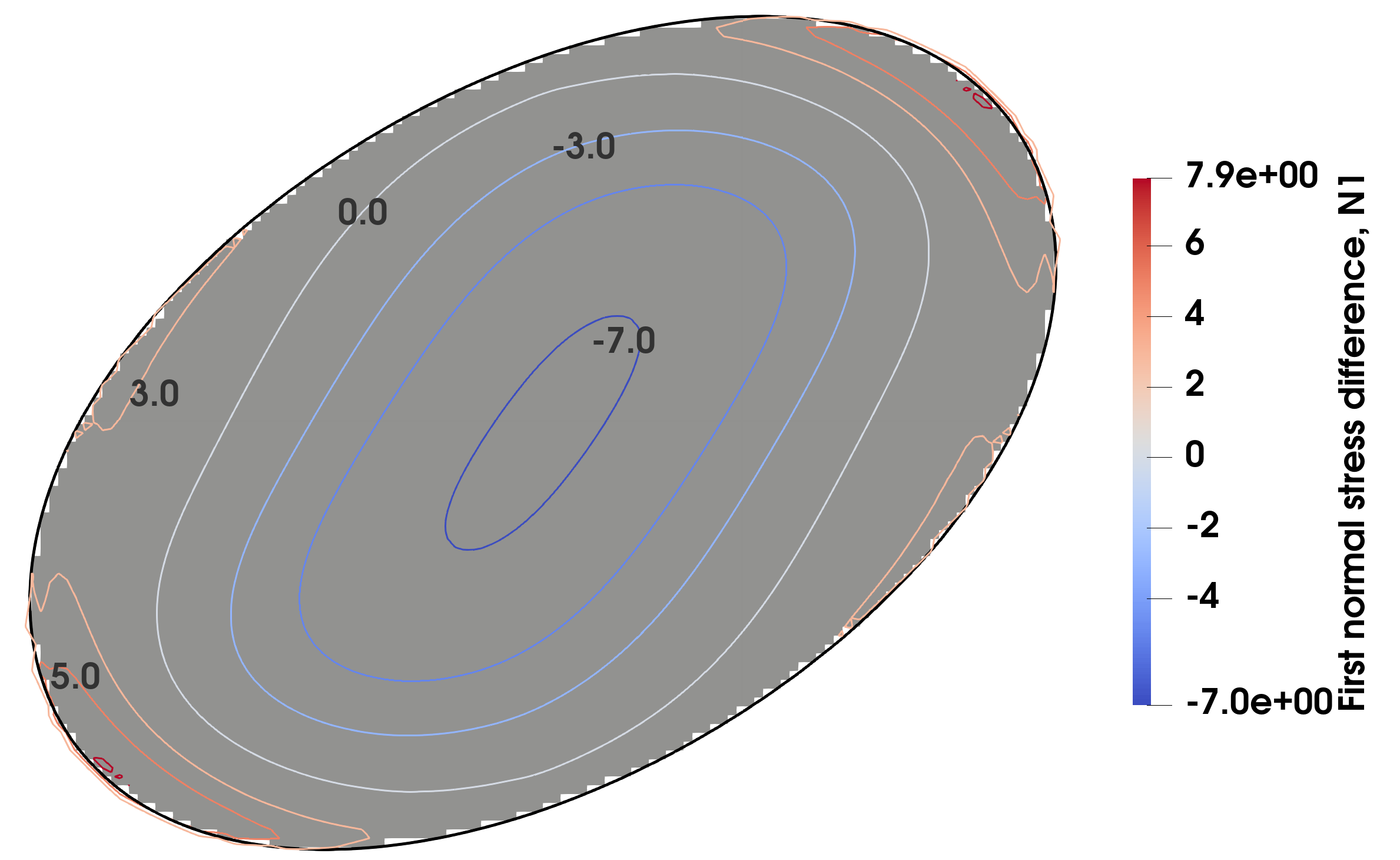}
            \label{subfig:N1_VN_OB}
        \end{subfigure}
    \end{minipage}\hfill
    \begin{minipage}{0.49\textwidth}
        \begin{subfigure}[b]{\textwidth}
            \caption{VN, Giesekus model.}
            \includegraphics[width=0.96\textwidth]{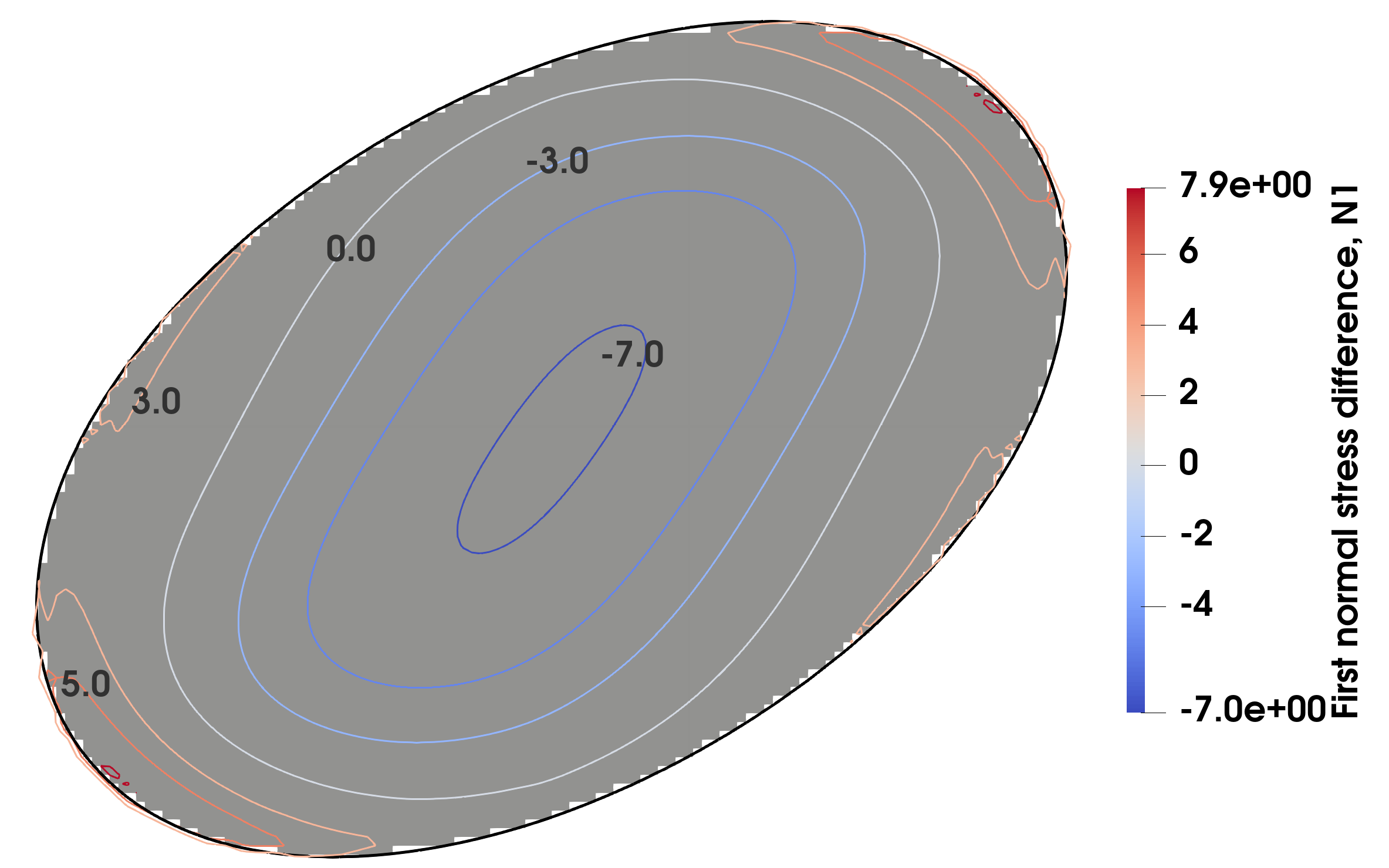}
            \label{subfig:N1_VN_Gi}
        \end{subfigure}
    \end{minipage}
    \caption{First normal stress difference, N1 contours for VN system for $\operatorname{Ca}=0.24$, $\operatorname{De}=0.4$ for Oldroyd-B and Giesekus model.}
    \label{fig:N1VN}
\end{figure}

\begin{figure}[h!]
    \centering

    \begin{minipage}{0.49\textwidth}
        \begin{subfigure}[b]{\textwidth}
            \caption{NV, Oldroyd-B model.}
            \includegraphics[width=0.96\textwidth]{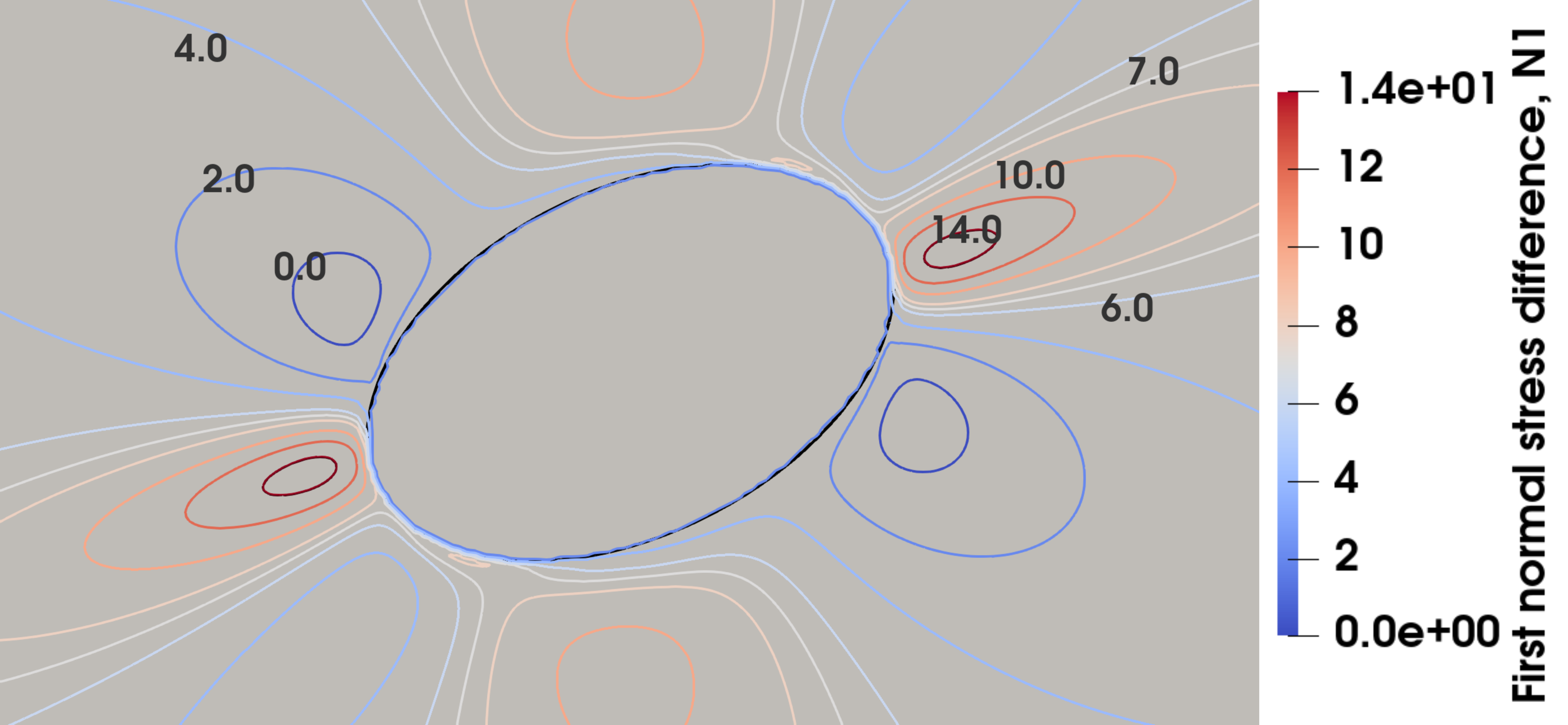}
            \label{subfig:N1_NV_OB}
        \end{subfigure}
    \end{minipage}\hfill
    \begin{minipage}{0.49\textwidth}
        \begin{subfigure}[b]{\textwidth}
            \caption{NV, Giesekus model.}
            \includegraphics[width=0.96\textwidth]{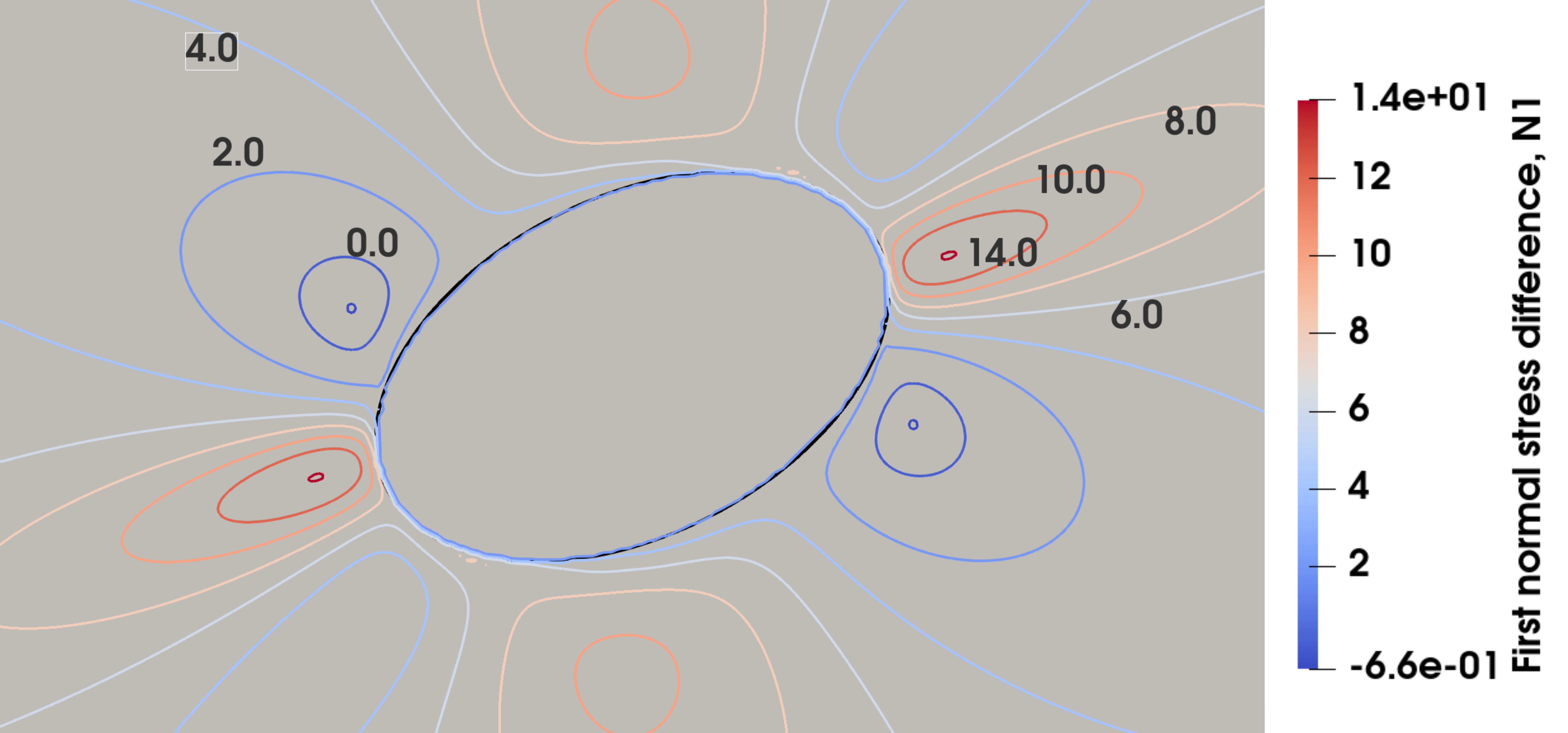}
            \label{subfig:2D_NV_gi}
        \end{subfigure}
    \end{minipage}
    \caption{First normal stress difference, N1 contours for NV system for $\operatorname{Ca}=0.24$, $\operatorname{De}=0.4$ for Oldroyd-B and Giesekus model.}
    \label{fig:N1NV}
\end{figure}

The similarities between the Oldroyd-B and Giesekus models in the VN system are confirmed by the two-dimensional plot of the first normal stress difference. \Cref{fig:N1VN} shows the contours of the first normal stress difference for $\operatorname{Ca}=0.24$ and $\operatorname{De}=0.4$ in the VN system. The first normal stress difference is computed from the polymer stress components as $N_1 = \tau_{p,xx} - \tau_{p,yy}$. In the VN case, the distribution is nearly identical. The first normal stress difference, $N_1$, has its maximum at the poles and lowest values near the equator. This $N_1$ distribution indicates that increasing $\operatorname{De}$ polymeric tensile stresses at the poles impact the stretching of the drop, as described in~\cite{Yue2005}.
\Cref{fig:N1NV} shows the corresponding contour maps for the NV cases. In the viscoelastic matrix fluid, the first normal stress difference in the Giesekus fluid is slightly smaller than in the Oldroyd-B fluid. The local extrema are positioned similarly to those in the Oldroyd-B fluid; however, their spatial extent is more confined. A maximum in $N_1$ is observed near the poles of the drop in both cases.
%

\subsection{3D drop deformation in simple shear flow}
Three-dimensional numerical studies analyzing the deformation of viscoelastic drops in simple shear have been extensively researched in the literature.
Khismatullin et al.~\cite{Khismatullin2006} employed the VOF method with a parabolic interface reconstruction (VOF-PROST), utilizing Oldroyd-B and Giesekus fluids for the viscoelastic phase. Aggarwal and Sarkar~\cite{Aggarwal2007, Aggarwal2008} applied a front-tracking Finite-Difference method, leveraging the Oldroyd-B constitutive equation for the viscoelastic fluid.
Verhulst et al.~\cite{Verhulst2009} adopted the VOF-CSF method and used both the Oldroyd-B and Giesekus models to characterize viscoelasticity. 
Afkhami et al.~\cite{Afkhami2009} applied a Finite-Difference VOF method paired with a piecewise linear interface reconstruction using the Oldroyd-B constitutive model for the viscoelastic phase.
Mukherjee and Sarkar~\cite{Mukherjee2009} utilized the front-tracking method, as described by Aggarwal and Sarkar~\cite{Aggarwal2007}, to conduct in-depth examinations of drop deformation with an Oldroyd-B viscoelastic phase.
Verhulst et al.~\cite{Verhulst2009, Verhulst2009b} employed the VOF method, as described by Khismatullin et al.~\cite{Khismatullin2006}, for comprehensive studies of the steady state and transient drop deformation, using the Oldroyd-B, Giesekus, Ellis, and multi-mode Giesekus constitutive equations.
The aforementioned studies have been referenced in more recent publications for the purposes of validation and comparison~\cite{Luo2018}.

In this section, we demonstrate the capability of our geometric VOF method to simulate three-dimensional drop deformations under shear flows. We explore parameter variations of both the Capillary and Deborah numbers. Specifically, we examine 3D drop deformations employing the Oldroyd-B constitutive equations for Deborah numbers extending up to $\operatorname{De} = 16$. This extends beyond the scope of most of the previous studies referenced and also underlines the efficacy of the high Deborah number stabilization implemented in our study.

\subsubsection{Problem description}
The setup for the three-dimensional case is based on the aforementioned two-dimensional study. The channel domain has the height $H = 8r$, the length $L=10 r$, and the depth $Z = 4r$. Initially, a spherical drop of radius $r$ is placed at the center of this domain. 
The (t)op and (b)ottom walls move with a constant velocity of $U_t$ = $-U_b$. 
Zero gradient boundary conditions for the pressure, stress, and generic constitutive variable are defined at the channel walls. Periodic boundary conditions are applied to the left and right sides of the channel. The simulation is initialized by a fully developed shear flow velocity field, while the other fields are initially zero. \textcolor{Reviewer22}{The RDF curvature model~\cite{Scheufler2023} is employed here for computing the mean curvature $\kappa$.}

The 3D simulation is governed by the dimensionless numbers defined in \cref{eq:dimensionlessnumbers}. For all simulations, a Reynolds number of $Re = 0.0003$ and a dimensionless viscosity ratio of $\Pi = 1$ are used. In the viscoelastic phase, the solvent viscosity is set equal to the polymer viscosity.
The analysis considers the drop deformation parameter, given by \cref{eq:deformationparameter}, along with the orientation angle $\theta$,  representing the alignment of the principal axes of the deformed drop with the flow direction.

\subsubsection{Results and discussion}
\begin{figure}[h!]
    \centering
    \includegraphics[height=0.4\textwidth]{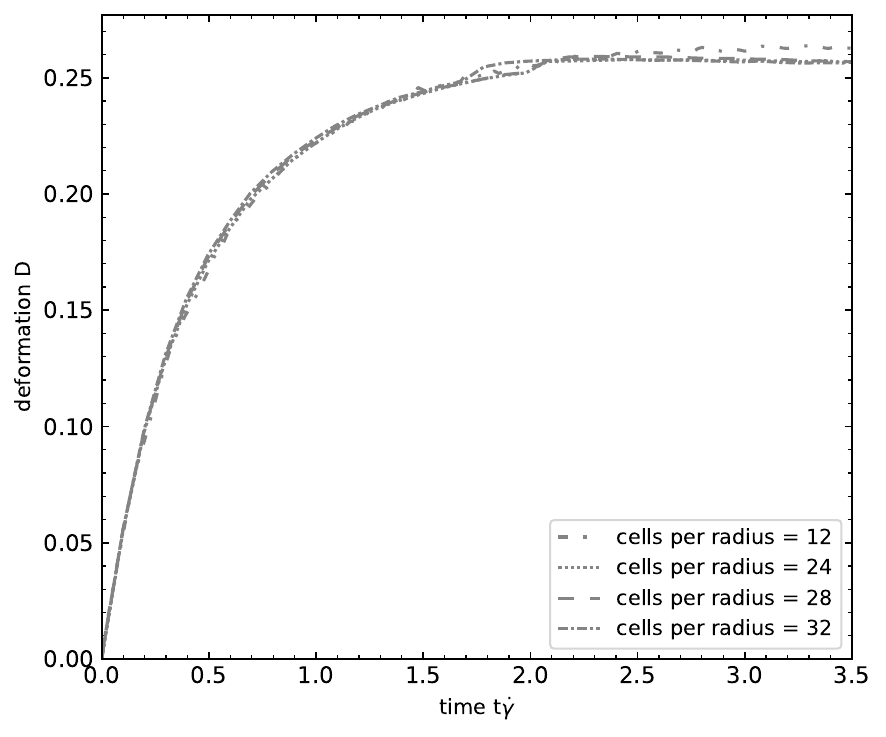}
    \caption{Drop deformation in 3D as a function of time with varying mesh refinement for the VN system at $\operatorname{Ca}=0.24$ and $\operatorname{De}=0.4$.}
    \label{fig:meshConvergence3D}
\end{figure}

In \cref{fig:meshConvergence3D}, a mesh convergence study is presented in 3D, examining drop resolutions ranging from $12$ cells per radius to $32$ cells per radius. The curve corresponding to a resolution of $12$ cells per radius is evidently underresolved. However, the curves for the higher resolutions closely align, indicating convergence towards a mesh-independent solution. The difference between the resolutions of the two finest meshes is marginal. Based on this observation, we consider the simulations with $32$ cells per radius to be sufficiently resolved, and this mesh resolution is adopted for the 3D drop deformation study. Note that the chosen 3D resolution of 32 cells per radius is smaller than the resolution used in the 2D simulations. This decision represents a compromise between accuracy and computational resources for the parameter variation.

\begin{figure}[h!]
    \centering

    \begin{minipage}{0.49\textwidth}
        \begin{subfigure}[b]{\textwidth}
            \caption{VN.}
            \includegraphics[width=\textwidth]{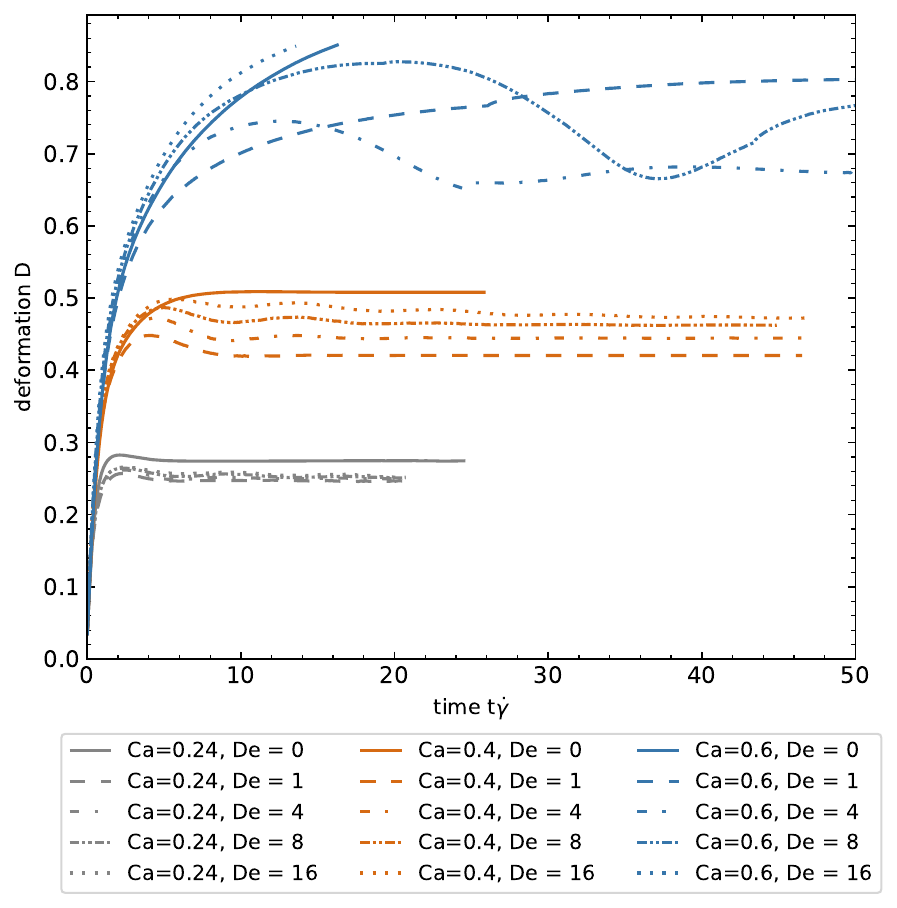}
            \label{subfig:3D_VN}
        \end{subfigure}
    \end{minipage}\hfill
    \begin{minipage}{0.49\textwidth}
        \begin{subfigure}[b]{\textwidth}
            \caption{NV.}
            \includegraphics[width=\textwidth]{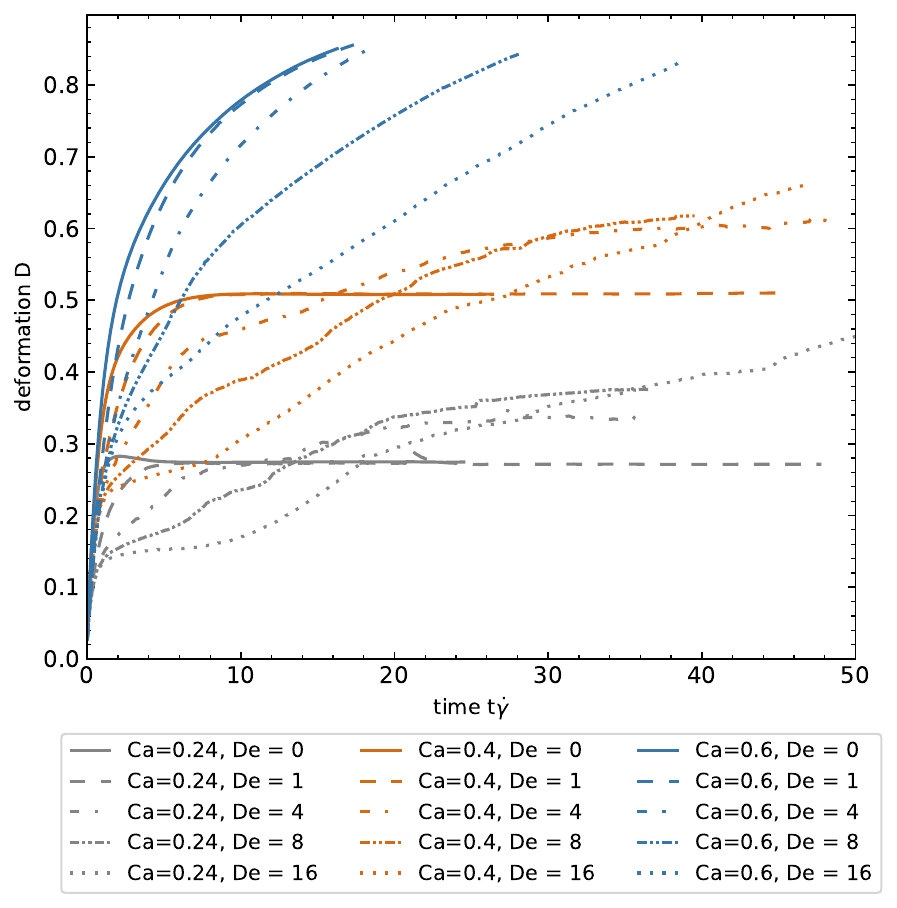}
            \label{subfig:3D_NV}
        \end{subfigure}
    \end{minipage}
    
    \caption{Three-dimensional droplet deformation as a function of time for NN, VN, and NV systems for varying Capillary numbers and Deborah numbers. Some of the simulations were terminated before $t\dot{\gamma} = 50$ because: (i) at $\operatorname{Ca} = 0.6$, the drops are strongly deformed and the computational domain's length constraint with periodic boundaries prevents running the simulations any further without risk of drop coalescence, or (ii) a steady state deformation was reached.}
    \label{fig:deformationTransient3D}
\end{figure}

%
\Cref{fig:deformationTransient3D} displays the transient deformation of a drop in 3D under varying Capillary numbers: $\operatorname{Ca} = 0.24$, $\operatorname{Ca} = 0.4$, and $\operatorname{Ca} = 0.6$. For each Capillary number, the Deborah number is varied from $\operatorname{De} = 0$ to $\operatorname{De} = 16$. The Oldroyd-B fluid is used in the viscoelastic phase in all simulations. The VN system results, shown in~\cref{subfig:3D_VN}, demonstrate that an increase in the Capillary number leads to stronger drop deformation. Significant drop dynamics are observed at $\operatorname{Ca} = 0.6$ when $\operatorname{De} > 1$, indicating oscillative deformation patterns in $D$. It appears that the oscillatory behavior may be damped up to at least $\operatorname{De} = 8$; however, the simulation time frame of $t\dot{\gamma} = 50$ is insufficient to determine the damping behavior conclusively. At $\operatorname{Ca} = 0.4$, damped oscillations in $D$ are observed on a much smaller scale, and within the given simulation period, a steady state deformation parameter is achieved for all Deborah numbers. When compared to the 2D results shown in Figure~\ref{fig:deformationTransient}, the 3D simulations exhibit larger drop deformations for all Capillary numbers. However, the overall qualitative behavior of the deformation remains very similar. Beginning from the Newtonian reference state, an increase in the Deborah number initially results in smaller deformation parameters, reaching a minimum around $\operatorname{De} = 2$. As the Deborah number rises, the deformation increases accordingly in the range up to $\operatorname{De} = 16$; the $D$ values remain smaller than the Newtonian reference state within this range.
\Cref{subfig:3D_NV} presents the results for the NV system, where again, deformations in 3D are observed to be larger than those in their 2D counterparts. Notably, in this system, the Deborah number has a more substantial impact on drop deformation compared to the VN system. Relative to the Newtonian reference state, the drop deformation parameter decreases when $\operatorname{De}$ is increased to $0.4$, marking the point of minimum deformation in 3D, which differs from the 2D case. As the Deborah number is further increased, the deformation parameter begins to rise. For $\operatorname{De} > 1$, the deformation exceeds the Newtonian reference state. The drop dynamics in the NV system appear to be less periodic than those in the VN system. We do not observe any damped oscillations, as seen in the VN system. The influence of the Deborah number on the drop dynamics is substantial, even at the lowest Capillary number, $\operatorname{Ca} = 0.24$. To thoroughly study the drop dynamics at higher Deborah numbers, simulation times greater than $t\dot{\gamma} = 50$ would be necessary. Furthermore, in such cases, it would be imperative to expand the size of the computational domain to consider the extensive stretching of the droplet over longer time intervals. These simulation setups require increased computational resources and go beyond the scope of the present work. Our simulations were terminated when the droplets approached the outer boundaries of the computational domain. Due to these constraints, the simulations with high Deborah numbers did not reach their maximum droplet deformation within the simulation time. For the highest Capillary number $\operatorname{Ca} = 0.6$, the drop deformation is so intense that a larger domain size becomes necessary, even for the Newtonian drop.

\begin{figure}[h!]
    \centering

    \begin{minipage}{0.49\textwidth}
        \begin{subfigure}[b]{\textwidth}
            \caption{VN.}
            \includegraphics[width=\textwidth]{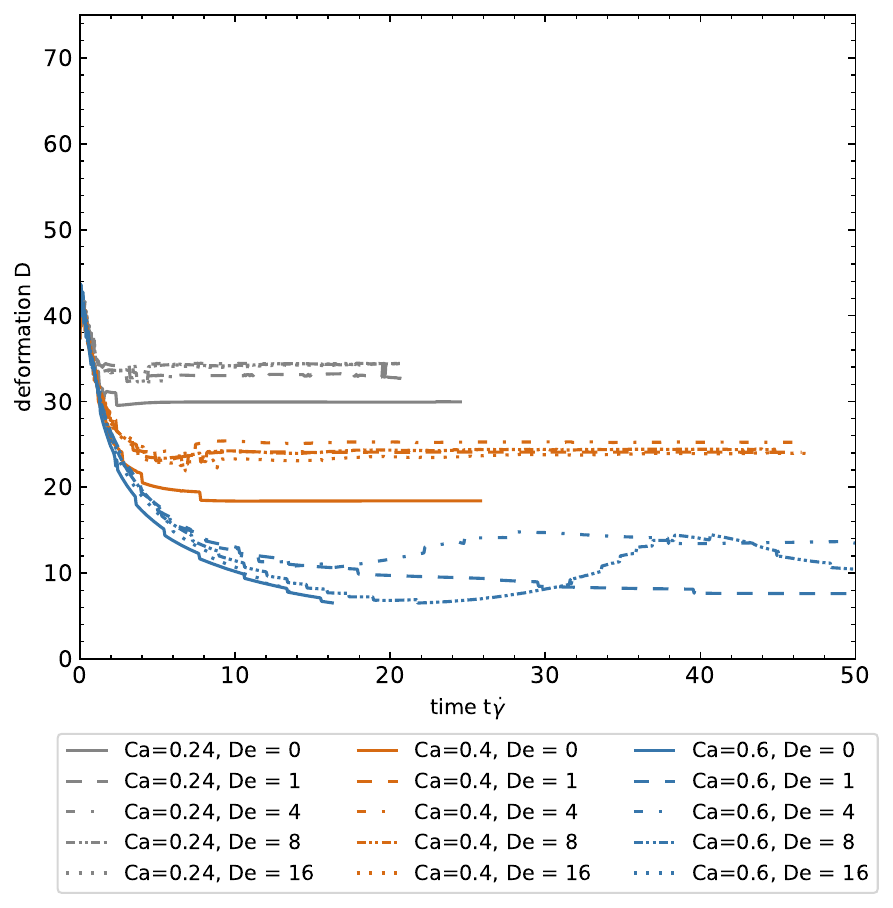}
            \label{subfig:3D_VN_angle}
        \end{subfigure}
    \end{minipage}\hfill
    \begin{minipage}{0.49\textwidth}
        \begin{subfigure}[b]{\textwidth}
            \caption{NV.}
            \includegraphics[width=\textwidth]{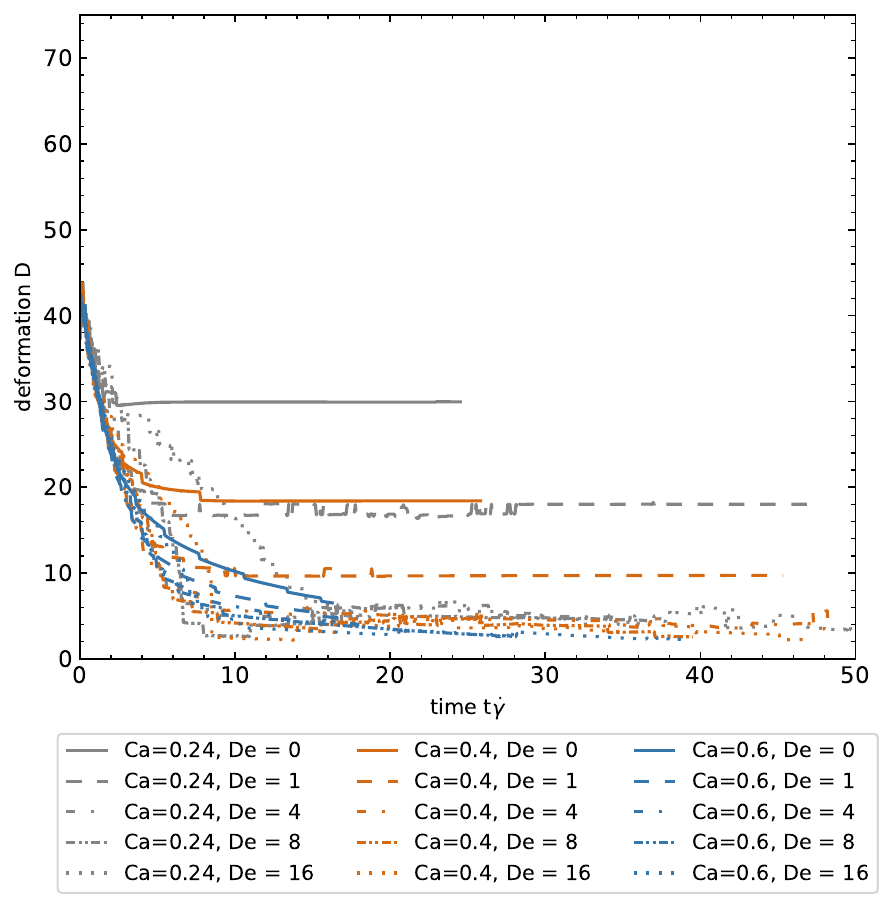}
            \label{subfig:3D_NV_angle}
        \end{subfigure}
    \end{minipage}

    \caption{Three-dimensional droplet orientation angle as a function of time for NN, VN, and NV systems for varying Capillary numbers and Deborah numbers.}
    \label{fig:orientationTransient3D}
\end{figure}

\Cref{fig:orientationTransient3D} shows the transient orientation angles of the drops in 3D simulations. In the VN system, the orientation angles in 3D are slightly smaller than their 2D counterparts, suggesting a more pronounced alignment with the flow direction in three dimensions. This discrepancy between 2D and 3D orientations becomes more pronounced as the Capillary number increases. Specifically, at a Capillary number of $\operatorname{Ca} = 0.6$, the orientation angles in 3D are about $10$ degrees smaller than those in 2D. As observed in the deformation parameter, the periodic droplet oscillations are also visible in the transient orientation angles. Therefore, these oscillations affect both the deformation and orientation of the drop. Regarding the variation of the Deborah number, a similar qualitative behavior to that in 2D drops is observed: relative to the Newtonian reference state, viscoelastic drops tend to exhibit stronger resistance to aligning with the flow direction, resulting in larger orientation angles. Starting from the Newtonian reference state, an increase in the Deborah number results in larger orientation angles, with a maximum $\theta$ at about $\operatorname{De} = 2$. As the Deborah number is further increased, the orientation angles begin to decrease, indicating a better alignment with the flow direction.
\Cref{subfig:3D_NV_angle} displays the orientation angles in the NV system. With respect to the Newtonian reference matrix fluid, an increase in the Deborah number leads to reduced drop orientation angles, meaning an improved alignment with the flow direction. The influence of the Deborah number on orientation angles is considerably more pronounced in the NV system than in the VN setup. At $\operatorname{De} = 1$, the orientation angles are approximately $10$ degrees smaller than those in the Newtonian matrix fluid in the $\operatorname{Ca} = 0.24$ and $\operatorname{Ca} = 0.4$ cases. A further increase of the Deborah number results in $\theta$ values below $10$ degrees, indicating a very strong impact of the flow in the viscoelastic matrix on the drop orientation. At $\operatorname{Ca} = 0.6$, the resistance of the drop is so small that the orientation angles approach values below $10$ degrees for all Deborah numbers.

\begin{figure}[h!]
    \centering

    \begin{minipage}{0.49\textwidth}
        \begin{subfigure}[b]{\textwidth}
            \caption{VN, $\operatorname{Ca}=0.24$.}
            \includegraphics[width=\textwidth]{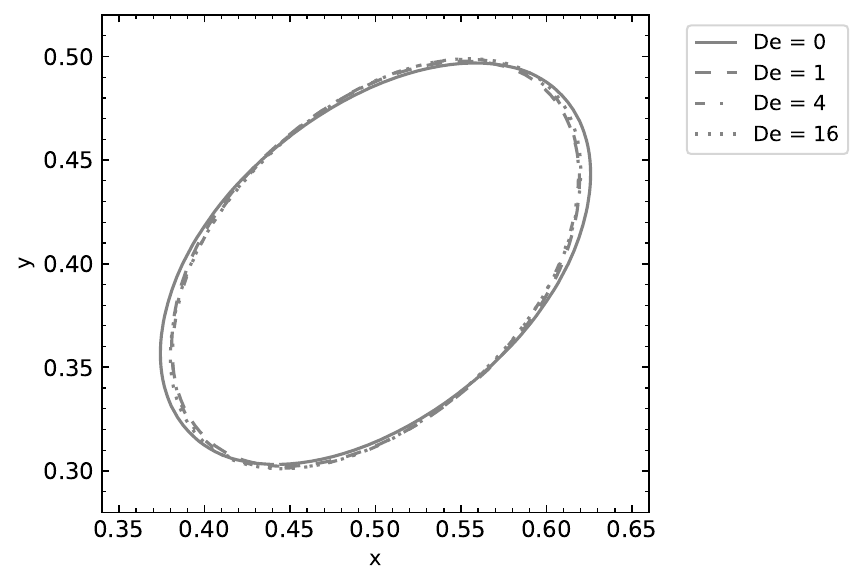}
            \label{subfig:3D_VN_1_tshape}
        \end{subfigure}
    \end{minipage}\hfill
    \begin{minipage}{0.49\textwidth}
         \begin{subfigure}[b]{\textwidth}
            \caption{NV, $\operatorname{Ca}=0.24$.}
            \includegraphics[width=\textwidth]{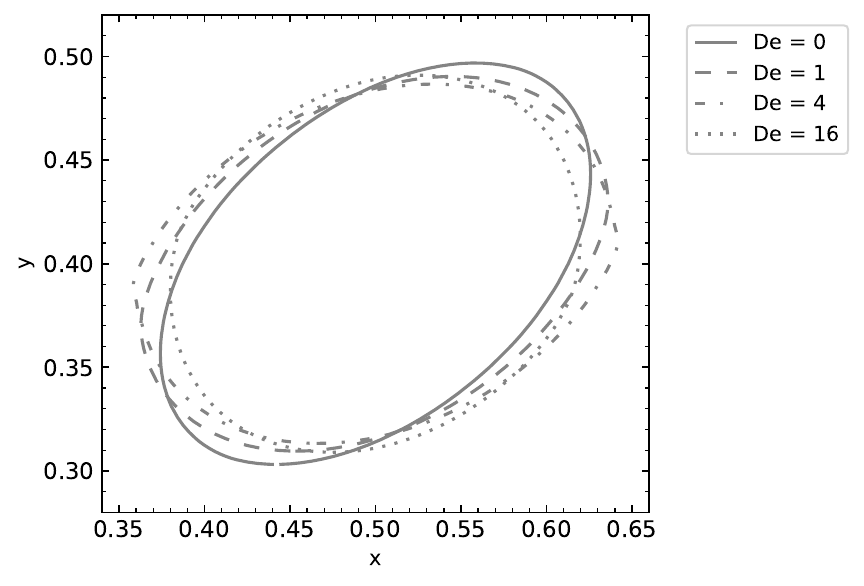}
            \label{subfig:3D_NV_1_tshape}
        \end{subfigure}
    \end{minipage}

    \begin{minipage}{0.49\textwidth}
        \begin{subfigure}[b]{\textwidth}
            \caption{VN, $\operatorname{Ca}=0.4$.}
            \includegraphics[width=\textwidth]{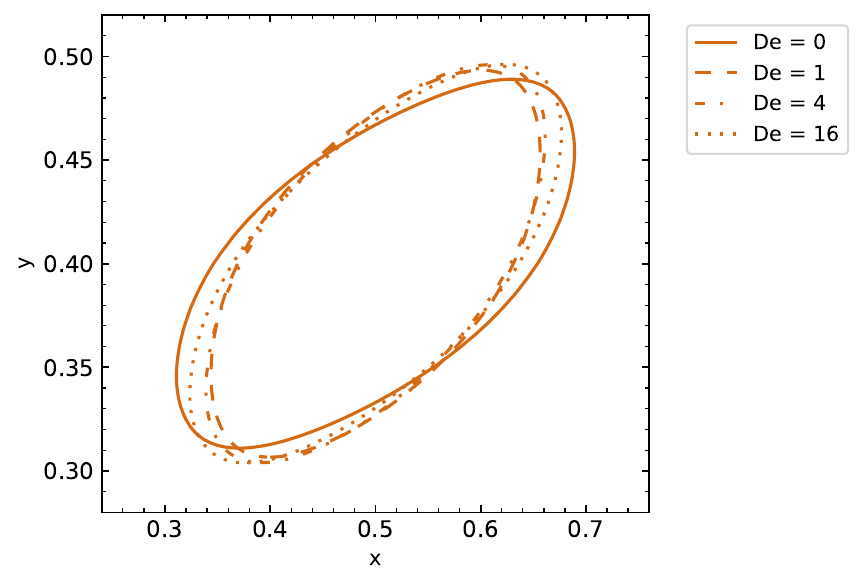}
            \label{subfig:3D_VN_2_tshape}
        \end{subfigure}
    \end{minipage}\hfill
    \begin{minipage}{0.49\textwidth}
        \begin{subfigure}[b]{\textwidth}
            \caption{NV, $\operatorname{Ca}=0.4$.}
            \includegraphics[width=\textwidth]{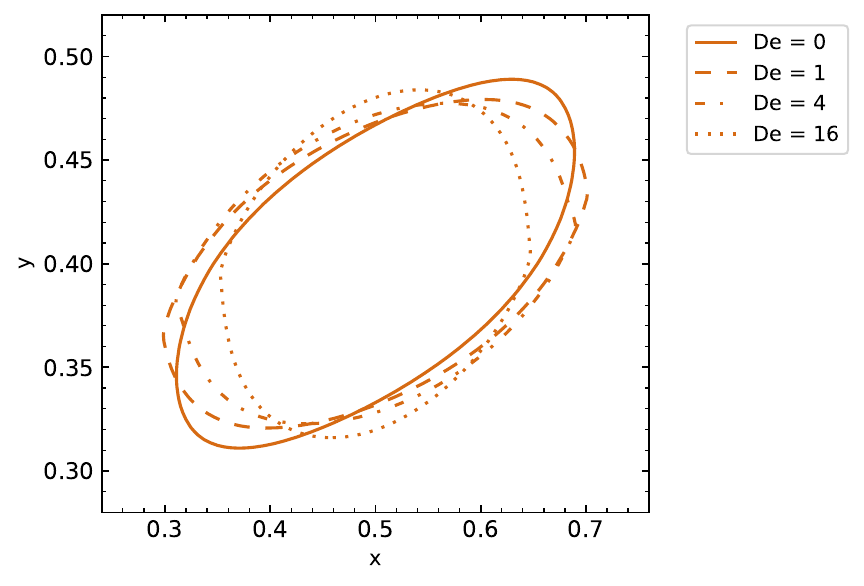}
            \label{subfig:3D_NV_2_tshape}
        \end{subfigure}
    \end{minipage}

    \begin{minipage}{0.49\textwidth}
        \begin{subfigure}[b]{\textwidth}
            \includegraphics[width=\textwidth]{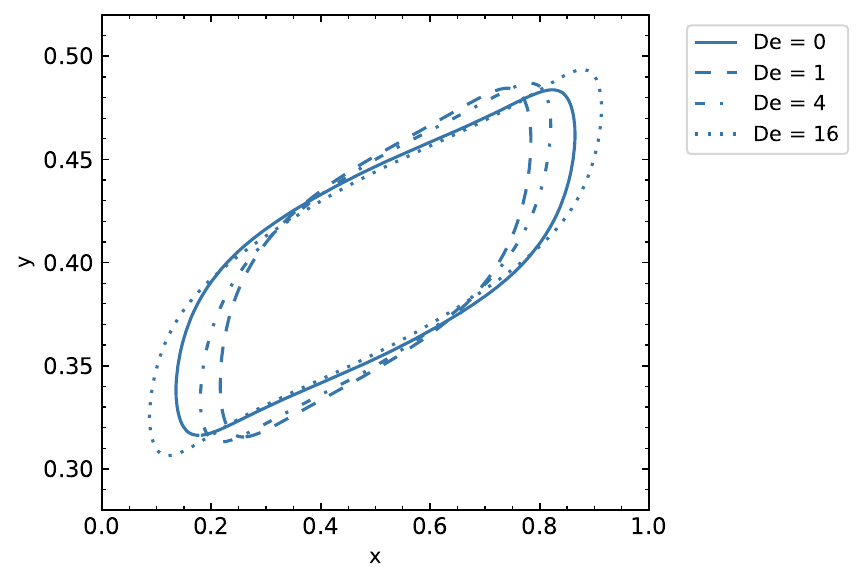}
            \caption{VN, $Ca=0.6$.}
            \label{subfig:3D_VN_3_tshape}
        \end{subfigure}
    \end{minipage}\hfill
    \begin{minipage}{0.49\textwidth}
        \begin{subfigure}[b]{\textwidth}
            \includegraphics[width=\textwidth]{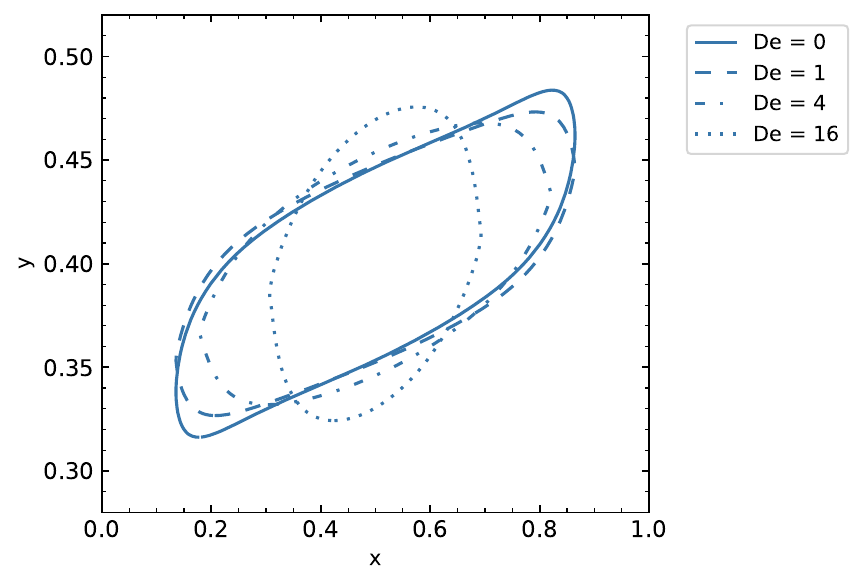}
            \caption{NV, $Ca=0.6$.}
            \label{subfig:3D_NV_3_tshape}
        \end{subfigure}
    \end{minipage}

    \caption{Contours of the drop shapes in a vertical cutting plane through the center of the drop in 3D at $t=10\,\dot{\gamma}^{-1}$.}
    \label{fig:3DdropShapes}
\end{figure}

\Cref{fig:3DdropShapes} depicts the contours of the drop shapes, captured in a vertical cross-section through the center of the three-dimensional drops at a fixed time, $t=10\,\dot{\gamma}^{-1}$. At small Deborah and Capillary numbers, the drops exhibit nearly elliptical shapes. With increasing Capillary numbers, the drops change into more elongated shapes. At $\operatorname{Ca} = 0.6$, we observe the characteristic dumbbell shapes~\cite{Chinyoka2005, Sheth1995}. Note that the drop shapes at $t=10\,\dot{\gamma}^{-1}$ have not reached a steady state. Actually, for certain configurations, a steady state may not exist.


\section{Summary and conclusion}\label{sec:conclusions}
A new geometric Volume-of-Fluid framework has been presented for simulating viscoelastic fluids on 3D general unstructured finite volume meshes, building on the `plicRDF-isoAdvector' method in OpenFOAM and the modular platform `DeboRheo' for computational rheology. This framework employs a change-of-variable formulation in the constitutive equations for viscoelastic fluids, which significantly enhances stability at moderate and high Deborah numbers. A comprehensive mathematical modeling approach for a broad range of rheological constitutive equations has been established within the VOF framework, utilizing conditional volume averaging and appropriate closure assumptions. Additionally, improvements in the reconstruction of viscosities at cell faces have been implemented to reduce local discretization errors and increase the accuracy of the simulations. An iterative segregated solution algorithm is proposed to solve the coupled system of equations.

The VOF framework has been extensively validated for transient single droplet deformation in shear flows in 2D and 3D. The Oldroyd-B and Giesekus models are considered, with the latter being evaluated across a range of mobility parameters. Our results match well with previous numerical studies in 2D~\cite{Chinyoka2005, Chung2008} and in 3D~\cite{Khismatullin2006, Aggarwal2007, Aggarwal2008, Mukherjee2009, Verhulst2009, Verhulst2009b}.
The present study extends the investigation to Deborah numbers up to $\operatorname{De} = 16$, which, to the best of our knowledge, is beyond the range explored in previous numerical research. No stability issues were observed with the square root conformation representation within this Deborah number range. Key observations from this study include how various combinations of Capillary and Deborah numbers correlate with the droplet deformation and orientation, as well as potential periodic droplet oscillations. At higher Deborah numbers, damped droplet oscillations are observed at lower Capillary numbers than those reported in previous numerical works, where limited ranges of Deborah numbers were considered. Despite the stability of the simulations, investigating higher Deborah numbers requires considerably more computational resources, as both the necessary domain size and the time scale for detailed droplet dynamics analysis increase substantially. Nevertheless, it would be valuable in future work to conduct a more systematic examination of critical combinations of Capillary and Deborah numbers beyond which steady-state deformation parameters and orientation angles can no longer be sustained.

\section*{Acknowledgement}
This work was supported by the German Research Foundation (DFG) – Project-ID 265191195 – SFB 1194. This support is gratefully acknowledged. Calculations for this research were conducted on the Lichtenberg High-Performance Cluster of the Technische Universität Darmstadt.

\section*{Data availability statement}
The data that support the findings of this study, including the case study, input data, primary and secondary data, as well as post-processing utilities, are publicly available online~\cite{researchData,DeboRheo}. The post-processing, based on Jupyter notebooks~\cite{jupyter}, simplifies the reproducibility of the results as described in the README.md file~\cite{researchData}.

\clearpage
\appendix

\section{Definitions of the tensors $\BT$ and $\OmegaT$}
\label{asec:definitionBOmega}
By employing the diagonalization $\CT = \QT \dprod \DT \dprod \trans{\QT}$, the tensor $\BT$ is computed as $\BT = \QT \dprod \tilde{\BT} \dprod \trans{\QT}$. In this formulation, the elements of the diagonal tensor $\tilde{\BT}$ are given by the tensor $\LT = \trans{\nabla \velocity} = \QT \dprod \tilde{\LT} \dprod \trans{\QT}$ as $\tilde{b}_{ii} = \tilde{l}_{ii}$. The tensor $\OmegaT$ is computed as $\OmegaT = \QT \dprod \tilde{\OmegaT} \dprod \trans{\QT}$, where $\tilde{\OmegaT}$ has zero diagonal entries $\tilde{\omega}_{ii} = 0$ and its off-diagonal elements are given by 
\begin{equation}
\label{diagonalizeCr82}
\tilde{\omega}_{{{ij}, \; {i \neq j}}} = \frac{d_{ii} {\tilde{l}_{{ij}, \; {i \neq j}}} + d_{jj} {\tilde{l}_{{ji}, \; {j \neq i}}}}{d_{jj} - d_{ii}}, \ i, j = 1, 2, 3.
\end{equation}
A comprehensive description of the local decomposition of the deformation terms in the convective derivative can be found in \cite{Niethammer2018}.


\section{Benchmark data for drop deformation and orientation}\label{sec:appendix}
\setcounter{table}{0}

\begin{table}[ht!]
    \centering
\caption{Comparison of (3D) drop deformation $D$ and orientation angle $\theta$ at time $t = 3~\dot{\gamma}^{-1}$, $t = 10 ~\dot{\gamma}^{-1}$ and $t_f$ (the final instance of simulations, corresponding to the results presented in~\cref{sec:results}). NV systems for $\operatorname{Ca} = 0.24$, $\operatorname{Ca} = 0.4$ and $\operatorname{Ca} = 0.6$ and varying Deborah numbers. $(^*)$ The deformation values do not correspond to the equilibrium state values.}
\begin{tabular}{@{}lllllllll@{}}
\toprule
   & \multicolumn{2}{l}{$t = 3~\dot{\gamma}^{-1}$} & & \multicolumn{2}{l}{$t = 10~\dot{\gamma}^{-1}$} & & \multicolumn{2}{l}{$t_f$} \\ \cmidrule(lr){2-3} \cmidrule(lr){5-6} \cmidrule(l){8-9} 
                          & $D$          & $\theta$        &  & $D$                              & $\theta$ &  & $D$                           & $\theta$ \\ \midrule
NV , $Ca=0.24$ , $De=0$   & 0.2805       & 29.60           &  & 0.2741                           & 29.93    &  & 0.2746                        & 29.95    \\
NV , $Ca=0.24$ , $De=0.4$ & 0.2637       & 25.62           &  & 0.2649                           & 24.16    &  & 0.2663                        & 25.57    \\ 
NV , $Ca=0.24$ , $De=1$   & 0.2529       & 21.53           &  & 0.2726                           & 16.71    &  & 0.2714                        & 18.03    \\
NV , $Ca=0.24$ , $De=2$   & 0.2249       & 20.90           &  & 0.2844                           & 7.43     &  & 0.2904                        & 8.47    \\
NV , $Ca=0.24$ , $De=4$   & 0.1897       & 23.59           &  & 0.2696                           & 5.02     &  & 0.3383                        & 4.86    \\
NV$^*$ , $Ca=0.24$ , $De=8$    & 0.1633       & 27.90           &  & 0.2354                           & 2.64     &  & 0.3755                        & 4.49    \\
NV$^*$ , $Ca=0.24$ , $De=16$  & 0.1481       & 28.51           &  & 0.1691                           & 16.43    &  & 0.4551                        & 3.29    \\
NV , $Ca=0.4$ , $De=0$    & 0.4543       & 22.68           &  & 0.5088                           & 18.40    &  & 0.5081                        & 18.42    \\
NV , $Ca=0.4$ , $De=0.4$  & 0.4298       & 21.14           &  & 0.4769                           & 16.25    &  & 0.4731                        & 16.35    \\
NV , $Ca=0.4$ , $De=1$    & 0.4083       & 19.40           &  & 0.5079                           & 9.68     &  & 0.5100                        & 9.75    \\
NV$^*$ , $Ca=0.4$ , $De=2$    & 0.3669       & 18.28           &  & 0.5096                           & 7.79     &  & 0.5625                        & 6.06    \\
NV$^*$ , $Ca=0.4$ , $De=4$    & 0.3146       & 19.79           &  & 0.4582                           & 5.68     &  & 0.6010                        & 4.18    \\
NV$^*$ , $Ca=0.4$ , $De=8$    & 0.2728       & 22.49           &  & 0.3889                           & 4.17     &  & 0.618                         & 2.57    \\
NV$^*$ , $Ca=0.4$ , $De=16$   & 0.2466       & 23.43           &  & 0.3050                           & 2.43     &  & 0.6618                        & 2.31    \\
NV$^*$ , $Ca=0.6$ , $De=0$    & 0.572        & 20.55           &  & 0.7771                           & 10.25    &  & 0.8512                        & 6.51    \\
NV$^*$ , $Ca=0.6$ , $De=0.4$  & 0.5420       & 19.62           &  & 0.7561                           & 9.40     &  & 0.8572                        & 4.70    \\
NV$^*$ , $Ca=0.6$ , $De=1$    & 0.5176       & 18.36           &  & 0.7714                           & 7.36     &  & 0.856                         & 4.68    \\
NV$^*$ , $Ca=0.6$ , $De=2$    & 0.4741       & 18.68           &  & 0.7695                           & 6.83     &  & 0.8500                        & 7.73    \\
NV$^*$ , $Ca=0.6$ , $De=4$    & 0.4188       & 19.35           &  & 0.7140                           & 6.15     &  & 0.8508                        & 3.96    \\
NV$^*$ , $Ca=0.6$ , $De=8$    & 0.3725       & 20.82           &  & 0.6028                           & 5.04     &  & 0.8442                        & 2.97    \\
NV$^*$ , $Ca=0.6$ , $De=16$   & 0.3433       & 21.77           &  & 0.4766                           & 4.62     &  & 0.8345                        & 2.28    \\ \bottomrule
\end{tabular}
\label{tab:NV_DandTheta}
\end{table}


\begin{table}[ht!]
    \centering
\caption{Comparison of (3D) drop deformation $D$ and orientation angle $\theta$ at time $t = 3~\dot{\gamma}^{-1}$, $t = 10~\dot{\gamma}^{-1}$ and $t_f$ (the final instance of simulations, corresponding to the results presented in~\cref{sec:results}). VN systems for $\operatorname{Ca} = 0.24$, $\operatorname{Ca} = 0.4$ and $\operatorname{Ca} = 0.6$ and varying Deborah numbers. $(^*)$ The deformation values do not correspond to the equilibrium state values.}
\begin{tabular}{@{}lllllllll@{}}
\toprule
   & \multicolumn{2}{l}{$t = 3~\dot{\gamma}^{-1}$} & & \multicolumn{2}{l}{$t = 10~\dot{\gamma}^{-1}$} & & \multicolumn{2}{l}{$t_f$} \\ \cmidrule(lr){2-3} \cmidrule(lr){5-6} \cmidrule(l){8-9} 
                          & $D$          & $\theta$        &  & $D$                              & $\theta$ &  & $D$                             & $\theta$ \\ \midrule
VN , $Ca=0.24$ , $De=0$   & 0.2805       & 29.60           &  & 0.2741                           & 29.93     &  & 0.2746                         & 29.95    \\
VN , $Ca=0.24$ , $De=0.4$ & 0.2578       & 32.43           &  & 0.2549                           & 31.03     &  & 0.2548                         & 31.01    \\ 
VN , $Ca=0.24$ , $De=1$   & 0.2550       & 34.27           &  & 0.2473                           & 32.99     &  & 0.2464                         & 32.68    \\
VN , $Ca=0.24$ , $De=2$   & 0.2577       & 34.08           &  & 0.2473                           & 33.00     &  & 0.2467                         & 33.02    \\
VN , $Ca=0.24$ , $De=4$   & 0.2604       & 33.24           &  & 0.2516                           & 34.34     &  & 0.250                          & 34.37    \\
VN , $Ca=0.24$ , $De=8$   & 0.2633       & 33.68           &  & 0.2556                           & 34.15     &  & 0.2512                         & 34.42    \\
VN , $Ca=0.24$ , $De=16$  & 0.2652       & 33.56           &  & 0.2588                           & 33.99     &  & 0.2519                         & 34.44    \\
VN , $Ca=0.4$ , $De=0$    & 0.4543       & 22.68           &  & 0.5088                           & 18.40     &  & 0.5081                         & 18.42    \\
VN , $Ca=0.4$ , $De=0.4$  & 0.4385       & 24.62           &  & 0.4491                           & 21.74     &  & 0.4478                         & 21.79    \\
VN , $Ca=0.4$ , $De=1$    & 0.4388       & 25.84           &  & 0.4198                           & 24.16     &  & 0.4205                         & 24.08    \\
VN , $Ca=0.4$ , $De=2$    & 0.4465       & 26.48           &  & 0.4201                           & 26.41     &  & 0.4235                         & 25.14    \\
VN , $Ca=0.4$ , $De=4$    & 0.4554       & 26.07           &  & 0.4414                           & 25.38     &  & 0.4447                         & 25.25    \\
VN , $Ca=0.4$ , $De=8$    & 0.4622       & 25.70           &  & 0.4663                           & 24.25     &  & 0.4624                         & 24.45    \\
VN , $Ca=0.4$ , $De=16$   & 0.4662       & 25.48           &  & 0.4878                           & 23.31     &  & 0.4878                         & 23.94    \\
VN$^*$ , $Ca=0.6$ , $De=0$    & 0.572        & 20.55           &  & 0.7771                           & 10.25     &  & 0.8512                         & 6.51    \\
VN$^*$ , $Ca=0.6$ , $De=0.4$  & 0.5516       & 21.57           &  & 0.7374                           & 11.20     &  & 0.8577                         & 5.42    \\
VN , $Ca=0.6$ , $De=1$    & 0.5550       & 22.42           &  & 0.6700                           & 13.08     &  & 0.8030                         & 7.62    \\
VN , $Ca=0.6$ , $De=2$    & 0.5670       & 22.78           &  & 0.6984                           & 13.70     &  & 0.6276                         & 15.43    \\
VN$^*$ , $Ca=0.6$ , $De=4$    & 0.5803       & 22.99           &  & 0.7375                           & 12.70     &  & 0.6748                         & 14.32    \\
VN$^*$ , $Ca=0.6$ , $De=8$    & 0.5904       & 22.45           &  & 0.7807                           & 11.49     &  & 0.7527                         & 9.83    \\
VN$^*$ , $Ca=0.6$ , $De=16$   & 0.6018       & 21.83           &  & 0.8108                           & 10.77     &  & 0.8492                         & 7.85    \\ \bottomrule
\end{tabular}
\label{tab:VN_DandTheta}
\end{table}

\section{Figures for drop deformation and orientation}\label{sec:appendix-figures}
\setcounter{figure}{0}

\begin{figure}[h!]
    \centering
    \begin{minipage}{0.49\textwidth}
        \begin{subfigure}[b]{\textwidth}
            \caption{VN,$\; \operatorname{Ca} = 0.24,\, 0.6$.}
            \includegraphics[width=\textwidth]{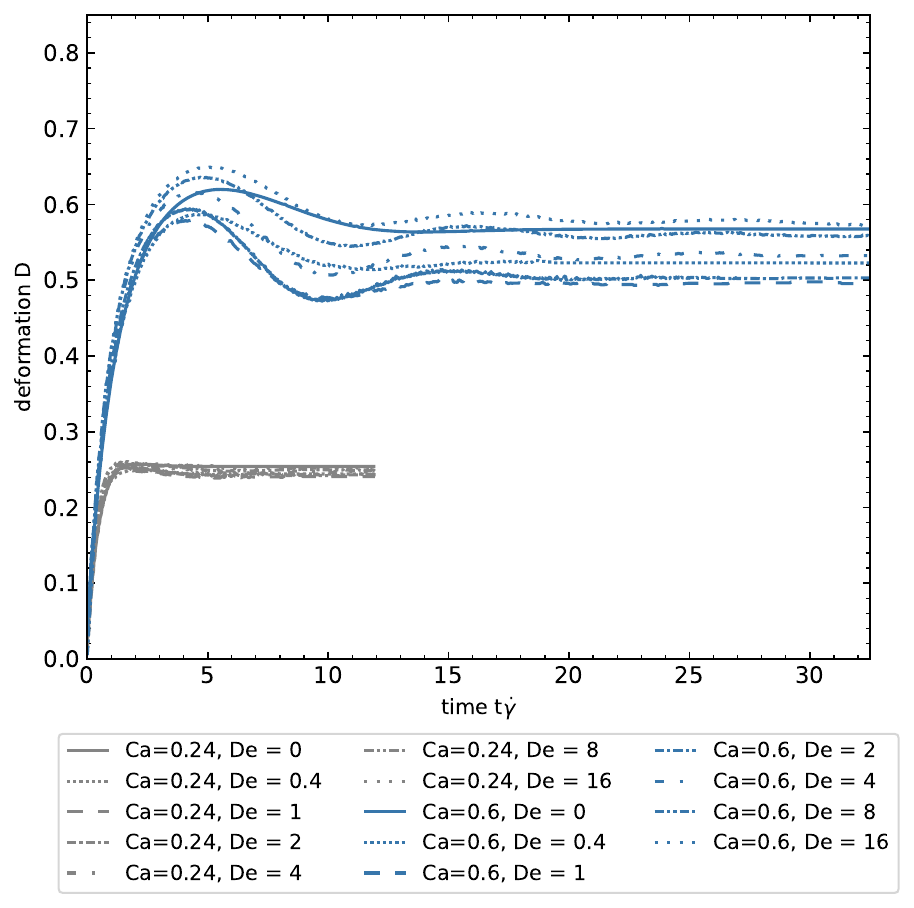}
            \label{subfig:2D_VN_ext}
        \end{subfigure}
    \end{minipage}\hfill
    \begin{minipage}{0.49\textwidth}
        \begin{subfigure}[b]{\textwidth}
            \caption{NV,\; $\operatorname{Ca} = 0.24,\, 0.6$.}
            \includegraphics[width=\textwidth]{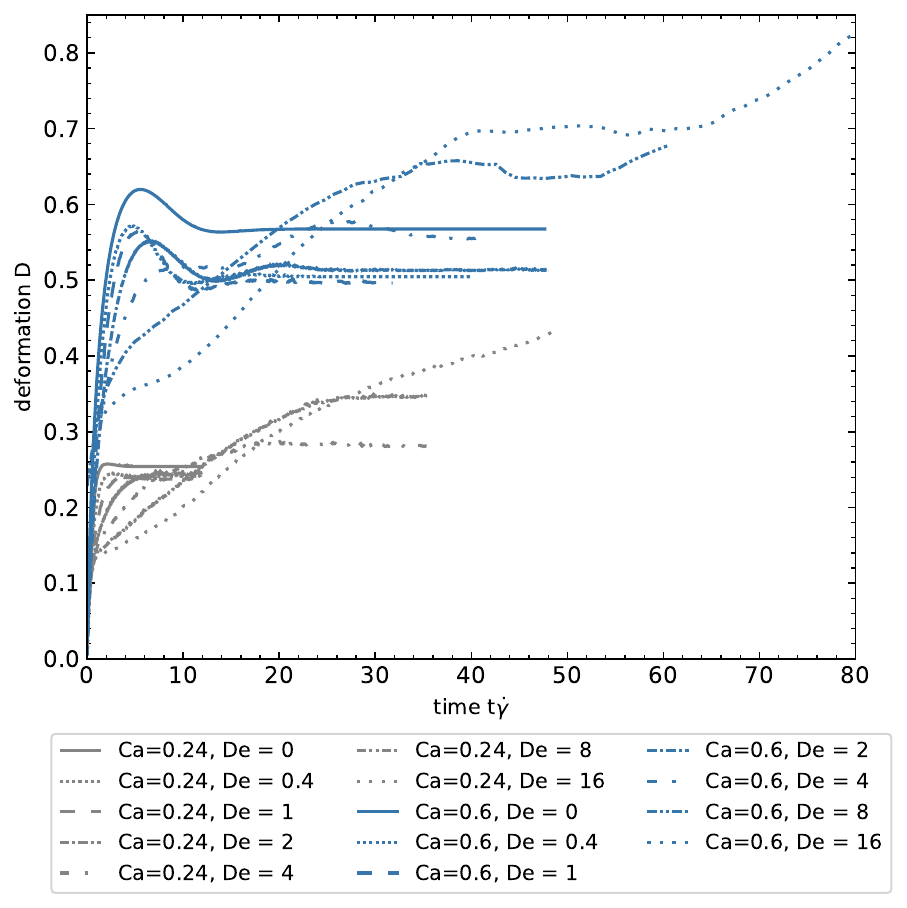}
            \label{subfig:2D_NV_ext}
        \end{subfigure}
    \end{minipage}
    \begin{minipage}{0.49\textwidth}
        \begin{subfigure}[b]{\textwidth}
            \caption{VN,$\; \operatorname{Ca} = 0.24,\, 0.6$.}            \includegraphics[width=\textwidth]{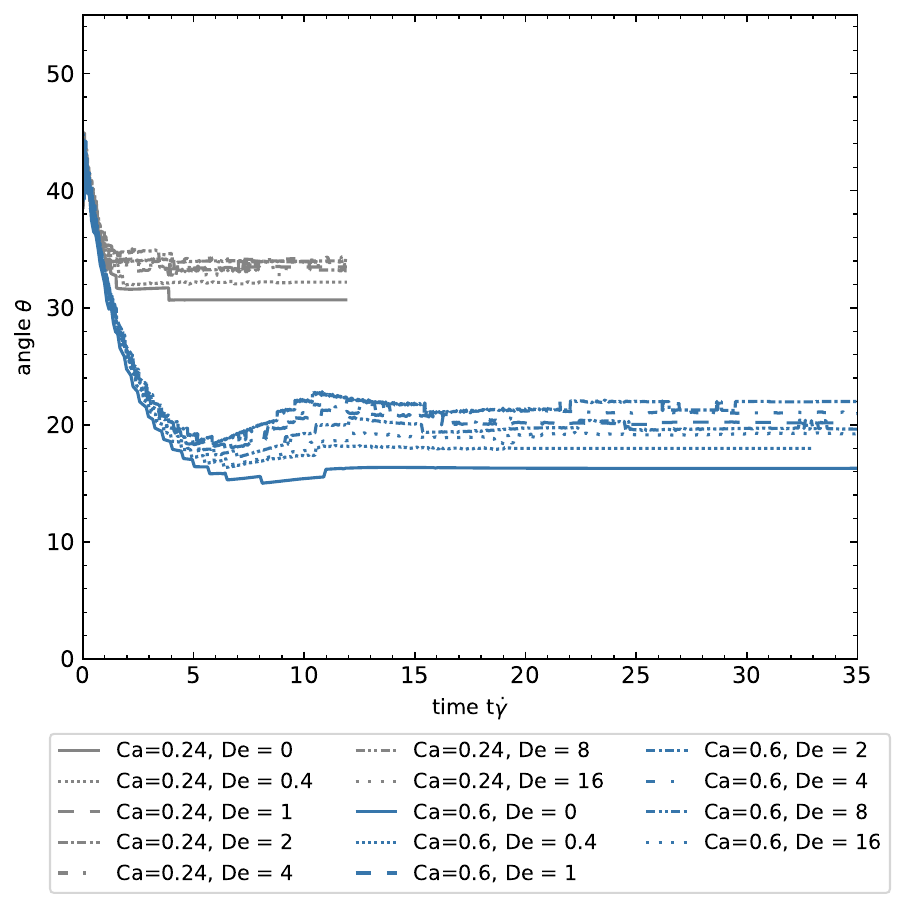}
            \label{subfig:2D_VN_angle_ext}
        \end{subfigure}
    \end{minipage}\hfill
    \begin{minipage}{0.49\textwidth}
        \begin{subfigure}[b]{\textwidth}
            \caption{NV,$\; \operatorname{Ca} = 0.24,\, 0.6$.}
            \includegraphics[width=\textwidth]{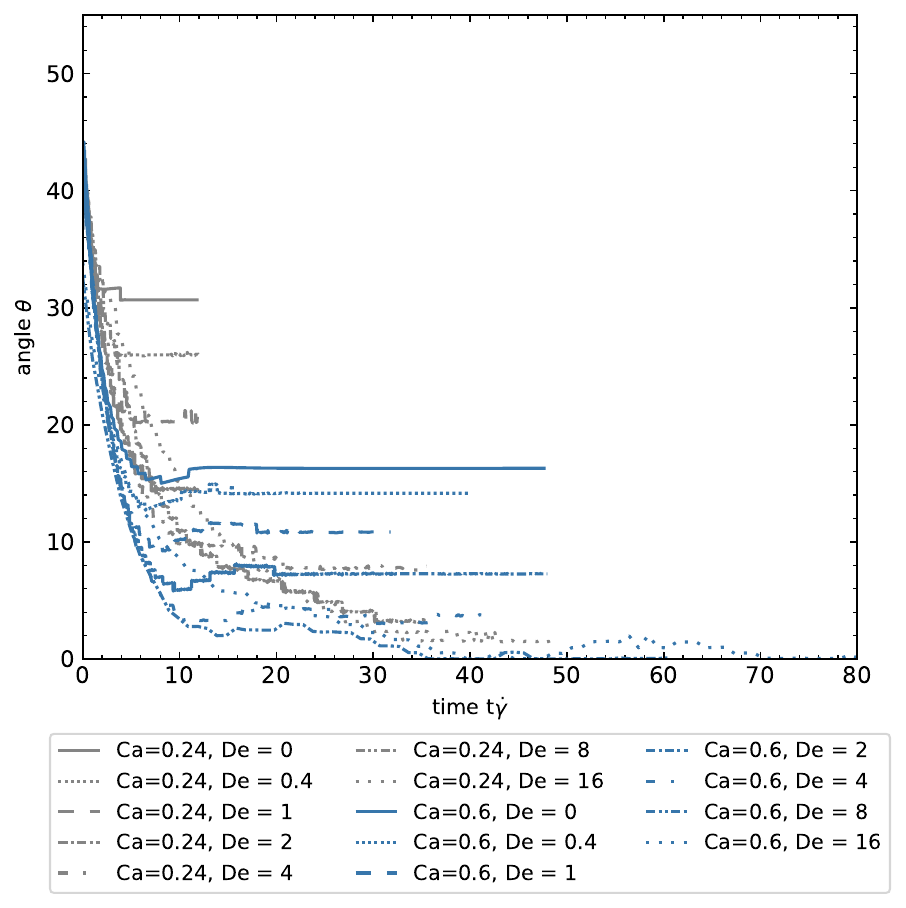}
            \label{subfig:2D_NV_angle_ext}
        \end{subfigure}
    \end{minipage}
    \caption{Drop deformation parameter $D$ and orientation angle $\theta$ as a function of time for the NN, VN, and NV systems at $\operatorname{Ca}=0.24$ and $\operatorname{Ca}=0.6$ and varying Deborah numbers.}
    \label{fig:deformationTransient_ext}
\end{figure}
%
\begin{figure}[h!]
    \centering

    \begin{minipage}{0.49\textwidth}
        \begin{subfigure}[b]{\textwidth}
            \caption{VN,$\; \operatorname{Ca} = 0.24$.}
            \includegraphics[width=\textwidth]{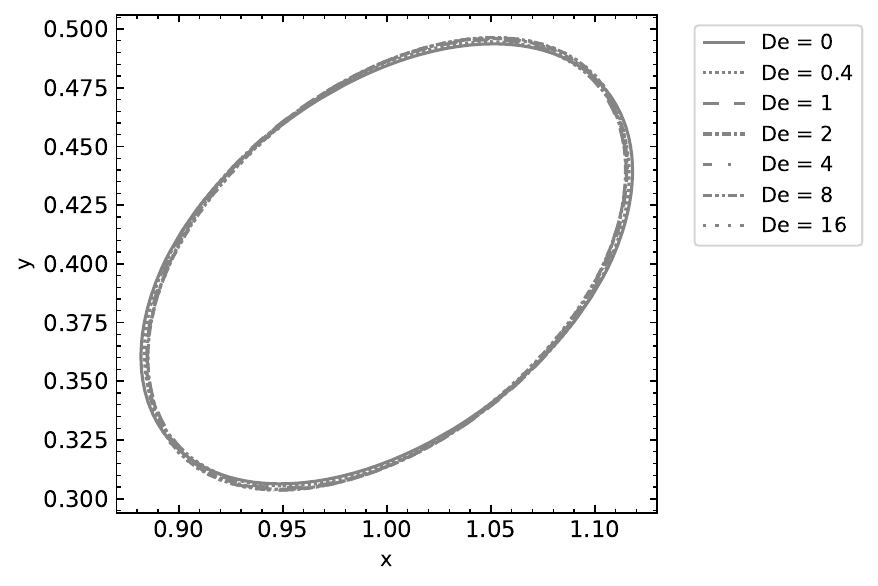}
            \label{subfig:2D_VN_1_tshape_ext}
        \end{subfigure}
    \end{minipage}\hfill
    \begin{minipage}{0.49\textwidth}
        \begin{subfigure}[b]{\textwidth}
            \caption{NV,$\; \operatorname{Ca} = 0.24$.}
            \includegraphics[width=\textwidth]{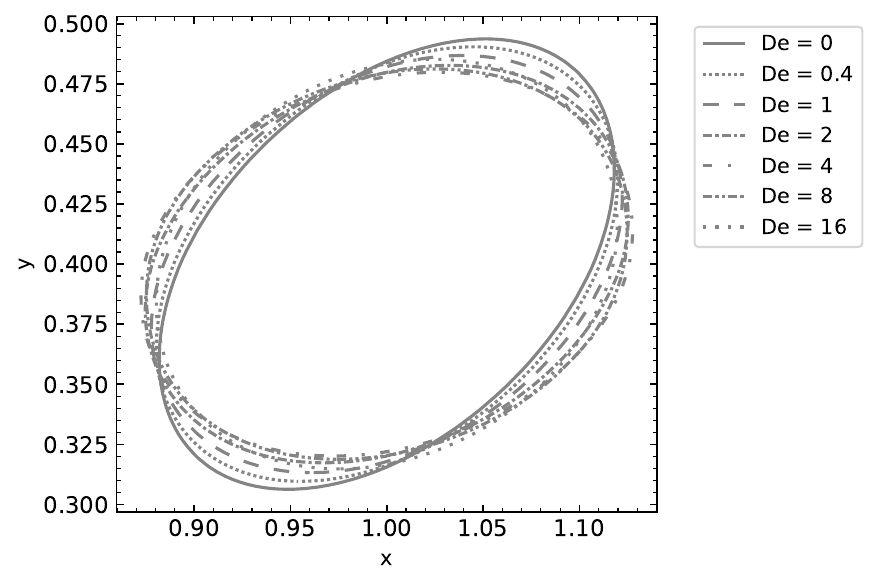}
            \label{subfig:2D_NV_1_tshape_ext}
        \end{subfigure}
    \end{minipage}

    \begin{minipage}{0.49\textwidth}
        \begin{subfigure}[b]{\textwidth}
            \caption{VN,$\; \operatorname{Ca} = 0.6$.}
            \includegraphics[width=\textwidth]{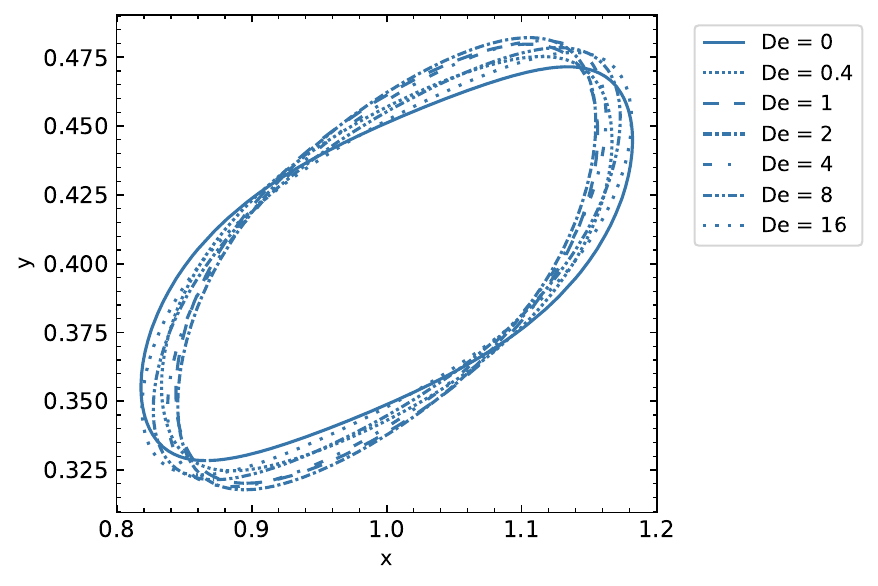}
            \label{subfig:2D_VN_2_tshape_ext}
        \end{subfigure}
    \end{minipage}\hfill
    \begin{minipage}{0.49\textwidth}
        \begin{subfigure}[b]{\textwidth}
            \caption{NV,$\; \operatorname{Ca} = 0.6$.}
            \includegraphics[width=\textwidth]{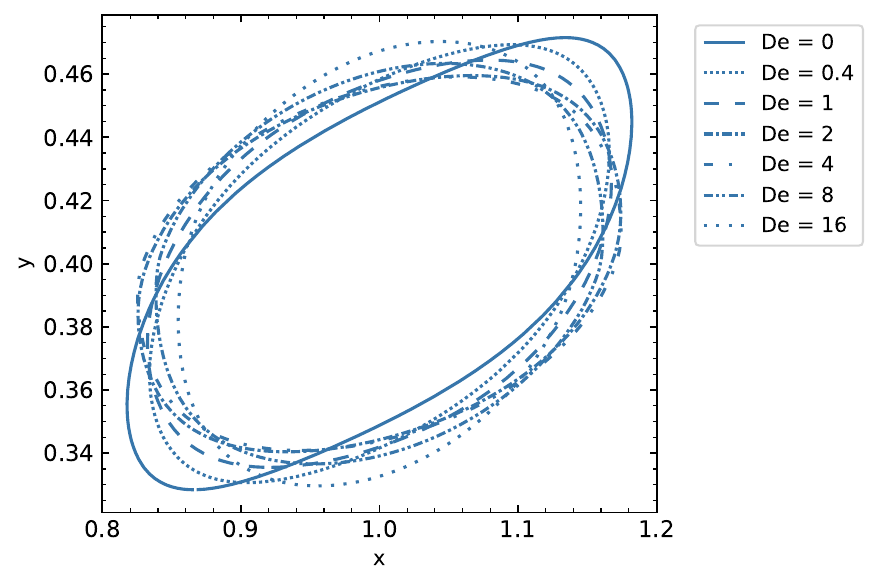}
            \label{subfig:2D_NV_2_tshape_ext}
        \end{subfigure}
    \end{minipage}

    \caption{Drop shapes at $t=10\,\dot{\gamma}^{-1}$.}
    \label{fig:dropShapes_ext}
\end{figure}
%
\begin{figure}[h!]
    \centering

    \begin{minipage}{0.49\textwidth}
        \begin{subfigure}[b]{\textwidth}
            \caption{VN,$\; \operatorname{Ca} = 0.24$.}
            \includegraphics[width=\textwidth]{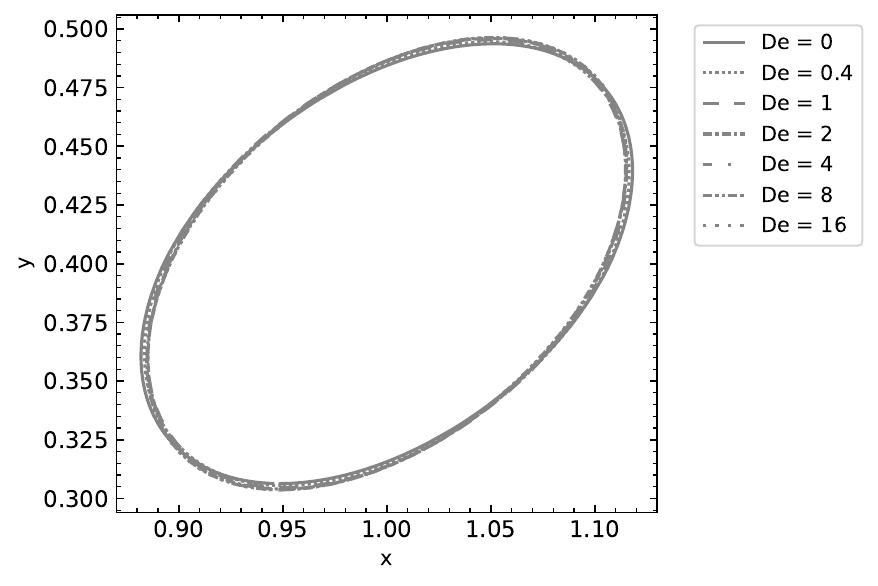}
            \label{subfig:2D_VN_1_eqshape_ext}
        \end{subfigure}
    \end{minipage}\hfill
    \begin{minipage}{0.49\textwidth}
        \begin{subfigure}[b]{\textwidth}
            \caption{NV,$\; \operatorname{Ca} = 0.24$.}
            \includegraphics[width=\textwidth]{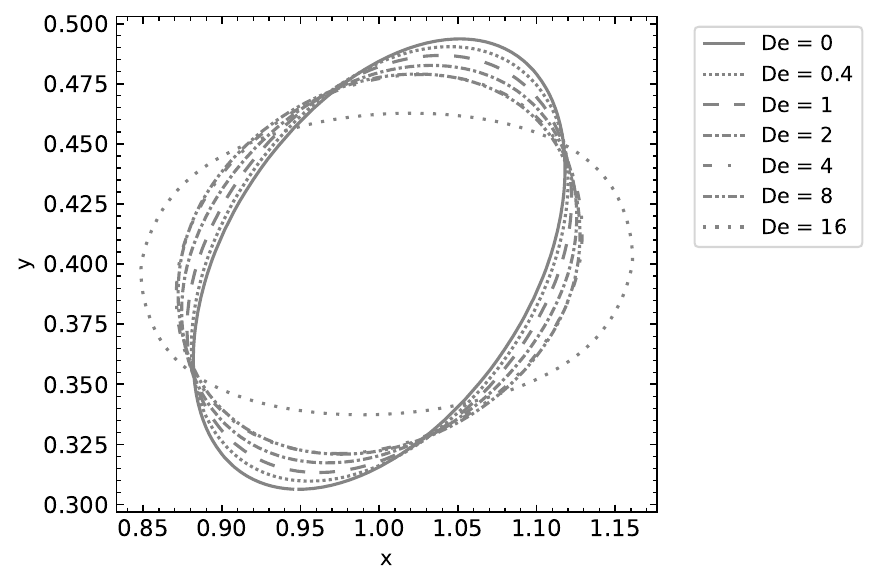}
            \label{subfig:2D_NV_1_eqshape_ext}
        \end{subfigure}
    \end{minipage}

    \begin{minipage}{0.49\textwidth}
        \begin{subfigure}[b]{\textwidth}
            \caption{VN,$\; \operatorname{Ca} = 0.6$.}
            \includegraphics[width=\textwidth]{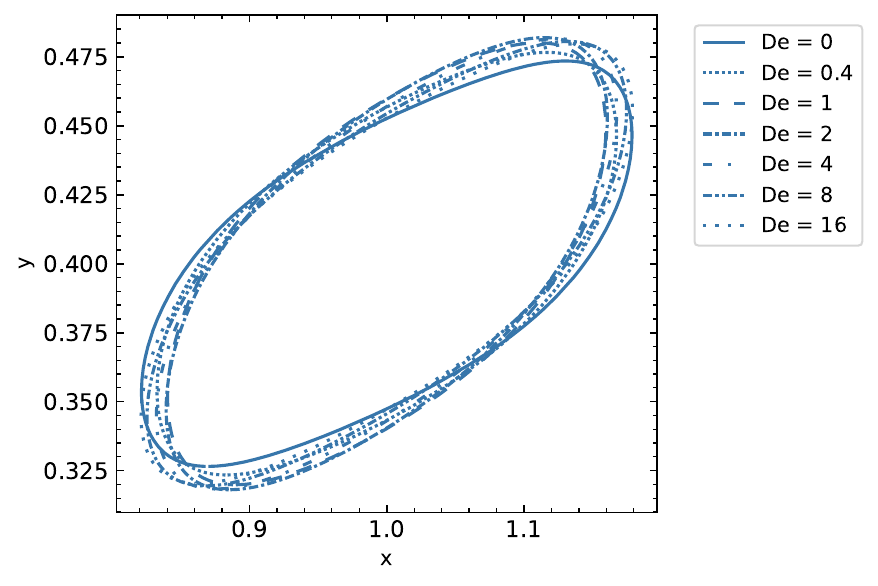}
            \label{subfig:2D_VN_2_eqshape_ext}
        \end{subfigure}
    \end{minipage}\hfill
    \begin{minipage}{0.49\textwidth}
        \begin{subfigure}[b]{\textwidth}
            \caption{NV,$\; \operatorname{Ca} = 0.6$.}
            \includegraphics[width=\textwidth]{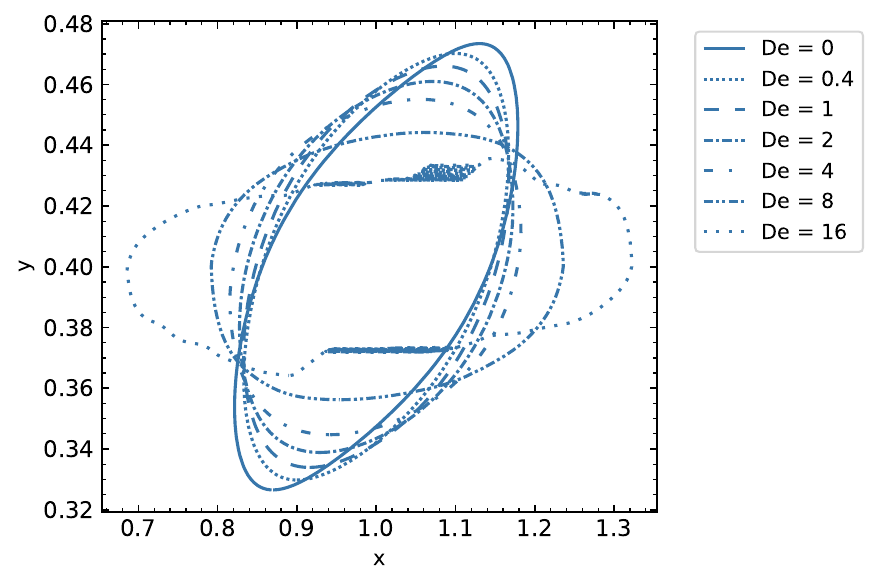}
            \label{subfig:2D_NV_2_eqshape_ext}
        \end{subfigure}
    \end{minipage}

    \caption{Drop shapes at the latest time corresponding to \cref{fig:deformationTransient_ext}.}
    \label{fig:equilibriumDropShape_ext}
\end{figure}
\begin{figure}
    \centering

    \begin{minipage}{0.49\textwidth}
        \begin{subfigure}[b]{\textwidth}
            \caption{VN.}
            \includegraphics[width=\textwidth]{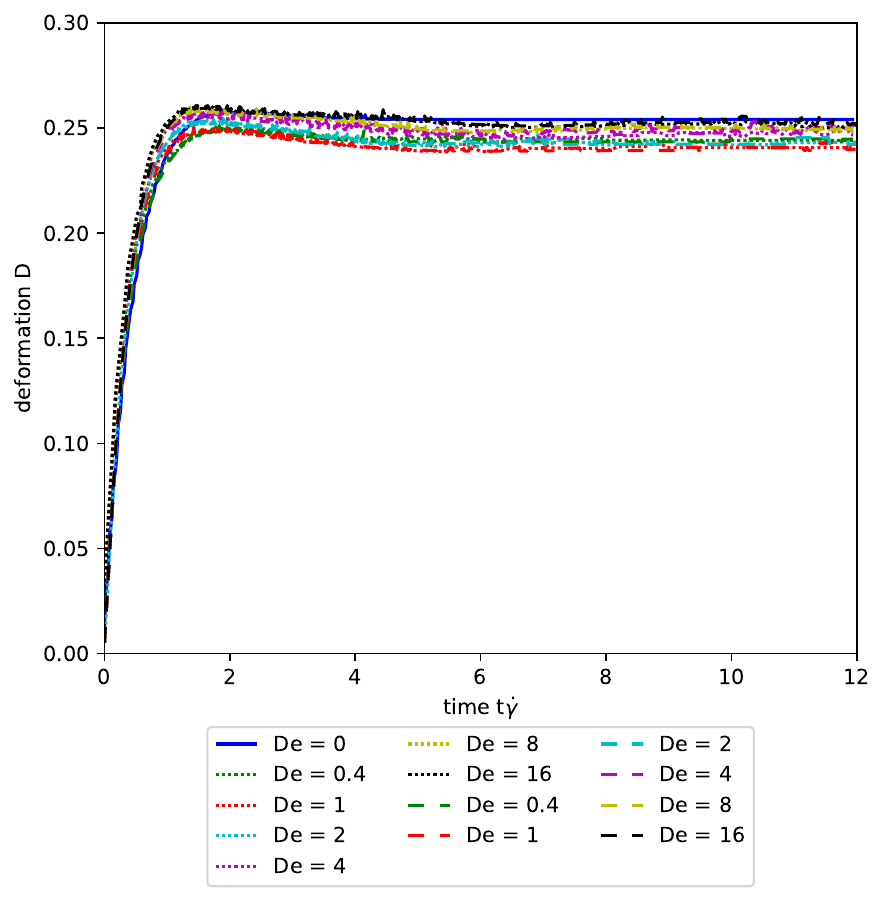}
            \label{subfig:2D_VN_RheoModelComp_ext}
        \end{subfigure}
    \end{minipage}\hfill
    \begin{minipage}{0.49\textwidth}
        \begin{subfigure}[b]{\textwidth}
            \caption{NV.}
            \includegraphics[width=\textwidth]{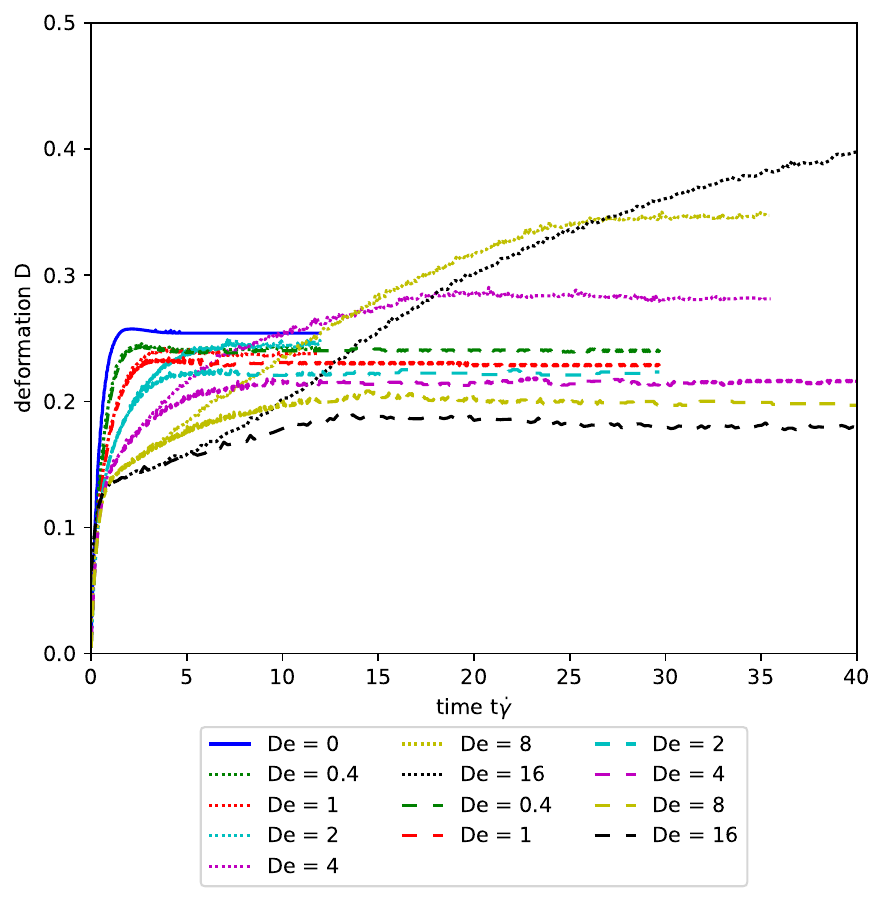}
            \label{subfig:2D_NV_RheoModelComp_ext}
        \end{subfigure}
    \end{minipage}
    \caption{Comparison of rheological models: Oldroyd-B (dotted) and the Giesekus model with $\alpha = 0.03$ (dashed) for droplet deformation as a function of time for NN, VN, and NV systems at $\operatorname{Ca}=0.24$ and varying Deborah numbers.}
    \label{fig:rheoModelComp_ext}
\end{figure}
%
\begin{figure}
    \centering

    \begin{minipage}{0.49\textwidth}
        \begin{subfigure}[b]{\textwidth}
            \includegraphics[width=\textwidth]{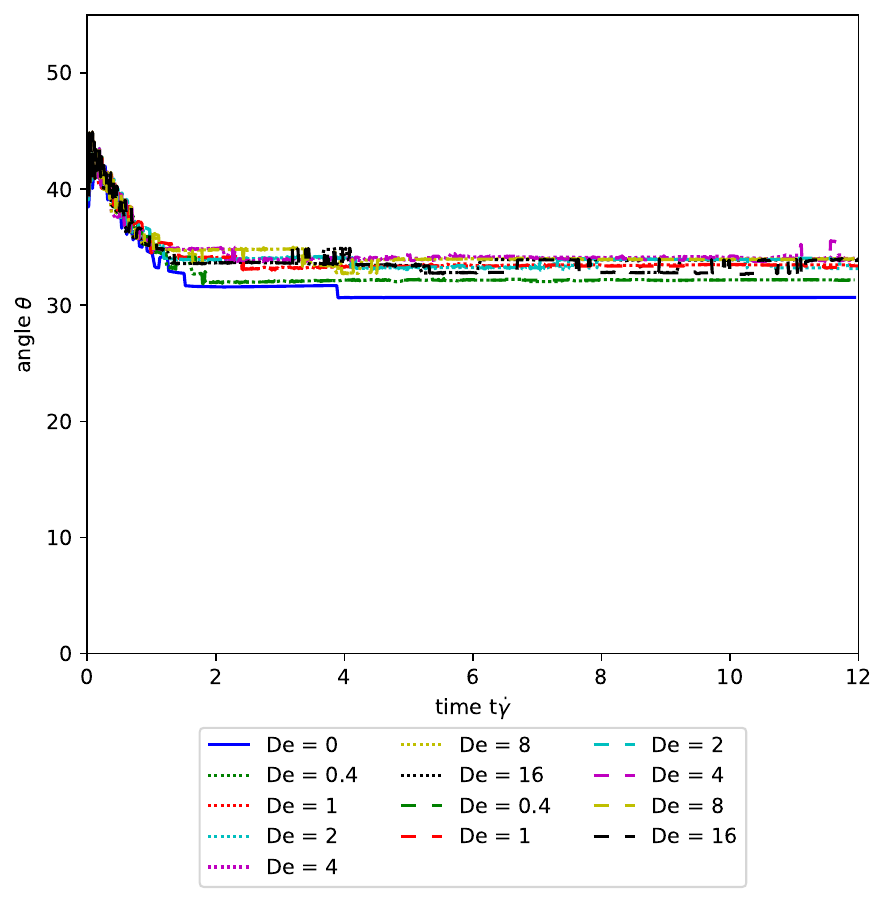}
            \caption{VN.}
            \label{subfig:2D_VN_RheoModelComp_angle_ext}
        \end{subfigure}
    \end{minipage}\hfill
    \begin{minipage}{0.49\textwidth}
        \begin{subfigure}[b]{\textwidth}
            \includegraphics[width=\textwidth]{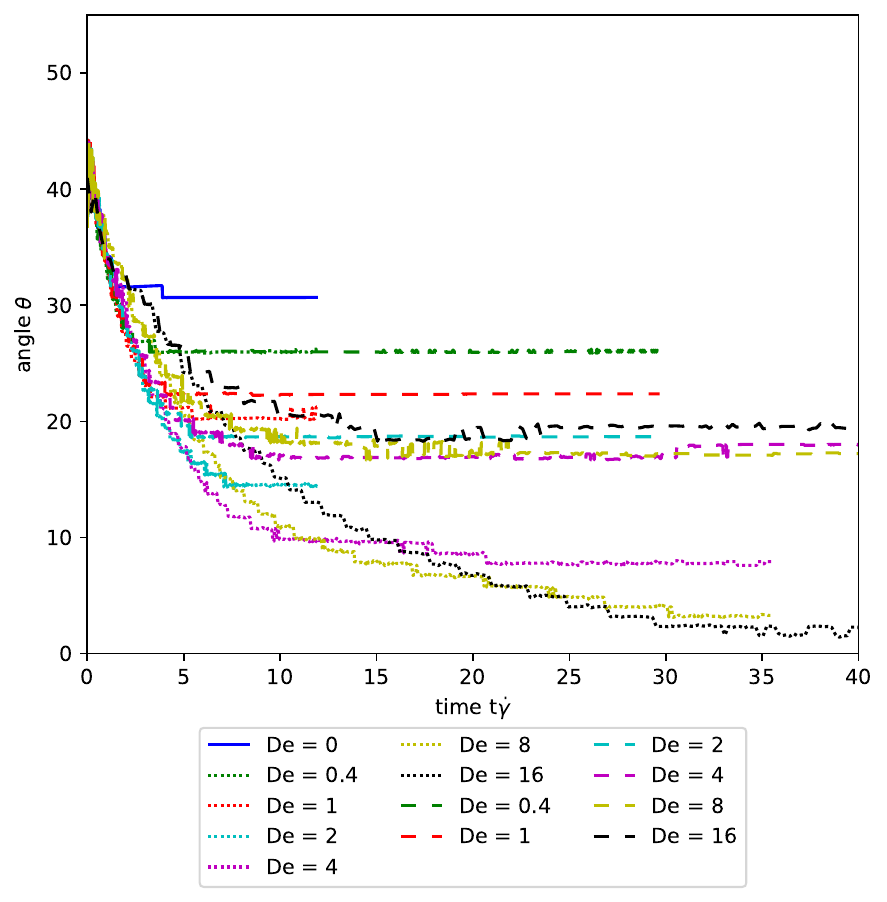}
            \caption{NV.}
            \label{subfig:2D_NV_RheoModelComp_angle_ext}
        \end{subfigure}
    \end{minipage}
    \caption{Comparison of rheological models: Oldroyd-B (dotted) and the Giesekus model (dashed) for droplet deformation angle as a function of time for NN, VN, and NV systems at $\operatorname{Ca}=0.24$ and varying Deborah numbers.}
    \label{fig:rheoModelComp_angle_ext}
\end{figure}
\begin{figure}[h!]
    \centering

    \begin{minipage}{0.49\textwidth}
        \begin{subfigure}[b]{\textwidth}
            \caption{VN.}
            \includegraphics[width=\textwidth]{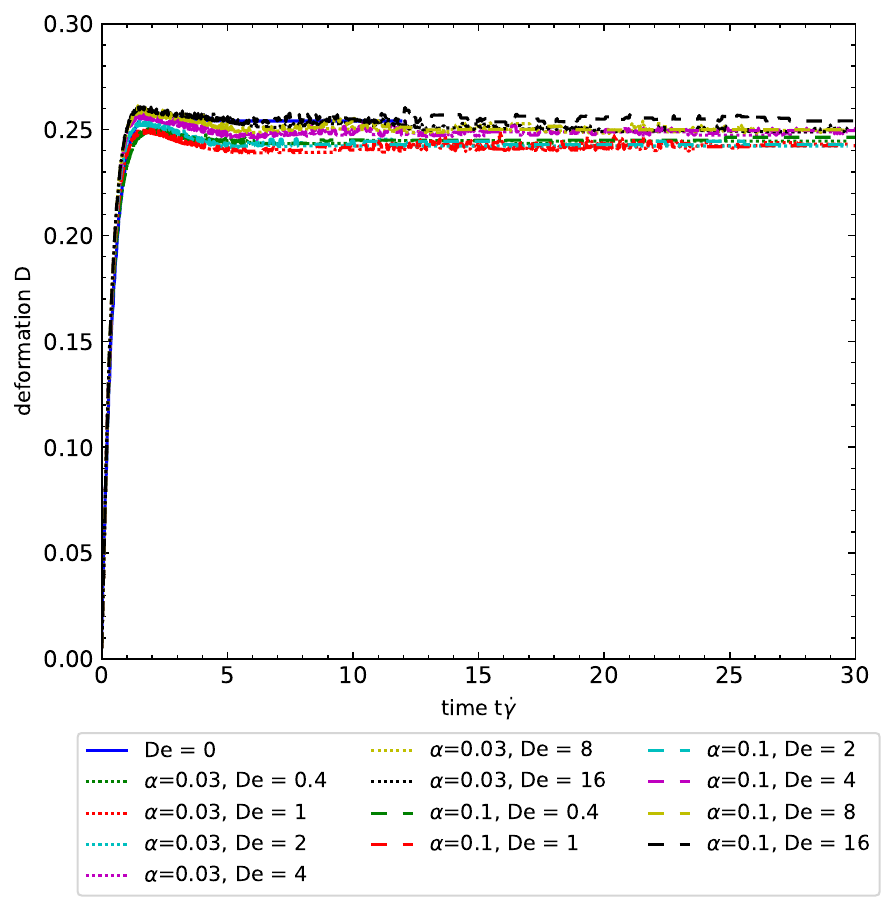}
            \label{subfig:2D_VN_Giesekus_alpha_ext}
        \end{subfigure}
    \end{minipage}\hfill
    \begin{minipage}{0.49\textwidth}
        \begin{subfigure}[b]{\textwidth}
            \caption{NV.}
            \includegraphics[width=\textwidth]{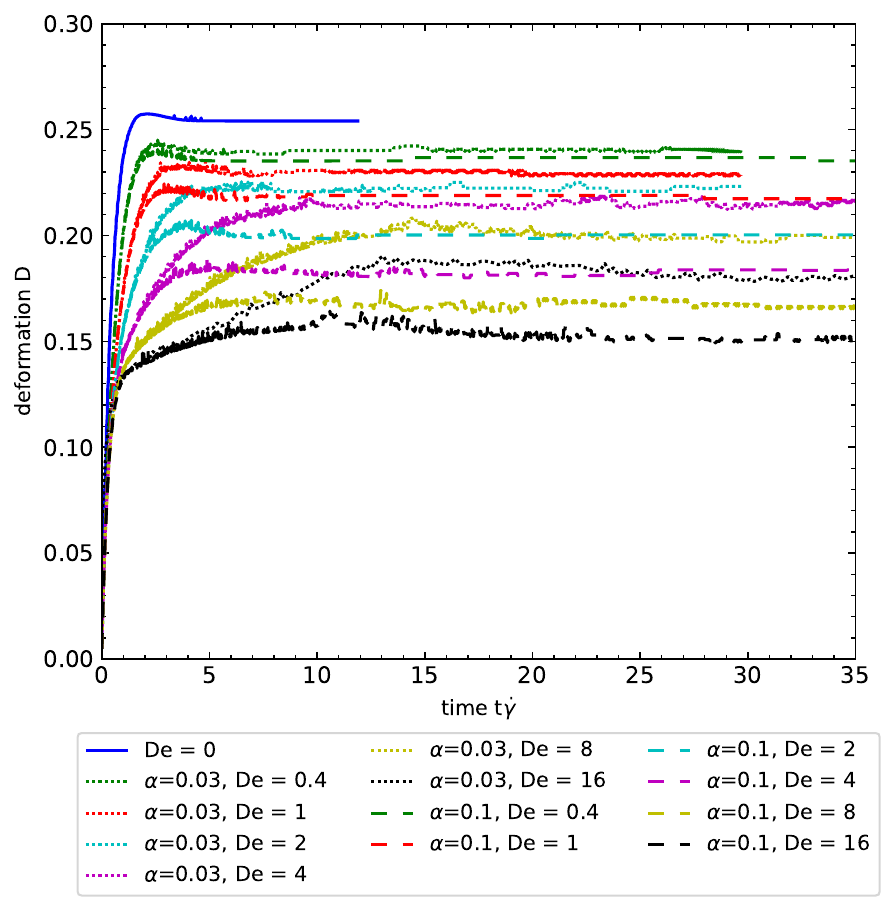}
            \label{subfig:2D_NV_Giesekus_alpha_ext}
        \end{subfigure}
    \end{minipage}
    \caption{Drop deformation as a function of time for NN, VN, and NV at $\operatorname{Ca}=0.24$. Comparison of the Giesekus fluid for the model parameters $\alpha = 0.03$ (dotted) and $\alpha = 0.1$ (dashed) at varying Deborah numbers.}
    \label{fig:giesekus_model_ext}
\end{figure}
\begin{figure}[h!]
    \centering

    \begin{minipage}{0.49\textwidth}
        \begin{subfigure}[b]{\textwidth}
            \caption{VN.}
            \includegraphics[width=\textwidth]{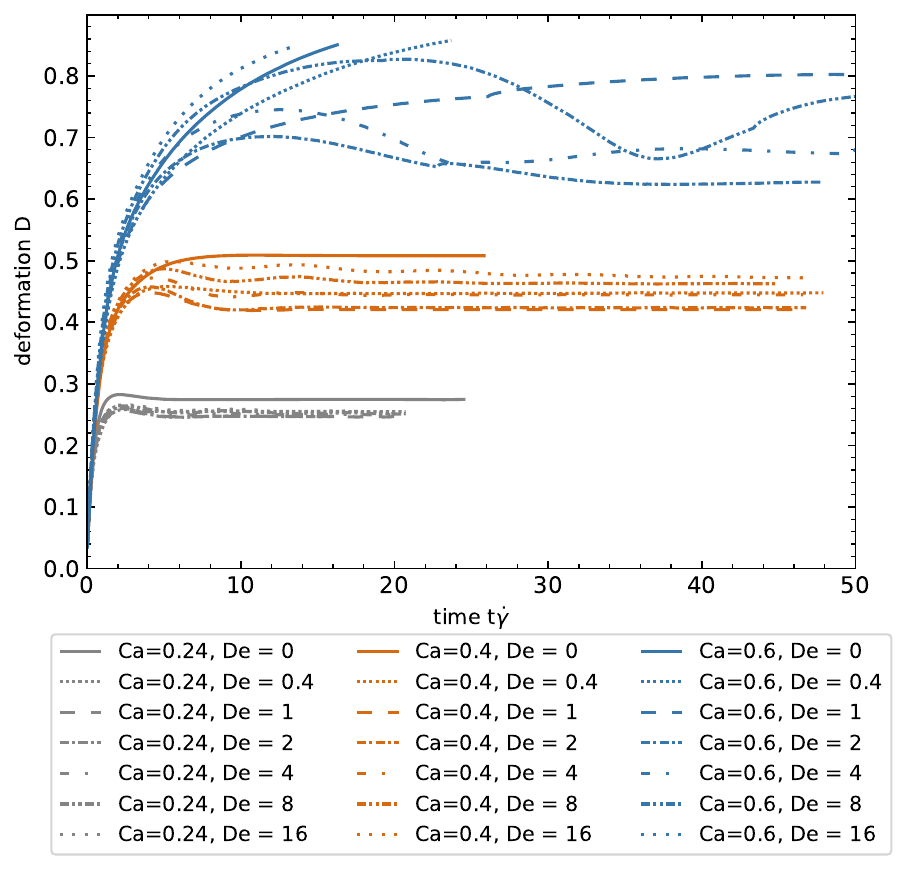}
            \label{subfig:3D_VN_ext}
        \end{subfigure}
    \end{minipage}\hfill
    \begin{minipage}{0.49\textwidth}
        \begin{subfigure}[b]{\textwidth}
            \caption{NV.}
            \includegraphics[width=\textwidth]{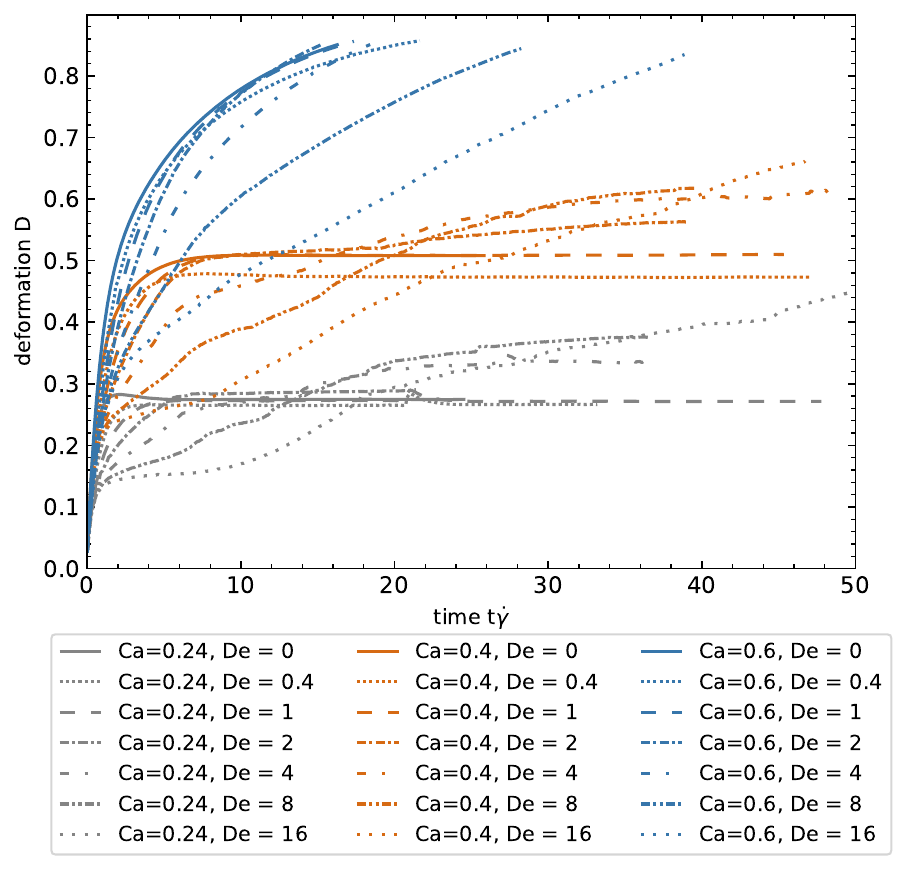}
            \label{subfig:3D_NV_ext}
        \end{subfigure}
    \end{minipage}
    
    \caption{Three-dimensional droplet deformation as a function of time for NN, VN, and NV systems for varying Capillary numbers and Deborah numbers. Some of the simulations were terminated before $t\dot{\gamma} = 50$ because: (i) at $\operatorname{Ca} = 0.6$, the drops are strongly deformed and the computational domain's length constraint with periodic boundaries prevents running the simulations any further without risk of drop coalescence, or (ii) a steady state deformation was reached.}
    \label{fig:deformationTransient3D_ext}
\end{figure}
\begin{figure}[h!]
    \centering

    \begin{minipage}{0.49\textwidth}
        \begin{subfigure}[b]{\textwidth}
            \caption{VN.}
            \includegraphics[width=\textwidth]{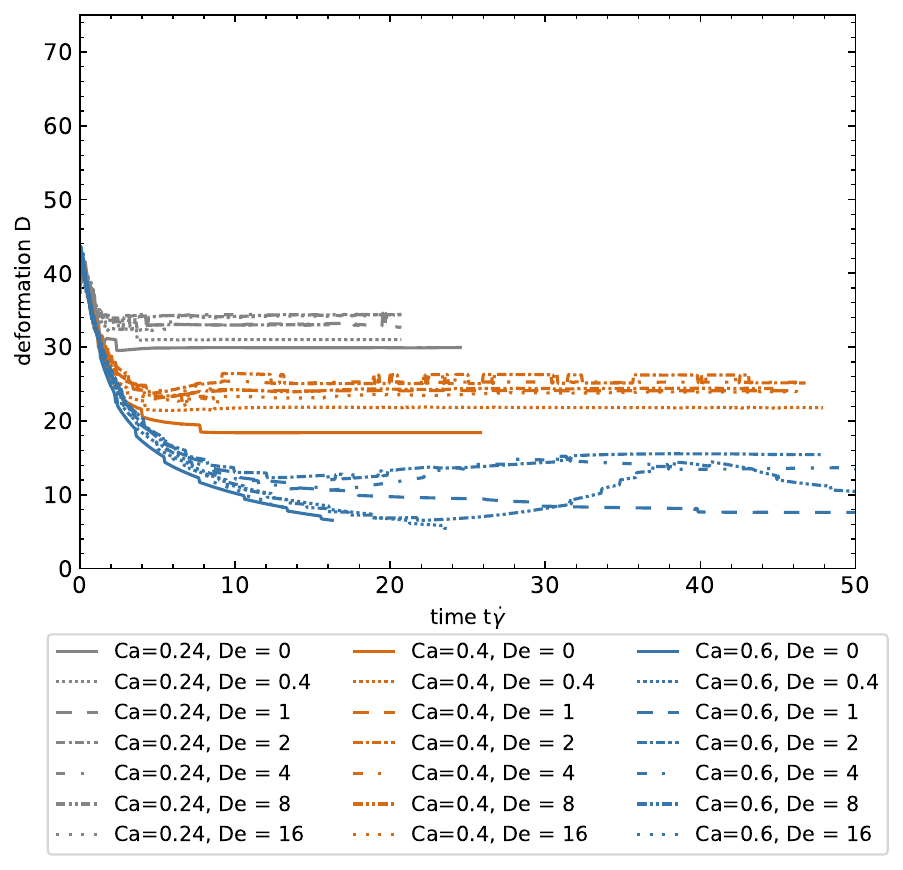}
            \label{subfig:3D_VN_angle_ext}
        \end{subfigure}
    \end{minipage}\hfill
    \begin{minipage}{0.49\textwidth}
        \begin{subfigure}[b]{\textwidth}
            \caption{NV.}
            \includegraphics[width=\textwidth]{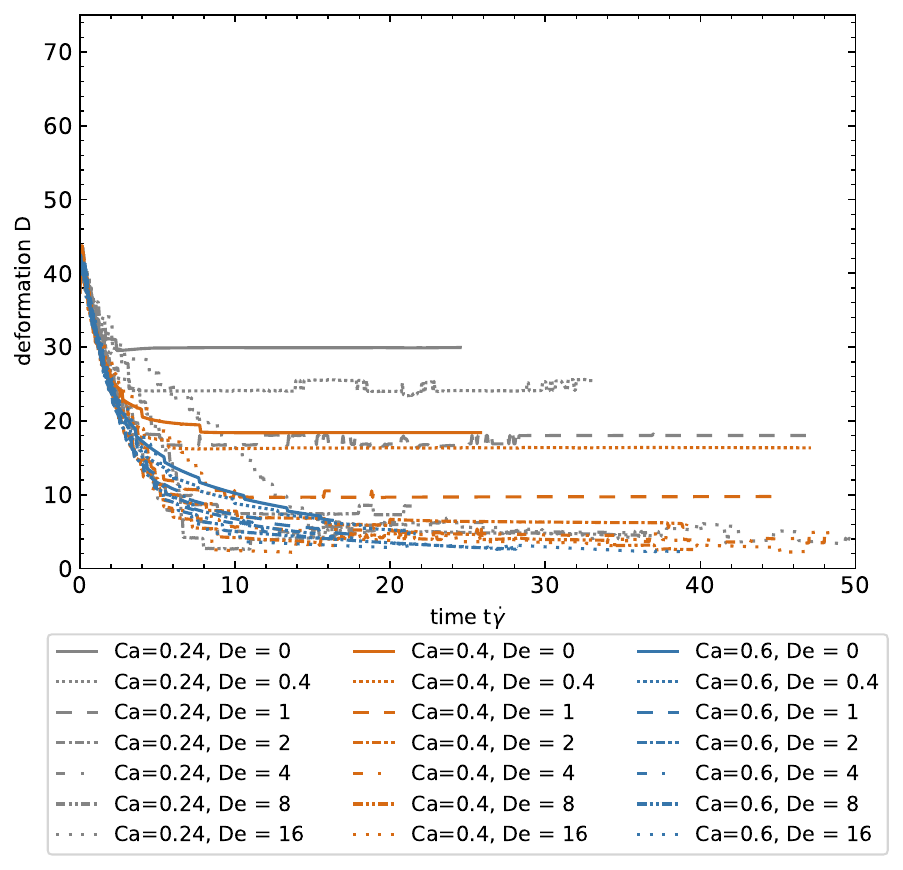}
            \label{subfig:3D_NV_angle_ext}
        \end{subfigure}
    \end{minipage}

    \caption{Three-dimensional droplet orientation angle as a function of time for NN, VN, and NV systems for varying Capillary numbers and Deborah numbers.}
    \label{fig:orientationTransient3D_ext}
\end{figure}

\clearpage

\nocite{*}

\bibliography{bibliography}%

\clearpage

\end{document}